\documentclass[preprint,aps,superscriptaddress,nofootinbib,showkeys,10pt]{revtex4}
\pdfoutput=1
\usepackage{graphicx}
\usepackage{amsmath,calligra,mathrsfs}
\usepackage{amsfonts}
\usepackage{amssymb}
\usepackage{physics}
\usepackage{tabularx}
\usepackage{multirow}
\usepackage{longtable}
\usepackage{array}
\usepackage{float}
\usepackage{xcolor, soul}
\usepackage{epstopdf}
\usepackage{graphicx}
\usepackage{amsmath}
\usepackage{amsfonts}
\usepackage{amssymb}
\usepackage{xcolor, soul}
\usepackage{epstopdf}
\usepackage{float}
\usepackage{subfigure}
\usepackage{float}
\usepackage{subfigure}
\usepackage{hyperref}
\hypersetup{
     colorlinks   = true,
     citecolor    = red,
     linkcolor    = blue,
     urlcolor     = blue,
}\DeclareGraphicsExtensions{.eps}
\def\beq{\begin{equation}}
\def\eeq{\end{equation}}
\def\bea{\begin{eqnarray}}
\def\eea{\end{eqnarray}}
\def\be{\begin{equation}}
\def\ee{\end{equation}}

\def\bse{\begin{subequations}}
\def\ese{\end{subequations}}

\graphicspath{{./figs/}}
\begin{document}
%\preprint{APS/123-QED}

\title{WIMPs, FIMPs, and Inflaton phenomenology via reheating, CMB and $\Delta N_{eff}$  \\
}% Force line breaks with \\
%\thanks{}%

\author{MD Riajul Haque}%
\email{riaj.0009@gmail.com}
\affiliation{Centre for Strings, Gravitation, and Cosmology, Department of Physics, Indian Institute of Technology Madras,
Chennai 600036, India}
\author{Debaprasad Maity}
\email{debu@iitg.ac.in}
\author{Rajesh Mondal}%
\email{mrajesh@iitg.ac.in}
\affiliation{%
	Department of Physics, Indian Institute of Technology Guwahati, Guwahati, Assam 781039, India}%

\date{\today}

\begin{abstract}
In this paper, we extensively analyzed the reheating dynamics after inflation and looked into its possible implication on dark matter (DM) and inflaton phenomenology. We studied the reheating through various possible channels of inflaton going into massless scalars (bosonic reheating) and fermions (fermionic reheating) via non-gravitational and gravity-mediated decay processes. We further include the finite temperature effect on the decay process. Along with their precise roles in governing the dynamics, we compared the relative importance of different temperature-corrected decay channels in the gradual process of reheating depending on the reheating equation of state (EoS), which is directly related to inflaton potential. Particularly, the universal gravitational decay of inflaton is observed to play a very crucial role in the reheating process for a large range of inflaton decay parameters. For our study, we consider typical $\alpha$-attractor inflationary models. We further establish the intriguing connection among those different inflaton decay channels and the CMB power spectrum that can have profound implications in building up a unified model of inflation, reheating, and DM. We analyze both fermion and scalar DM with different physical processes being involved, such as gravitational scattering, thermal bath scattering, and direct inflaton decay. Gravitational decay can again be observed to play a crucial role in setting the maximum limit on DM mass, especially in the FIMP scenario, which has already been observed earlier in the literature\cite{Haque:2021mab}. Depending on the coupling strength, we have analyzed in detail the production of both FIMP and WIMP-like DM during reheating and their detailed phenomenological implications from the perspective of various cosmological and laboratory experiments. 

\end{abstract}
\maketitle
\section{Introduction}
 Reheating is a phenomenon that has been studied quite extensively over the years. It is the phase that bridges the two paradigms of cosmology, namely, inflation \cite{r4,r5,r6} and the standard Big Bang. While inflation sets the uniform initial condition for all the causally disconnected patches of exponentially large homogeneous space of the size of our present universe, the subsequent phase fills the spaces with visible hot matters homogeneously distributed through the process called reheating \cite{r1,r2,r3,b1}. 
In the standard scenario, reheating is the physical process through which the inflaton field decays into the visible matter fields. The endpoint of the reheating is when the standard radiation-dominated universe begins, which sets the proper initial conditions for Big Bang nucleosynthesis (BBN). From these chronological cosmological events, it is obvious that the state of our present universe must be non-trivially dependent on the process of reheating.  Depending on the nature of inflaton and its decay, reheating dynamics can be effectively described by parameters such as the equation of state of inflaton (EoS) and its coupling with other fields.  
Investigation of this phase is still an ongoing effort since its theoretical inception proposed in \cite{r1,r2,r3,b1}.  Since there is no way to directly probe this phase with the present experimental techniques, it is important instead to understand this phase through various possible physical processes and look for direct/indirect process-dependent observables which can be probed directly/indirectly in future experiments. Our present study comes with two main objectives: firstly, study the reheating through various possible decay channels; namely, i) inflaton $(\phi)$ decaying into bosonic radiation through $g^r_1 \phi s^2,g^r_2 \phi^2 s^2 $ interaction, ii) inflaton decaying into fermionic radiation through $h^r \phi f\bar{f}$ interaction, iii) inflaton decaying into all fundamental fields through minimal s-channel graviton exchange interaction, $\mathcal (1/M_P)(h_{\mu\nu}T^{\mu\nu}_\phi+h_{\mu\nu}T^{\mu\nu}_{s/f})$ \cite{Tang:2017hvq,Holstein:2006bh}, and identify the region of their dominance in the parameter space of bosonic coupling $g^r_i$ and fermionic coupling $h^r$. $T^{\mu\nu}$ is the energy-momentum tensor and $h_{\mu\nu}$ is the tensor metric perturbation. To this end, we would like to point out that in the context of gravity-mediated decay, the effect of non-minimal Ricci curvature $(R)$ coupling $\xi \phi^2 R$ has been considered in the reference \cite{Clery:2022wib}. However,  the contribution of such a term has been shown to be negligible for dimensionless coupling $\xi < 1$ and hence will be ignored in this paper. On top of those couplings, we further included the finite temperature effect on the decay process and compared it with the zero-temperature case for all the cases. Our second objective is to study the dark matter (DM) production during reheating considering different physical processes via gravitational scattering, thermal bath scattering, and direct inflaton decay to DM via $g^x_1\phi S^2,g^x_1 \phi^2 S^2$ interactions for scalar DM $S$, and $h^x \phi \bar F F$ interaction for fermionic DM $F$.     

From the phenomenological perspective, DM is assumed to be an integral part of the visible standard model components in quantum field theoretic framework \cite{Arkani-Hamed:2008hhe,Feng:2010gw,Pospelov:2007mp,Tenkanen:2016twd,Heikinheimo:2016hid,Hall:2009bx,Chu:2013jja,Blennow:2013jba,Elahi:2014fsa,Mambrini:2015vna,Nagata:2015dma,Chen:2017kvz,Bernal:2018qlk,Bernal:2018ins,Garcia:2020eof,Garcia:2021gsy,Hochberg:2014dra,Hochberg:2014kqa,Hochberg:2015vrg,Bernal:2015xba,Falkowski:2011xh,LopezHonorez:2010eeh}. In this framework it is just the DM mass and the cross-section which are shown to be sufficient to explain the current abundance of DM. However, a large number of attempts over the years to detect\cite{XENON100:2012itz,XENON:2018voc,PandaX-II:2016vec,LUX:2016ggv,BOSS:2013rlg} such particle appeared to go in vain. Therefore, going beyond the existing framework of both experimental and theoretical approaches to understanding dark matter may seem to be important  \cite{Capozziello:2006dj,Capozziello:2011et,Boehmer:2007kx,Nojiri:2017ncd}. Towards this endeavor recent proposal of graviton mediated DM production  \cite{Donoghue:1994dn,Choi:1994ax,Mambrini:2021zpp,Bernal:2021kaj,Barman:2021ugy, Ema:2015dka,Ema:2016hlw,Ema:2018ucl,Garny:2015sjg,Garny:2017kha} has been shown to have some promising universal features, and it cannot be ignored in any DM studies. Interestingly such gravitational production has been observed to set a limit on the maximum possible value of the DM mass \cite{Haque:2021mab,Clery:2021bwz}. In the standard DM literature, two distinct DM production mechanisms exist in the early universe. For the standard WIMP (Weakly Interacting Massive Particle) scenario, the DM is assumed to be in thermal equilibrium with the radiation bath. During the course of background evaluation, DM particles became thermally decoupled from the bath (known as the Freeze-out mechanism), and the present value of abundance is achieved \cite{fo1,fo2,fo3,fo4,fo5,fo6,fo7,fo8}. In the second type, known as FIMP (Feebly Interacting Massive Particle) scenario, the DM is assumed to remain out of equilibrium with the radiation bath and produced due to decay of other fields throughout. During the course of background evaluation, the decay channel ceases to produce DM at some point (known as the Freeze-in mechanism), and the present value of abundance is achieved \cite{Maity:2018dgy,Haque:2019prw,Bernal:2020qyu,f1,f2,f3,f4,f5,f6}. In this paper, we will explore those mechanisms in the context of the early universe with a non-trivial reheating phase. {\it Apart from understanding the very nature of DM, such studies actually open up the possibilities of looking for the signature of reheating and, most importantly, the nature of inflaton through the physics of DM.} 

Most of the DM phenomenological studies were confined to the early radiation-dominated universe \cite{fo1,fo2,fo3,fo4,fo5,fo6,fo7,fo8}. DM physics during reheating has gained significant interest only recently \cite{Maity:2018dgy,Garcia:2020eof,Garcia:2020wiy,Garcia:2021gsy,Maity:2018exj,Haque:2020zco,Giudice:2000ex,Barman:2022tzk,Bhattiprolu:2022sdd,Harigaya:2014waa,Harigaya:2019tzu,Okada:2021uqk,Ghosh:2022fws}. Primary motivation of this attempt has two-folds: a) to construct a unified framework where inflaton is assumed to be an integral part of DM model building, and b) explore the physics of reheating and its impact on DM physics. Finally, analyze and constrain the inflaton, reheating and DM parameters through the constraint on extra relativistic degrees of freedom in terms of $\Delta N_{\mbox{eff}}$ at the time of Big-Bang Nucleosynthesis (BBN)  \cite{Cyburt:2015mya,Knapen:2017xzo,Nollett:2014lwa,Paul:2018njm}, and different DM searches in the astrophysical/cosmological/laboratory experiments\cite{Leane:2018kjk,Bergstrom:2013jra,Calore:2022stf,Hess:2021cdp,CTA:2020qlo}.         
Keeping those two-fold motivations in mind, we study in detail the parameter space wherein both WIMP and FIMP-type mechanisms can be realized. In this study, we will further see how universal gravitational DM production during the initial stage of the reheating phase plays an important role in constraining the DM parameters.  
WIMP production during reheating has been considered very recently in \cite{Arias:2019uol,Bernal:2022wck}. However, detailed studies taking into account the physics of inflation and the subsequent reheating processes are still lacking. In this paper, we fill this gap and study in detail its constraints and significance in the context of present and future experiments. 

We organize our paper as follows: In section \ref{section2}, we will discuss the basic setup for the perturbative reheating processes for different decay channels and their connection with the inflationary parameters.
As mentioned earlier, we include the effect of finite temperature corrections in the decay widths for various decay channels. In section \ref{section3}, we discuss in detail the reheating dynamics due to two different bosonic decay channels. We identify the parameter regions with respect to reheating EoS, where reheating can be successfully achieved depending upon the strength of the different decay channels under consideration. We further discuss the possible constraints on those inflationary coupling parameters with respect to the CMB observations. In section \ref{bgdfer}, we discuss in detail the fermionic reheating scenario where the reheating occurs due to the inflaton decaying into fermionic radiation. Similar to the bosonic reheating case, we analyze and constrain the inflaton-fermion coupling parameter through the perspective of inflationary observable and CMB constraints.
In section \ref{section5}, We consider various possible scenarios corresponding to DM production. We elaborately discuss DM production from direct inflaton decay and thermal radiation separately. In section \ref{section6} and \ref{section7}, we discuss their possible constraints from the perspective of various theoretical and experimental bounds. Finally, we conclude with some future directions.    

\section{Perturbative reheating: general set up}{\label{section2}}
In the first half of our present paper, we will discuss in detail the single-sector reheating, which has been studied earlier for some special cases \cite{Garcia:2020wiy,Garcia:2021gsy,Garcia:2020eof,Ahmed:2021fvt,Ahmed:2022qeh,Ahmed:2022tfm}. However, we will perform a comprehensive study on this with many significantly new results. In the second half, we include dark matter production and explore different possible production mechanisms and their observable possibilities in a unified framework.

\begin{figure}[t!] 
 	\begin{center}
 		\includegraphics[width=13.0cm,height=2.4cm]{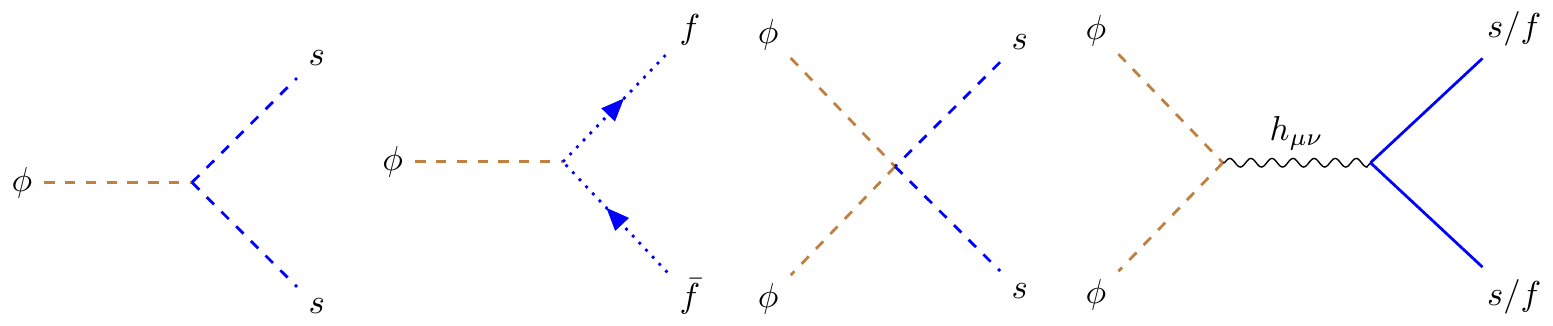}\quad
 	\caption{Fynmann diagram for all possible interactions between inflaton ($\phi$) and radiation (R)}
 		\label{fynmanrad}
 	\end{center}
 \end{figure}

After the end of inflation, the inflaton field starts to oscillate with a decaying amplitude. During this phase, the inflaton field transfers its energy into massless fields termed as radiation ($R$) through various decay channels, $\phi\rightarrow ss/\bar{f}f$ and $\phi\phi\to ss$, where we assume explicit coupling between inflation and daughter fields. In addition, there exists universal gravitational coupling between inflaton and daughter field through s-channel graviton $(h_{\mu\nu})$ exchange interaction, $\mathcal (1/M_P^2)h_{\mu\nu}T^{\mu\nu}$, which has recently been shown to influence the reheating dynamics \cite{Haque:2022kez,Clery:2021bwz}. In our present analysis, we shall include such universal contributions as well. When we talk about  gravitational scattering, we mainly consider the process $\phi\phi\rightarrow h_{\mu\nu}\rightarrow ss/ff$. In Fig.\ref{fynmanrad}, we have shown the Feynman diagram for all possible interactions between inflation ($\phi$), and radiation ($R$). We consider inflaton coupled with scalar $(s)$ and fermion $(f)$ as massless radiation for the case of bosonic and fermionic reheating, respectively. Compared to the direct decay with free coupling parameters, decay due to gravitational interaction is universal and hence will always be present. Setting all the direct inflaton coupling to zero implies purely gravitational scattering, named as gravitational reheating (GR), which has been studied in detail by two of the present authors in \cite{Clery:2021bwz,Haque:2022kez}. Therefore, we consider following general interaction Lagrangian,
\begin{equation}
   \mathcal{L}_{int}=\frac{h_{\mu\nu}}{M_P}(T^{\mu\nu}_\phi+T^{\mu\nu}_{s/f})+g_1^r\phi s^2+g_2^r\phi^2s^2+h^r\phi\bar ff,
\end{equation}
where $g_1^r$ is the coupling parameter for the trilinear interaction with mass dimension unity,  $g_2^r$ is the dimensionless coupling parameter for quartic interaction and $h^r$ is the Yukawa coupling.

%%%%%%%%%%%%%%%%%%%%%%%%%%%%%%%%%%%%%%%%%%%%%
\subsection{Finite temperature decay widths and Boltzmann equation}
%%%%%%%%%%%%%%%%%%%%%%%%%%%%%%%%%%%%%%%%%%%%
The standard assumption of any reheating studies is that radiation is always thermalized among its constituents throughout the entire process. Hence, inflaton decay products must encounter a finite temperature thermal bath which will modify the decay width \cite{Drewes:2015coa,Mukaida:2012qn,Mukaida:2012bz,Drewes:2013iaa}. Such finite temperature effect has already been discussed in \cite{3,Adshead:2019uwj,Drewes:2014pfa} for the special cases with a matter-dominated reheating phase. Our goal is to generalize those studies for the arbitrary inflaton equation of state (EoS) $w_\phi$, which has not been studied earlier. For the bosonic decay of inflaton, the thermal effect introduces Bose enhancement factor into the decay width, which makes bosonic reheating efficient compared to the zero temperature one. On the other hand, for fermionic decay of inflaton, the thermal bath induces an additional Pauli blocking factor into the decay width, which makes fermionic reheating less efficient compared to the zero-temperature one. Including the finite temperature effect, we list up the following decay widths\footnote{In our study, we have not taken the oscillation effect of the zero mode inflaton \cite{Garcia:2020wiy} in the decay width formula. However, even taking the oscillation effect in the production rate, our illustrated results  would be the same, only there would be a minute modification in the fermionic gravitational dark matter production (for detailed discussion, see Appendix-\ref{modifieddecay}).} for various decay/scattering channels as 
\cite{Garcia:2020wiy,Ahmed:2022tfm,Adshead:2019uwj,Adshead:2016xxj,Drewes:2019rxn}
%This effective coupling parameter ($g^r_{eff}/h^r_{eff}$) is a function of the original coupling parameter ($g^r/h^r$) and the EoS $w_\phi$
\begin{eqnarray}{\label{t1}}
&& \Gamma_{s/f} = \left\{\begin{array}{ll} 
\Gamma_{\phi\rightarrow ss}&=\frac{(g^r_1)^2}{8\pi m_\phi(t)}(1+2f_B(m_\phi/2T))\,, ~~~~~~~\mbox{for}~~ g^r_1 \phi s^2\\
\Gamma_{\phi\phi\rightarrow ss}&=\frac{(g^r_2)^2\rho_\phi(t)}{8\pi m^3_\phi(t)}(1+2f_B(m_\phi/T))\,, ~~~~~~~\mbox{for}~~ g^r_2 \phi^2 s^2\\
\Gamma_{\phi\rightarrow \bar f f}&=\frac{(h^r)^2}{8\pi}m_\phi(t)(1-2f_F(m_\phi/2T))\,, ~~~\mbox{for}~~ h^r \phi \bar f f\\
\end{array} \right. \\
&& \Gamma^{gr}_{\phi\phi\rightarrow{ss}} = \frac{\rho_\phi m_\phi}{1024\pi M^4_p}(1+2f_B(m_\phi/T))\,,~~~\mbox{\cite{Barman:2021ugy,Ahmed:2022tfm}} \\
%\left(1+\frac{m^2_b}{2m^2_\phi}\right)\sqrt{1-\frac{m^2_b}{2m^2_\phi}}\\
&& \Gamma^{gr}_{\phi\phi\rightarrow{ff}}=\frac{\rho_\phi m^2_f}{4096\pi M^4_pm_\phi}(1-2f_F(m_\phi/T))\,,~~~\mbox{\cite{Barman:2021ugy,Ahmed:2022tfm}}
\end{eqnarray}
where $f_{B/F}(z)=\frac{1}{e^{z}\mp 1}$ are the equilibrium Bose-Einstein (B)$(-)$ and Fermi-Dirac (F)$(+)$ distribution function. The last two decay width expressions are the gravity-mediated inflaton decay to other fields. In these expressions, we ignore the effect of thermal mass correction due to self-interaction. $m_\phi(t)$ corresponds to time-dependent inflaton mass defined as $m^2_\phi(t)= \partial^2 V(\phi)/\partial\phi^2$ for a generic inflaton potential $V(\phi)$. It is obvious from the expression that when the temperature of the radiation bath is greater than the inflaton mass, i.e. ($ T_{rad}> m_\phi(t)$), the Bose-enhancement or Pauli-blocking is effective. Due to the Bose enhancement, the decay rate is enhanced for the Bosons, and due to Pauli blocking, the decay rate is suppressed for the Fermions.\\
%%%%%%%%%%%%%%%%%%%%%%%%%%%%%%%%%%%%%%%%%%%%%%%%%
Since the radiation particles are massless, the total decay width for the gravitational sector is mainly associated with the scalar particles. For fermionic particles, the decay width for gravitational interaction $\propto  m^2_f$, and that is the reason for not taking the contribution from the fermionic particles in the thermal bath for gravitational production. In addition, in the case of vector boson production, we need to introduce the mass term to break the conformal invariance; thus, massless vector boson production is impossible. Therefore, the total gravitational scattering rate to radiation production will be simply $\Gamma^{gr}_{\phi\phi\rightarrow RR}=\Gamma^{gr}_{\phi\phi\rightarrow ss}$. One important point is to note that we consider only one type of scalar particle in the radiation bath, and all other particles are assumed to be fermions or vector bosons, so their gravitational production is unimportant. It has already been established that for inflaton equation of states $w_\phi > 0.65$, the gravitational scattering can alone reheat the universe (see, for instance, reference \cite{Haque:2022kez}). 
In this paper, we will include the explicit inflaton coupling with radiation and do a comprehensive combined analysis to figure out the complete parameter space for successful single-sector reheating. 
The general set of Boltzmann equations for single sector reheating are  \cite{Giudice:2000ex,Chung:1998rq}
%%%%%%%%%%%%%%%%%%%%%%%%%%%%
\begin{eqnarray}
&& \dot{\rho_\phi}+3H(1+w_\phi)\rho_\phi+(\Gamma_{s/f}+\Gamma^{gr}_{\phi\phi\rightarrow{RR}})(1+w_\phi)\rho_\phi=0 ,  \label{B1}\\
&&\dot{\rho}^r_{s/f}+4H\rho^r_{s/f}-\Gamma_{s/f}(1+w_\phi)\rho_\phi=0\, \label{B2}\\
&&\dot{\rho}_{gr}^{r}+4H\rho_{gr}^{r}-\Gamma^{gr}_{\phi\phi\rightarrow{RR}}(1+w_\phi)\rho_\phi=0 . \label{B3}
\end{eqnarray}
%%%%%%%%%%%%%%%%%%%%%%%%%%%
Where $\rho_\phi$ is the inflaton energy density. $\rho^r_{s/f}$ corresponds to radiation energy density from direct inflaton decay to either scalar $(s)$ or fermion $(f)$ via the coupling parameters $(g_i^r,h^r)$ respectively. $\rho^r_{gr}$ is the radiation energy density produced through universal gravitational scattering. For solving this set of equations numerically, we define dimensionless comoving variables $ \Phi={\rho_\phi A^{3(1+w_\phi)}}/{(m^{end}_\phi)^4}$, and $R_{rad}={\rho_{rad} A^4}/{(m^{end}_\phi)^4}$. Where $R_{rad}$ is the radiation produced from either direct decay or gravitational scattering from inflaton. The derivative is taken with respect to the cosmic time defined through the Friedmann–Lemaitre–Robertson–Walker (FLRW) metric $ds^2 = -dt^2 + a(t)^2(dx^2 + dy^2 + dz^2)$. The normalized scale factor is defined as $A={a}/{a_{end}}$, with $a_{end}$ as the scale factor at the end of inflation, and the Hubble expansion
parameter $H = {\dot A}/A$.  At the end of inflation i.e., $A=1$, $\rho_\phi={3 V(\phi_{end})}/{2}$ and the energy densities of all the other components are set to be, $\rho^r_{s/f/gr}=0$. Hence the appropriate initial condition for the above set of Boltzmann equations is,
\begin{equation}\label{incond}
\Phi(A=1)=\frac{3}{2}\frac{V(\phi_{end})}{(m^{end}_\phi)^4}~~;~~
R_b(A=1)=R_f(A=1)=R_{gr}(A=1)=0\,.
\end{equation}
Where $(\phi_{end}, m_{\phi}^{end})$ are the values of the inflaton field and its mass at the end of  inflation, which we define later.
%%%%%%%%%%%%%%%%%%%%%%%%%%%%%%%%%%%%%%%%%%%%% 
 \subsection{Relating reheating and inflationary parameters through CMB}
The connection between inflation and reheating parameters are established through the initial conditions Eq.\ref{incond}. We consider the well known $\alpha$-attractor $E$-model \cite{Kallosh:2013hoa,Ferrara:2014cca,Ueno:2016dim} inflaton potential,
 \bea
\label{pot1}
V(\phi)=\Lambda^4\left(1-e^{-\sqrt{\frac{2}{3\alpha}}\frac{\phi}{M_p}}\right)^{2n}\,,
\eea
where $\Lambda$ is the mass-scale fixed by the CMB power spectrum, which is typically of the order $8\times 10^{15}$ GeV, and the parameter $(\alpha,n) $ controls the shape of the potential. Through out our analysis we have taken $\alpha=1$, although our analysis is not much sensitive within the allowed range of $\alpha$ values from Planck and BICEP/$Keck$ combined results \cite{Ellis:2021kad,Chakraborty:2023ocr}. Using the slow roll parameters $\epsilon(\phi)=({M^2_p}/{2})({V (\phi)^\prime}/{V(\phi)})^2$ and $\eta(\phi)=M^2_p ({V (\phi)^{\prime\prime}}/{V (\phi)})$, one obtains physically measurable quantities namely scalar spectral index $(n_s)$ and the tensor to scalar ratio $(r)$ defined at a particular pivot scale $k$,
\begin{equation}{\label{eq18}}
n_s=1-6\epsilon(\phi_k)+2\eta(\phi_k)\quad;\quad\quad r=16\,\epsilon (\phi_k) \,.
\end{equation}
Another important quantities are inflationary e-folding number $(N_k)$ and the Hubble constant $H_k$ defined for a particular pivot scale as,  
\begin{equation}{\label{hk}}
\begin{aligned}
&N_k=\int^{a_{end}}_{a_k} d(\ln a)=\frac{1}{M_p}\int^{\phi_{end}}_{\phi_{k}}\frac{d\phi}{\sqrt{2\epsilon}}\,,\\
& H_k= \sqrt{\frac {V(\phi_k)}{3 M_p^2}} = \frac{\pi M_p\sqrt{r\,A_s}}{\sqrt 2}\,,
\end{aligned}
\end{equation}
where $A_s$ represents the amplitude of the inflation fluctuation. Throughout our analysis, we assume the central value of $A_s\sim 2.1\times 10^{-9}$ from Planck \cite{Planck:2018jri}. Here $\phi_k$ and $\phi_{end}$ are the values of the inflaton field at the point of horizon crossing for a particular pivot scale and at the end of the inflation, respectively. $\phi_{end}$ is obtained from the ending condition of the inflation $\epsilon(\phi_{end}) =1$. The field and the potential value at the end of inflation take the following form
\begin{equation}{\label{phiend}}
 \phi_{end}=\frac{\sqrt{3\alpha}}{2n}M_p\ln{\left(\frac{2n+\sqrt{3\alpha}}{\sqrt{3\alpha}}\right)}\quad,\quad V(\phi_{end})=\Lambda^4\left(\frac{2n}{2n+\sqrt{3\alpha}}\right)^{2n} .
\end{equation}
Combining Eqs.\ref{eq18}-\ref{phiend}, one can find the field value $\phi_k$ and the mass scale $\Lambda$ as,
\begin{eqnarray}
 \phi_k &=&\sqrt{\frac{3\alpha}{2}}M_p \ln{\left[1+\frac{4n+\sqrt{16n^2+24\alpha n(1-n^k_s)(1+n)}}{3\alpha(1-n^k_s)}\right]}, \\
    \Lambda&=&M_p\left(\frac{3\pi^2rA_s}{2}\right)^{1/4}\left[\frac{2n(2n+1)+\sqrt{4n^2+6\alpha(1+n)(1-n_s)}}{4n(1+n)}\right]^{n/2} .
 \end{eqnarray}
 After the end of the inflation, inflaton field oscillates around its minima, and the reheating phase begins. At the minimum of the potential, we expand the inflaton potential (Eq.\ref{pot1}) in the limit $\phi\ll M_p$ as
\begin{equation}
 V(\phi)\simeq\Lambda^4\beta^{2n}\phi^{2n},
\end{equation}
where $\beta=\sqrt{{2}/{3\alpha M_p^2}}$. The field-dependent mass becomes, 
\begin{equation}
 m^2_{\phi}=V^{''}(\phi_0)\simeq2n(2n-1)\Lambda^4\beta^2\left(\frac{{}V(\phi_0)}{\Lambda^4}\right)^{1-\frac{1}{n}} .
\end{equation}
Using the envelope approximation at any instant of time, the envelope value of the field $\phi_0$ must satisfy $V(\phi_0)\simeq \rho_\phi(t)$ \cite{Garcia:2020wiy}. Under this approximation, the inflaton mass can be written as
\begin{equation}
m^2_{\phi}(t)=2n(2n-1)\beta^2\Lambda^{4/n}\rho_\phi(t)^{\frac{n-1}{n}}.
\end{equation}
Using the virial theorem, one can further calculate the equation of states (EoS) $w_\phi$ as a function of the power of the inflation potential \cite{Garcia:2020wiy}
\begin{eqnarray}
    w_\phi=\frac{n-1}{n+1}.
\end{eqnarray}

Through the background dynamics and entropy conservation, we can connect the inflationary parameters defined above with the reheating parameters. 
%%%%%%%%%%%%%%%%%%%%%%%%%%%%%%%%%%%%%%%%%%
The reheating period is effectively described by very few parameters viz. reheating temperature ($T_{re}$), reheating e-folding number ($N_{re}$), and the equation of state  of inflation $\omega_{\phi}$. In general, the end of reheating is marked at the point when the condition $\rho_\phi(A_{re})=\rho_r(A_{re})$ is satisfied. Where $\rho_r$ is the total radiation density constituted of all massless daughter fields.
% This condition is equivalent to $\Gamma_{\phi}(A_{re})= H({A_{re}})$ for $w_\phi<\frac{1}{3}$. For $w_\phi<\frac{1}{3}$, reheating is never happened if  $\Gamma_{\phi}(A_{re})= H({A_{re}})$ condition is not satisfied but this is not necessary for  $w_\phi>\frac{1}{3}$. This is because, when $w_\phi>\frac{1}{3}$, the inflaton energy density $\rho_\phi$ is decays faster than radiation density $\rho_r$, so we will definitely get a point where $\rho_\phi(A_{re})=\rho_r(A_{re})$ is satisfied. So in our case, the reheating condition is 
%\begin{equation}{\label{rc1}}
% \rho_{\phi}(A_{re})=\rho_r({A_{re}})
%\end{equation}
The reheating temperature $T_{re}$ can be expressed in terms of the radiation temperature $T_{rad}$ as
\begin{equation}{\label{tt1}}
 T_{re}=T_{rad}\,(A_{re})=\left(\frac{30}{\pi^2g_\star(T_{re})}\right)^{\frac{1}{4}}\rho^{1/4}_r(A_{re}),
\end{equation}
where $g_\star(T_{re})$ is the effective number of relativistic degrees of freedom at the end of reheating, and we take $g_\star(T_{re})=100$ though out the paper. Combining the above two equations, we can get the one-to-one correspondence between the coupling constant ($g^r_i$,$h^r$) and reheating temperature $T_{re}$, where $i= 1,2$ corresponding two different inflaton-Boson couplings described in the introduction.
One can further obtain a constraint relation between  reheating temperature $T_{re}$ and the present CMB temperature $(T_0)$ by considering a physical assumption that from the end of reheating to the present time, the co-moving entropy density of the universe is conserved. This condition leads to \cite{Cook:2015vqa},
\begin{equation}{\label{rhc}}
 T_{re}=\left(\frac{43}{11 g_\star({T_{re}})}\right)^{\frac{1}{3}}\left(\frac{a_0T_0}{k}\right)H_k e^{-N_k}e^{-N_{re}}. 
\end{equation}
 Where $k/a_0=0.05~ \mbox{Mpc}^{-1}$ is the CMB pivot scale, the present CMB temperature $T_0=2.725$ K and $a_0$ is the present day scale factor. Combining the above Eq.\ref{rhc} and \ref{tt1}, we can essentially obtain an indirect connection between the coupling constant ($g^r_i$,$h^r$) and the inflationary spectral index $n_s$, which parameterizes the anisotropies in the CMB fluctuations. In the next section, we moved our discussion to the possible constraints on the perturbative reheating scenario from the fragmentation of the inflaton that leads to self-resonance. \\
 %%%%%%%%%%%%%%%%%%%%%%%%%%%%%%
 \section{The effect of self-resonance in reheating parameter space}{\label{sreffect}}
Homogeneous oscillation of the inflation condensate can be unstable, leading to self-resonance. In Figs. \ref{couplingbound} and \ref{boundfermion}, we have shown the region in the $w_\phi$ versus coupling parameter space where self-resonance may be important. It has been pointed out that self-resonance can be sufficient to reheat the universe (except $w_\phi=0$) without any coupling to the other fields with inflaton \cite{Lozanov:2016hid,Lozanov:2017hjm}, however, that strictly depends on the values of $\alpha$. When $\alpha<<1/6$, self-resonance is efficient, and the RD universe is established within less than an e-fold of expansion after inflation end. However, for $\alpha>1/6$, self-resonance is not very efficient, and it takes many e-folds to give rise to radiation dominated universe. In the reference \cite{Lozanov:2017hjm}, the authors provide an estimation for the number of e-folds calculated from the end of inflation to the beginning of the radiation domination for the $\alpha$-attractor E-model,
\begin{eqnarray}{\label{npr1}}
    \Delta N_{sr}
    %ln\left(\frac{a_{br}}{a_{end}}\right)=ln (A_{br})
    \simeq&\left\{\begin{array}{ll}
         & \frac{n+1}{3} \ln\left[\frac{\sqrt{6\alpha}}{2d\delta}\frac{k}{\Delta k}\frac{\lvert 4-2n\rvert}{n+1}\right]~~~~\mbox{for}~~n>1~(n\neq 2) \\
         & \ln\left[\frac{\sqrt{6\alpha}}{2d\delta}\right]~~~\mbox{for}~~n=2 ,
    \end{array}\right.
\end{eqnarray}
where $\Delta k/k$ is the fractional width of the resonance band, $d$ is its dimensionless "strength" and  $\delta=0.126$ is a dimensionless number which is independent of $n$. This bound can be used as an upper bound on the transition duration between the inflation end and the radiation-dominated states of the universe.\\
Self-resonance can significantly modify the dynamics for which a comprehensive analysis is required. Here, we have shown the region where it can play its role. If the perturbative reheating is completed with e-folding number $<  \Delta N_{sr}$, our analysis will not be affected. Otherwise, self-resonance will significantly affect the parameter space, which needs to be taken into account, and we defer this for our future study. Using the bound from Eq. \ref{npr1}, in Figs. (\ref{couplingbound}), (\ref{Fig5}), (\ref{boundfermion}), and (\ref{Fig10}) we have shown regions where the self-resonance effect is important. We have found that self-resonance is effective when $0<w_\phi\leq0.75$ for $\alpha=1$.
\section{Bosonic Reheating: Dynamics and constraints }{\label{section3}}
\subsection{Dynamics: probing different decay channels}
During reheating, if the dominant decay channel of the inflaton is through massless Bosons, we call it bosonic reheating. For our phenomenological purpose, we discuss two different non-gravitational bosonic channels $\phi\to ss$ and $\phi\phi\to ss$ of inflaton along with gravitational scattering process $\phi\phi\rightarrow h_{\mu\nu}\rightarrow ss$.
The total radiation component will be composed of all these thermalized decay products produced from different decay channels. 
Most of the studies on this single sector reheating were done at zero temperature, except for a few very special cases \cite{3,Adshead:2019uwj}. Therefore, a more realistic consideration would be to take into account finite temperature correction due to radiation baths. We compare the results with the zero temperature case in all the figures to quantify such effects. The detailed analytic calculation for both zero and finite temperature cases and the description of the dynamics are given in Appendix-\ref{bosonanawth}. 

\begin{figure}[t]\centering
          \includegraphics[width=17.00cm]
         % {Scalarboundfinal.pdf}
        {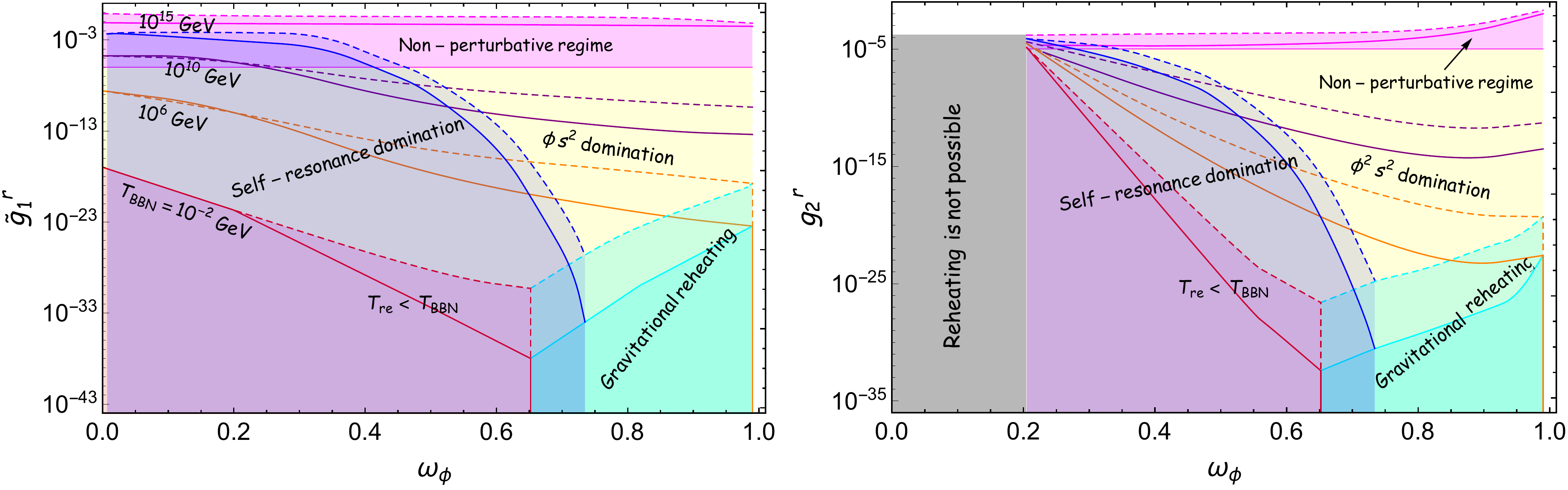}
          \caption{ We have plotted the Variation of the dimensionless bosonic coupling $\tilde{g}^r_1=g^r_1/{m^{end}_\phi}$ (for $\phi s^2$ left figure) and $g^r_2$ (for $\phi^2 s^2$ right figure) as a function of $w_\phi$. Dashed and solid lines correspond to without and with the thermal effect being taken into account in the decay process respectively.  The yellow and pink shaded regions indicate the explicitly coupling-dominated region where the decay channel controls the reheating temperature. The light-cyan region corresponds to gravitational reheating. The light-red region corresponds to $T_{re}<T_{BBN}\simeq 10$ MeV. The light-gray region corresponds to the no reheating region where inflaton energy density falls faster than radiation energy density, and successful reheating 
 is not impossible. The pink region corresponds to the non-perturbative regime where bounds on coupling.
 $\tilde g_1^r\geq \left({V^{1/2}_{end} m^{end}_\phi}/({24M_p}\phi^2_{end})\right)^{1/2}$ and $g_2^r\geq ({V^{1/8}_{end}}/{\phi_{end}})\left({V^{1/2}_{end}(m^{end}_\phi)^3}/({\sqrt{2}M_p}\phi^4_{end})\right)^{1/4}$ are obtained from resonance condition of Mathieu equation for scalar field \cite{Kofman:1994rk,Kofman:1997yn,Maity:2018qhi}. The light-blue region corresponds to the self-resonance domination region.}
          \label{couplingbound}
          \end{figure}

We solve the Boltzmann equations for the general inflaton equation of state $w_\phi$ and scan the entire inflaton-scalar coupling $(g^r_i)$ parameter space and figure out one of our most significant results shown in Fig.\ref{couplingbound}. The parameter space $(w_{\phi}, g^r_i)$ can be generically divided into five regions marked in different colors: i) Light-cyan region is where reheating is entirely controlled by the gravity-mediated decay channel (gravitational reheating), ii) Yellow region is controlled by mostly inflaton-Boson coupling, iii) Light-red region is where successful reheating cannot be achieved as reheating temperature $T_{re} < T_{BBN}$, iv) Pink region where initial parametric resonance will be the important and v) Light-blue region where the self-resonance of the inflaton will be important. In our present paper, we ignore both effects and defer for our future study. The system has two competing effects due to direct and gravity-mediated inflaton decay. Based on their relative dominance, we observe three distinct regions of coupling $g^r_i$ where reheating evolution will be different: 1) {\bf Case-I:} Entire reheating dynamics will be dominated by direct inflaton decay, 2) {\bf Case-II:} Both the decay processes will partially dominate the reheating dynamics, 3) {\bf Case-III:} Entire reheating dynamics will be dominated by gravity mediated decay (gravitational reheating \cite{Haque:2022kez}). These three cases immediately suggest the existence of two critical coupling values for every individual inflaton-scalar radiation coupling $g^r_i$, which set the phase boundaries among those three cases in $(w_{\phi}, g^r_i)$ plane. If the coupling $g^r_i > \mathscr{G}^{1,\,th}_{ci}\simeq \mathscr{G}^1_{ci}$, the reheating evolution will be according to {case-I}. Where, $\mathscr{G}^1_{ci}$ is computed without thermal effect, for it turned to be same for both with and without thermal effect (see detailed derivation in the Appendix-\ref{bosonanawth}). The critical coupling strength $\mathscr{G}^1_{ci}$ is found to be,
\begin{eqnarray}{\label{gcr1}}
&& \mathscr{G}^{1,\,th}_{ci} = \left\{\begin{array}{ll}
&\left[\frac{3(5+3w_\phi)(H_{end}m^{end}_\phi)^2}{128M^2_p(1+15w_\phi)}\frac{\left[\left(\frac{9+15w_\phi}{8}\right)^{\frac{-8}{1+15w_\phi}}-\left(\frac{9+15w_\phi}{8}\right)^{\frac{-9-15w_\phi}{1+15w_\phi}}\right]}{\left[\left(\frac{8}{3(1-w_\phi)}\right)^{\frac{3(w_\phi-1)}{(5+3w_\phi)}}-\left(\frac{8}{3(1-w_\phi)}\right)^{\frac{-8}{5+3w_\phi}}\right]}\right]^{1/2}~~~\mbox{for}~~ g^r_1 \phi s^2\,,\\
&\left[\frac{(9w_\phi-1)(m^{end}_\phi)^4}{64M^4_p(1+15w_\phi)}\frac{\left[\left(\frac{9+15w_\phi}{8}\right)^{\frac{-8}{1+15w_\phi}}-\left(\frac{9+15w_\phi}{8}\right)^{\frac{-9-15w_\phi}{1+15w_\phi}}\right]}{\left[\left(\frac{9(1-w_\phi)}{8}\right)^{\frac{9(w_\phi-1)}{1-9w_\phi}}-\left(\frac{9(1-w_\phi)}{8}\right)^{\frac{-8}{1-9w_\phi}}\right]}\right]^{1/2}~~~~~~~~~\mbox{for}~~ g^r_2 \phi^2 s^2\,.
\end{array}\right.
\end{eqnarray}
 If we lower the couplings $g_i^{r}$ below $\mathscr{G}^{1,\,th}_{ci}$, the gravitational scattering starts to reveal its presence in the early phase of the reheating process, and the complete takeover happens if the non-gravitational coupling strength is lower than a new critical coupling which we denoted as $\mathscr{G}^{2,\,th}_{ci}$. Therefore, if coupling strength in between $\mathscr{G}^{2,\,th}_{ci}<g_i^{r}<\mathscr{G}^{1,\,th}_{ci}$, the reheating evolution will be according to {case-II}.   
 The expressions of $\mathscr{G}^{2,\,th}_{ci}$ for different interaction are calculated as,
\begin{eqnarray}
&&\mathscr{G}^{2,\,th}_{ci}=\left\{
\begin{array}{ll}
&\left[\left(\frac{9(1+w_\phi)H^3_{end}m^{end}_\phi}{512\pi(1+15w_\phi)}\right)\left(\frac{4\pi\epsilon^{1/4}(m^{end}_\phi)^2(3w_\phi+1)}{3(1+w_\phi)M^2_pH_{end}}\right)^{4/3}(A^{gr}_{re})^{-2-6w_\phi}\right]^{3/8}~~\mbox{for}~~~g^r_1\phi s^2\\\
&\left[\left(\frac{9(1+w_\phi)H^3_{end}m^{end}_\phi}{512\pi(1+15w_\phi)}\right)\left(\frac{8\pi\epsilon^{1/4}(m^{end}_\phi)^4(5w_\phi-1)}{9(1+w_\phi)M^4_pH^3_{end}}\right)^{4/3}(A^{gr}_{re})^{2-10w_\phi}\right]^{3/8}~~\mbox{for}~~~g^r_2\phi^2 s^2\,,
\end{array}\right.
\end{eqnarray}
where, $A^{gr}_{re}$ is the scale factor defined at reheating end for the gravitational reheating scenario (see, for this instance, Eqn.\ref{tgr}).

Finally, as pointed out just above, if $g_i^{r} < \mathscr{G}^{2,\,th}_{ci}$, the reheating evolution will be according to {case-III}, which we call gravitational reheating. Detailed analysis on this possibility have been discussed in \cite{Haque:2022kez}. We will now dwell on these three cases and discuss their thermal histories in detail:

\textbf{\bf Case-I: Coupling strength $\bf g_i^{r}>\mathscr{G}^{1,\,th}_{ci}$:} In this regime, direct decay of inflaton into radiation controls the entire reheating process. In the left panel of Fig.\ref{Casemplot}, we showed the evolution of the different energy components with the coupling parameter. Since the radiation bath of temperature $T_{rad}$ is produced from the decay products of homogeneous inflaton background, the typical energy of the bath particles will be of the order of inflaton mass $m_{\phi}$. And hence for the condition $T_{rad} > m_\phi (t)$,  the thermal effect will be dominant. For any reheating dynamics, there exist two important energy scales of importance, and those are maximum radiation temperature $(T_{rad}^{max})$ and the reheating temperature $(T_{re})$. Given the inflation model under consideration, we have two free parameters namely, the inflaton equation of state $(\omega_{\phi})$ and the inflaton-Boson coupling $(g^r_i)$, where "$i$" stands for two different bosonic decay channels mentioned earlier. Depending upon the evolution of $(T_{rad}, m_{\phi})$, and consequently the behavior of thermal effect, we have observed rich reheating histories. In the following section, we lay bare the detailed discussions on those for different cases in different temperature regimes.

%%%%%%%%%%%%%%%%%%%%%%%%%%%%%%%%%%%%%%%%%%
\underline{When $T_{rad}^{max}>m^{end}_\phi$}: This is the situation which typically occurs for large value of inflaton-scalar coupling mostly in the pink region of Fig.\ref{couplingbound}. 
Since the maximum radiation temperature $T_{rad}^{max}$ is greater than the inflaton mass $m_\phi^{end}$ defined at the end of inflation, the thermal effect influences the reheating dynamics significantly. Details of this finite temperature effect will be further controlled by the parameters $(w_{\phi},\, g^r_i)$ and the time-dependent inflaton mass $m_{\phi} (t)$. We will first discuss the situation when $T_{rad} > m_{\phi} (t)$ throughout the entire period of reheating. However, important to remember that such a condition does not satisfy though out the entire reheating parameter range, and this can be observed from Fig.\ref{couplingbound}. 

It is observed that in this region the ratio $T_{rad}/m_\phi(t)$ varies as $A^{(9w_\phi-1)/2} (A^{(11w_\phi-3)/2})$ for $\phi\to ss (\phi\phi\to ss)$. Such variation of the ratio indicates that there exists a critical value $\omega_{\phi}^c = 1/9$ for the decay channel $\phi\to ss$ and $\omega_{\phi}^c = 3/11$ for the decay channel  $\phi\phi\to ss$, above which $T_{rad}>m_\phi(t)$ condition is always maintained. With this condition, the finite temperature decay widths can be approximated as,
\begin{eqnarray}\label{decayth}
&&\Gamma_\phi=\left\{
\begin{array}{ll}
&\Gamma_{\phi\rightarrow ss}=\frac{4(g^r_1)^2T_{rad}}{8\pi m^2_\phi(t)}\,,\\
&\Gamma_{\phi\phi\rightarrow ss} =\frac{2(g^r_2)^2}{8\pi}\frac{\rho_\phi(t)T_{rad}}{m^4_\phi(t)}\,,
\end{array}\right.
\end{eqnarray}
%%%%%%%%%%%%%%%%%%%%%%%%%%%%%%%%%%%%%%%%%
%%%%%%%%%%%%%%%%%%%%%%%%%%%%%%%%%%%%%%%%
and with this the reheating temperatures are estimated for two different decay channels as,
\begin{eqnarray}\label{reheatth}
&&T_{re}=\left\{
\begin{array}{ll}
&\left[\frac{3M^2_p(1+w_\phi)H_{end}}{4\pi\epsilon(1+3w_\phi)(m^{end}_\phi)^2}(g^r_1)^2A_{re}^{-\frac{3}{2}(1-3w_\phi)}\right]^{1/3}~~\mbox{for}~~ g^r_1\phi s^2\,,\\
&\left[\frac{9M^4_p(1+w_\phi)H^3_{end}}{8\pi\epsilon(5w_\phi-1)(m^{end}_\phi)^4}(g^r_2)^2 A_{re}^{\frac{-3(3-5w_\phi)}{2}}\right]^{1/3}~~\mbox{for}~~ g^r_2\phi^2 s^2\,.
\end{array}\right.
\end{eqnarray}
Where, $A_{re}$ is the normalized scale factor at the end of reheating (detailed derivations and expressions can be found in Appendix-\ref{bosonanawth}).
To this end we would like to point out an interesting fact associated with the maximum 
\begin{longtable}{|c|c|c|c|c|c|c|} 
 \caption{The evolution of bath temperature for bosonic reheating:}\label{tempevboson} \\
               \hline
            \multirow{3}{*}{\text{Channel}}& \multicolumn{2}{|c|}{
                 $T\ll m_\phi(t)$ (\text{Without thermal effect})
 }& \multicolumn{2}{|c|}{ $T\gg m_\phi(t)$ (\text{With thermal effect})}
        \\       
   \cline{2-5}
         & Non-gravitational & Gravitational  & Non-gravitational & Gravitational\\
               \hline
                 $\phi\rightarrow ss$ &$A^{-\frac{3(1-w_\phi)}{8}}$&$A^{-1}$&$A^{-\frac{(1-3w_\phi)}{2}}$&$A^{-1}$\\\hline
                  $\phi\phi\rightarrow ss$ &$A^{-\frac{9(1-w_\phi)}{8}}$&$A^{-1}$&$A^{-\frac{(3-5w_\phi)}{2}}$&$A^{-1}$\\\hline 
\end{longtable}
\noindent
radiation temperature $T^{max}_{rad}$, which generically satisfies $T^{max}_{rad} > T_{re}$. We observed the existence of a critical value of $w^t_{\phi}=(1/3,\,3/5)$ for two different bosonic decay channel $\phi\to ss$ and $\phi\phi \to ss$ (see, for instance, Eqs.\ref{the1} and \ref{the2}). If $w_{\phi} <w_t$, the maximum radiation temperature $(T_{rad}^{max})$ satisfies the usual condition mentioned above, 
 \bea
T_{rad}^{max}\simeq T_{s}^{r,\,max}=\left\{
\begin{array}{ll}
	&\left[\frac{3M^2_p(1+w_\phi)H_{end}}{4\pi\epsilon(1+3w_\phi)(m^{end}_\phi)^2}(g^r_1)^2\bigg\{\left(\frac{2}{1-3w_\phi}\right)^{\frac{3w_\phi-1}{3+9w_\phi}}-\left(\frac{2}{1-3w_\phi}\right)^{\frac{-2}{3+9w_\phi}}\bigg\} \right]^{1/3}~~~~~\mbox{for}~~~g^r_1\phi s^2\,,\\
&\left[\frac{9M^4_p(1+w_\phi)H^3_{end}}{8\pi\epsilon(5w_\phi-1)(m^{end}_\phi)^4}{(g^r_2)}^2\bigg\{\left(\frac{2}{3-5w_\phi}\right)^{\frac{5w_\phi-3}{5w_\phi-1}}-\left(\frac{2}{3-5w_\phi}\right)^{\frac{-2}{5w_\phi-1}}\bigg\}\right]^{1/3}~~~~~\mbox{for}~~~g^r_2\phi^2 s^2\,.\\
\end{array} \right.
\eea
Where $T_{s}^{r,\,max}$ indicates the maximum radiation temperature. Surprising result emerges, however, for $w_{\phi}> w^t_{\phi}$ case, for which the evolution of radiation and background conspire in such a manner that at the end of reheating maximum radiation temperature becomes equal to the reheating temperature, $T_{rad}^{max}\simeq T_{re}$ (for better visualization, see the left most plot of Fig.\ref{Casemplot}). Such behavior has been observed before considering the phenomenological expression of the decay rate as a function of temperature \cite{Co:2020xaf}. The implication of this specific case could be interesting to study.\\
Again, when EoS stays within $0\leq w_\phi<w_\phi^c$, due to initial high radiation temperature thermal correction will have a significant effect. As the reheating progresses, such effect diminishes with the complete takeover by the zero temperature dynamics at a certain value of scale factor $A_c$, which depends on the inflaton equation of state as follows,\begin{eqnarray}\label{acc}
 && A_{c}=\left\{
 \begin{array}{ll}
 &\left(A_{max}\right)^{\frac{1-3w_\phi}{1-9w_\phi}}\left(\frac{m^{end}_\phi}{(\rho^{r,\,max}_{s}/\epsilon)^{1/4}}\right)^{-\frac{2}{(1-9w_\phi)}}~~~~~\mbox{for}~~~g^r_1\phi s^2\,,\\
  &\left(A_{max}\right)^{\frac{3-5w_\phi}{3-11w_\phi}}\left(\frac{m^{end}_\phi}{(\rho^{r,\,max}_{s}/\epsilon)^{1/4}}\right)^{-\frac{2}{3(1-11w_\phi)}}~~~~\mbox{for}~~g^r_2\phi^2 s^2\,,
 \end{array}\right.
 \end{eqnarray}
 where $A_{max}$ and $\rho_{s}^{r,\,max}$ are defined in Eqs.\ref{amaxth},\ref{a2cth}. After this crossover happens, the radiation energy density simply follows Eqs.\ref{s1}. In the DM studies, we will see such an intermediate scale will have non-trivial dependence on its abundance. We find the associated reheating temperature as
 \begin{eqnarray}
&&T_{re}=\left\{\begin{array}{ll}
&\left(\frac{6M^2_p(1+w_\phi)H_{end}}{8\pi\epsilon(5+3w_\phi) m^{end}_\phi}(g^r_1)^2\right)^{1/4}A^{-\frac{3}{8}(1-w_\phi)}_{re}~~~~~\mbox{for } g^r_1\phi s^2\,,\\
&\left(\frac{9 M^4_p(1+w_\phi)H^3_{end}}{4\pi\epsilon(9w_\phi-1)( m^{end}_\phi)^3}(g^r_2)^2\right)^{1/4} A^{\frac{9(w_\phi-1)}{8}}_{re}~~~~~\mbox{for } g^r_2\phi^2 s^2\,,\\
\end{array}\right.
\end{eqnarray} 
where
\begin{eqnarray}
&&A_{re}=\left\{\begin{array}{ll}
&\left(\frac{4\pi(5+3w_\phi)H_{end}m^{end}_\phi}{(1+w_\phi)(g^r_1)^2}\right)^{\frac{2}{3+9w_\phi}}~~~~~~~~\mbox{for } g^r_1\phi s^2\,,\\
&\left(\frac{4\pi(9w_\phi-1)(m^{end}_\phi)^3}{3(1+w_\phi)M^2_pH_{end}(g^r_2)^2}\right)^{\frac{2}{3(5w_\phi-1)}}~~~~~\mbox{for } g^r_2\phi^2 s^2\,.
\end{array}\right.
\end{eqnarray}

In reheating model building, gravitational contribution to the radiation sector is universally present along with the non-gravitational one. Therefore, it is natural to expect that the $T^{max}_{rad}$ for a given reheating model can not assume an arbitrarily low value. In fact, due to universal gravitational contribution, there exists a lower limit on $T_{rad}^{max}$, which is set by the gravitation reheating $T_{gr}^{r,\,max} \simeq 10^{11}\to 10^{12}$ GeV \cite{Clery:2021bwz,Haque:2022kez}. The small variations are due to different values of $\omega_{\phi}$. Therefore, the minimum possible value of $T^{max}_{rad}$ simply turns out as $T^{r, max}_{gr}$. In the following discussion, we now consider the regime where $T_{rad}^{max}>T_{gr}^{r,\,max}$ but less than $m_\phi^{end}$. 

\underline { When $ T_{gr}^{r,\,max}<T_{rad}^{max}<m^{end}_\phi$}: The parameter region (see Fig.\ref{couplingbound}) wherein this condition is satisfied belong to the yellow region. However, for this case, initially  $T_{rad}^{max}<m_\phi^{end}$, and hence there is no thermal effect initially, and the ratio $T_{rad}/m_\phi \propto A^{- (3-27w_\phi)/8}$ ($A^{-(9-33 w_\phi)/8}$) for $\phi\to ss (\phi\phi\to ss)$ respectively. Thus, for $w_\phi<w_\phi^c$ (defined before), the finite temperature effect will never be significant. Consequently, the reheating dynamics will be the same as that of the zero temperature, and details of such dynamics are described in the Appendix-\ref{bosonanawth} (for example, the radiation energy density evolves following Eq.\ref{s1}). However, if the inflaton equation of state satisfies the condition $w_\phi>w_\phi^c$, the finite temperature effect ($T_{rad} > m_{\phi}$) is expected to occur at  some intermediate radiation temperature $T_c=T_{rad}(A_c)$ with the scale factor $A_c$ during reheating. We have
\begin{eqnarray}\label{ac}
 && A_{c}=\left\{
 \begin{array}{ll}
 &\left(A_{max}\right)^{\frac{1-w_\phi}{1-9w_\phi}}\left(\frac{m^{end}_\phi}{(\rho^{r,\,max}_{s}/\epsilon)^{1/4}}\right)^{-\frac{8}{3(1-9w_\phi)}}~~~~~\mbox{for}~~~g^r_1\phi s^2\,,\\
  &\left(A_{max}\right)^{\frac{3(1-w_\phi)}{3-11w_\phi}}\left(\frac{m^{end}_\phi}{(\rho^{r,\,max}_{s}/\epsilon)^{1/4}}\right)^{-\frac{8}{3(1-11w_\phi)}}~~~~\mbox{for}~~g^r_2\phi^2 s^2\,,
 \end{array}\right.
 \end{eqnarray}
 where $A_{max}$ and $\rho_{s}^{r,\,max}$ are defined in Eqs.\ref{amaxnthboson}, \ref{mm2}. After this crossover happens, the radiation energy density simply follow the Eqs.\ref{the1} and \ref{the2}. In the DM studies, we will see such an intermediate scale will have non-trivial dependence on its abundance.

For case-I, above two temperature regimes will be possible. To this end let us point out an another important situation that deserves detailed discussion is a special case when $T_{rad}^{max}=T^{r,\,max}_{gr}$, which indicates that the initial phase of reheating must be dominated by the graviton mediated inflaton decay. And such situation arises only when non-gravitational inflaton coupling $g^r_i < \mathscr{G}^{1,\,th}_{ci}$. This condition, therefore, belongs to the other two cases of coupling ranges mentioned before. 

\begin{figure}
          \begin{center}
          \includegraphics[width=15.0cm,height=5.0cm]{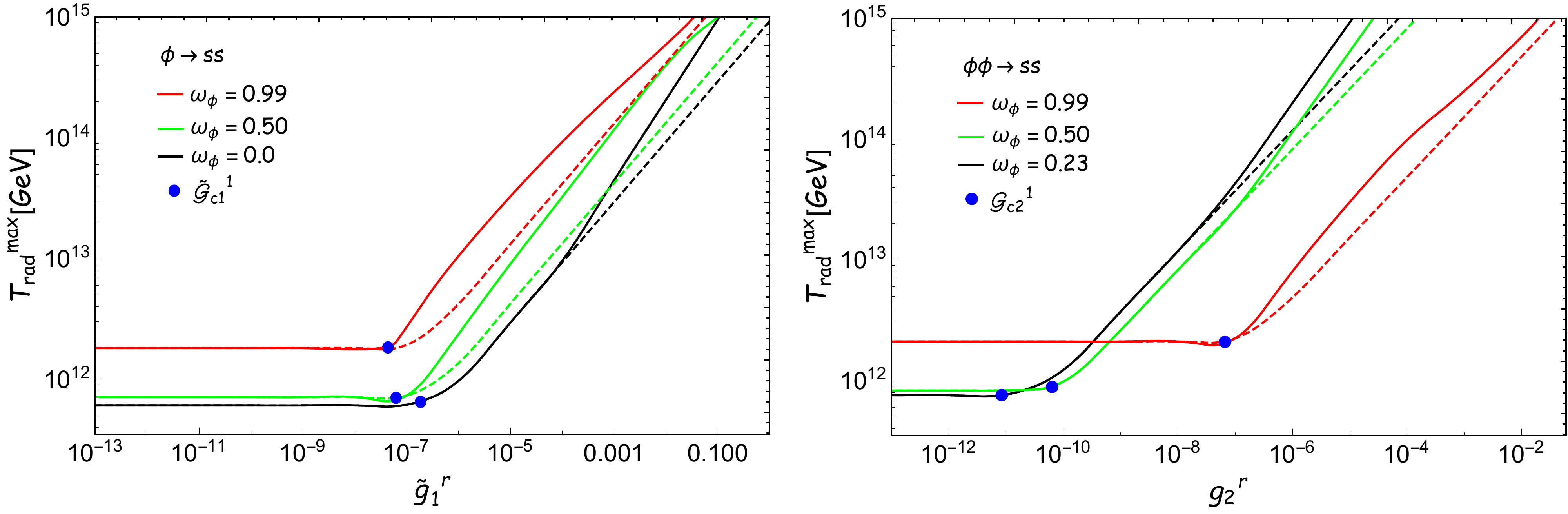}
         \caption{\scriptsize Variation of maximum radiation temperature $T_{rad}^{max}$ as function of dimensionless coupling parameter $\tilde g^r_1=\frac{g}{m^{end}_\phi}$,$g^r_2$, for three different inflaton equation of state $w_\phi=0.0(0.23 ),0.50,0.99$ for the $\phi\rightarrow ss$ (left-most), and $\phi\phi\rightarrow ss$ (right-most) model.}
          \label{Tmax}
          \end{center}
      \end{figure}
 %%%%%%%%%%%%%%%%%%%%%%%%%%%%%%%%%%%%%%%%%%%%%%%%%%%%%%%%%%%%%     
      \begin{figure}
         \begin{center}
\includegraphics[width=0015.50cm,height=007.5cm]
          {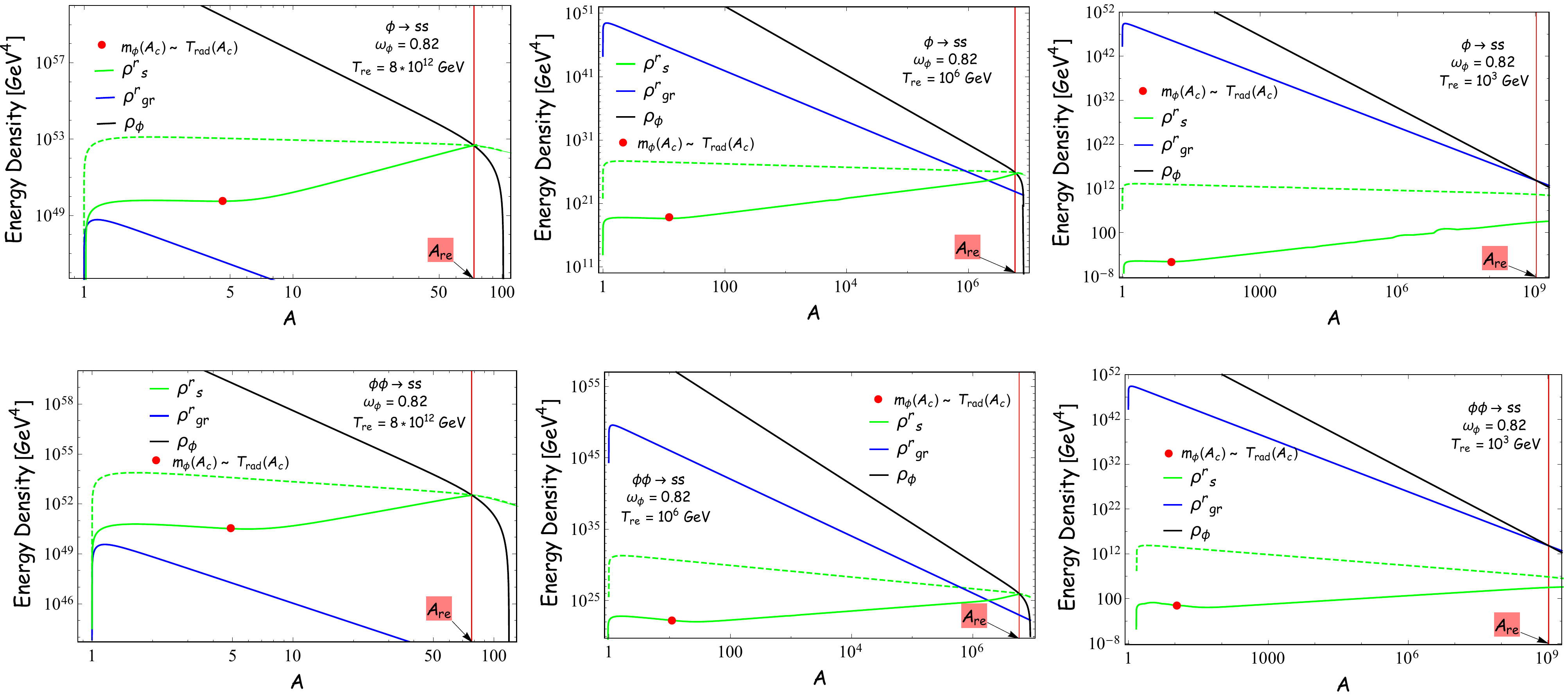}
          \caption{ Evolution of inflaton and radiation energy density as a function of normalized scale factor $A=a/a_{end}$ for both  $\phi\rightarrow ss$ and $\phi\phi\to ss$ with and without thermal effect (solid line for with thermal effect and dashed line for without thermal effect). \textbf{Left panel:} Coupling is in the range of  $g_i^{r}>\mathscr{G}^{1,\,th}_{ci}$. \textbf{Middle panel:} Coupling is in the range of $\mathscr{G}^{2,\,th}_{ci}/\mathscr{G}^{2}_{ci}<g_i^{r}<\mathscr{G}^{1,\,th}_{ci}$. \textbf{Right panel:} Coupling is in the range of $g_i^{r}<\mathscr{G}^{2,\,th}_{ci}/\mathscr{G}^{2}_{ci}$. For all these cases, we have considered the inflaton equation of state $w_\phi=0.82$. }
          \label{Casemplot}
          \end{center}
      \end{figure}
%%%%%%%%%%%%%%%%%%%%%%%%%%%%%%%%%%%%%%%%%%%%%%%%%%%%%%%%
\textbf{Case-II: Coupling strength in between $\bf \mathscr{G}^{2,\,th}_{ci}<g_i^{r}<\mathscr{G}^{1,\,th}_{ci}$ :}  
In this coupling range, gravitational interaction drives the dynamics of reheating at the beginning. The gravitational production is nearly instantaneous and happens just at the beginning of reheating. In fact, this case is true for the entire range of coupling with $g_i^{r} < \mathscr{G}^{1,\,th}_{ci}$, and $T_{rad}^{max}=T^{r,\,max}_{gr}$ condition always holds. However, since the non-gravitational bosonic coupling is non-zero, during the reheating process gravitational coupling, non-gravitational coupling, and thermal effect of the produced radiation undergo interesting interplay among themselves. Let us illustrate the following two different possibilities of thermal history in this context,
 \begin{itemize}
  \item  The thermal effect may start dominating the early phase when gravity-mediated decay controls the reheating. During this phase the ratio behaves $T_{rad}/m_\phi(A)\propto A^{3w_\phi-1}$. Hence, for $w_\phi>1/3$, it is clear from this ratio that the thermal effect cannot be ignored. And we found a particular value of scale factor $A_c^g$, after which Bose enhancement starts affecting the dynamics,
  \begin{eqnarray}
A_c^g=\left[A_{max}\left(\frac{m^{end}_\phi}{(\rho^{r,
{max}}_{gr}/\epsilon)^{1/4}}\right)\right]^{\frac{1}{1+3w_\phi}}\,,
\end{eqnarray}
where, $A_{max}$ is the scale factor at which $T_{rad} = T_{rad}^{max}=T^{r,\,max}_{gr}$, and maximum radiation energy density $(\rho_{gr}^{r,\,max})$ is obtained from gravitational decay (see, for instance, the last expression of Eq.\ref{amaxnthboson} and Eq.\ref{gm}). 
After this point, the radiation energy density simply varies as $A^{-4}$. However, as reheating proceeds towards the end, there is another crossover from gravitational decay domination to non-gravitational decay domination, which will happen at the value of scale factor, defined as
\begin{eqnarray}
 &&A_{gr\rightarrow ngr}=\left\{
 \begin{array}{ll}
      & \left(\frac{2M^2_p(1+w_\phi)H_{end}}{8\pi(1+3w_\phi)(m^{end}_\phi)^2(\mathcal{C}(w_\phi))^3}(g^r_1)^2\right)^{\frac{-2}{3(1+3w_\phi)}}~~~~~~\mbox{For $g^r_1\phi s^2$\,,} \\
      & \left(\frac{4(3M^2_pH^2_{end})^2M^2_p(1+w_\phi)H_{end}}{8\pi(5w_\phi-1)(m^{end}_\phi)^4(\mathcal{C}(w_\phi))^2}(g^r_2)^2\right)^{\frac{2}{3(1+5w_\phi)}}~~~\mbox{For $g^r_2\phi^2 s^2$}\,,
 \end{array}\right.
 \end{eqnarray}
 where $\mathcal{C}(w_\phi)=\left(\frac{9(1+w_\phi)H^3_{end}}{1024\pi(1+3w_\phi)\epsilon M^2_p}\right)^{1/3}$. The reheating temperature assumes the same form as Eq.\ref{reheatth}. For better visualization of the evolution of different energy components see the middle panel of Fig.\ref{Casemplot}. In summary, the reheating dynamics can be read off as follows: at the beginning gravitational sector dominates the porcess with radiation temperature varies as $T_{rad}\propto A^{-1}$ $\bf{\to}$ as reheating proceeds the thermal effect starts to play its role but with the same temperature evolution $T_{rad}\propto A^{-1}$ $\to$ non-gravitational coupling takes over the process, and the radiation temperatures vary as $T_{rad}\propto A^{-\frac{1}{2}+\frac{3w_\phi}{2}} \left(A^{-\frac{3-5w_\phi}{2}}\right)$ for the decay process $\phi\rightarrow ss$ ($\phi\phi\rightarrow ss$)).
\item There may be a situation where the thermal effect will be important only during non-gravitational production. For this case, reheating proceeds from gravity-mediated decay to explicit inflaton decay domination, and the transition occurs at the scale factor,
 \begin{eqnarray}{\label{gr2ngr}}
 &&A_{gr\to ngr}=\left\{\begin{array}{ll}
      & \left(\frac{\rho^{r,max}_{gr}}{\rho^{r,max}_s}\frac{\left(\frac{9+15w_\phi}{8}\right)^{\frac{8}{1+15w_\phi}}}{\left(\frac{8}{3(1-w_\phi)}\right)^{\frac{3(w_\phi-1)}{5+3w_\phi}}}\right)^{\frac{2}{5+3w_\phi}}~~\mbox{ for $g^r_1\phi s^2$}\,, \\
      &\left(\frac{\rho^{r,max}_{gr}}{\rho^{r,max}_s}\frac{\left(\frac{9+15w_\phi}{8}\right)^{\frac{8}{1+15w_\phi}}}{\left(\frac{9(1-w_\phi)}{8}\right)^{\frac{9(w_\phi-1)}{1-9w_\phi}}}\right)^{\frac{2}{9w_\phi-1}}~~\mbox{ for $g^r_2\phi^2 s^2$}\,,
 \end{array}\right.
 \end{eqnarray}
 where $\rho_{gr}^{r,\,max}$ and $\rho_{s}^{r,\,max}$ are defined in Eqs.\ref{mm2}, \ref{gm}. Once reheating process starts to dominate by the explicit inflation coupling, the scale factor beyond which the thermal effect starts working is followed by Eq.\ref{ac}. Finally, the decay channel defines the reheating temperature (see, for instance, Eq.\ref{reheatth}). In summary, dynamics can be described as follows: reheating proceeds through gravity-mediated decay with no finite temperature effect ($T_{rad}\propto A^{-1}$) $\to$ non-gravitational decay dominates the phase with negligible thermal effect with radiation temperature varies as $T_{rad}\propto A^{-\frac{3}{8}(1-w_\phi)} \left(A^{-\frac{9}{8}(1-w_\phi)}\right)$ for decay process $\phi\rightarrow ss$ ($\phi\phi\rightarrow ss$)) $\to$ non-gravitational coupling domination with significant thermal effect ($T_{rad}\propto A^{-\frac{1}{2}+\frac{3w_\phi}{2}}$ ($A^{-\frac{3-5w_\phi}{2}}$)  for $\phi\rightarrow ss$ ($\phi\phi\rightarrow ss$)).
  \end{itemize}
 \textbf{Case III: when $\bf g_i^{r} < \mathscr{G}^2_{ci}$:} The bath temperature always falls as $A^{-1}$, and thermal effect is observed to play no role throughout. The gravity-mediated decay of inflaton controls the entire dynamics of reheating, and the scenario is termed as gravitational reheating, ant that will occur only for $w_\phi>0.65$ (see the light-cyan region of Fig.\ref{couplingbound}). The reheating temperature is defined when $\rho_\phi=\rho_{gr}^r$, and the condiction  gives
 \begin{equation}{\label{tgr}}
 T^{gr}_{re}=\left(\frac{9H^3_{end}m^{end}_\phi(1+w_\phi)}{512\epsilon\pi(1+15w_\phi)(A^{gr}_{re})^4}\right)^{1/4},\, A^{gr}_{re}=\left(\frac{512\pi M^2_p(1+15w_\phi)}{3H_{end}m^{end}_\phi(1+w_\phi)}\right)^{\frac{1}{3w_\phi-1}}\,,
\end{equation}
 where $A^{gr}_{re}$ is the normalized scale factor at the end of gravitational reheating. In the right panel of Fig.\ref{Casemplot}, we have shown the dynamical behaviour of different energy components, and in Table-\ref{tempevboson}, showing the evolution of the bath temperature with $A$ for non-gravitational reheating and gravitational reheating. 
 
 In the subsequent section, we will focus on  the possible constraints on the inflaton coupling strengths depending on the CMB ($n_s$) and reheating parameter $T_{re}$.
 %%%%%%%%%%%%%%%%%%%%%%%%%%%%%%%%%%%%%%%%%%%%%%%%%%%%%%%
\subsection{Inflaton phenomenology: Constraining reheating and bosonic decay parameters:}
For illustration, we consider five different values of the inflaton equation of state $w_\phi=(0,\,0.2,\,0.5,\,0.82,0.99)$. For each $w_\phi$, we have plotted :(i) $T_{re} $ vs $n_{s}$, (ii) ${g_i^r}$ vs $n_s$, (iii) $ g_i^r$ vs $T_{re}$. We compare the results with and without the thermal effect for all cases.
%%%%%%%%%%%%%%%%%%%%%%%%%%%%%%%%%%%%%%%%%%%
\begin{figure}[t]\centering
     \includegraphics[width=1\linewidth]{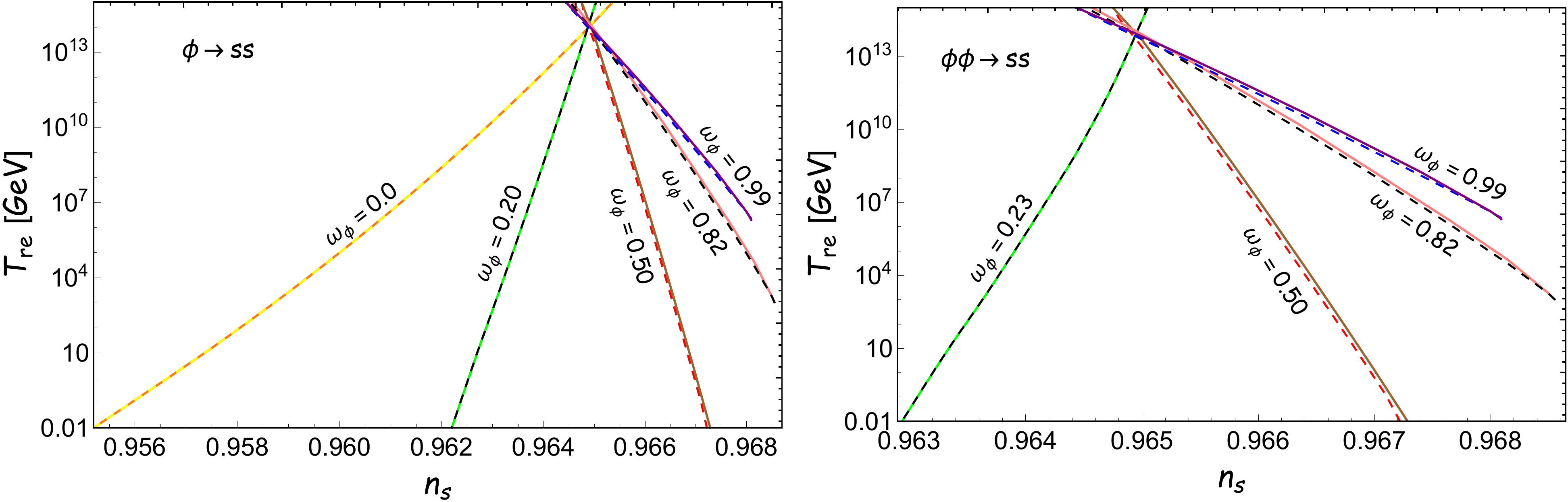}

      \caption{Variation of reheating temperature $T_{re}$ as a function of spectral index $n_s$ for $\alpha-$attractor model ($\alpha=1$) with $w_\phi=(0,\,0.20\,0.50,\,0.82,\,0.99)$. The plot is on the left side for the bosonic reheating with $\phi\to ss$ process and on the right side for the process $\phi\phi\to ss$. The solid lines are for considering the thermal effect, and the dashed lines are for without the thermal effect.}\label{ntre}
\end{figure}
%%%%%%%%%%%%%%%%%%%%%%%%%%%%%%%%%%%%%%

i) \underline{Reheating temperature ($\mathbf{T_{re}}$) in terms inflationary (CMB) parameter ($\mathbf{n_s}$)}: From the Fig.\ref{ntre} we observe that the evolution of reheating temperature in terms of the inflationary scalar spectral index ($n_s$) is insensitive to the finite temperature correction of inflaton decay width. Such behavior of the reheating temperature has already been reported in \cite{3}. The generic feature is that for $w_\phi<1/3$, (see Fig.\ref{ntre}), $T_{re}$ increases with increasing $n_s$, and as a consequence the reheating e-folding number $N_{re}$ decreases with $n_s$. This indicates the existence of a maximum scalar spectral index $n^{max}_s$ corresponding to the maximum reheating temperature $T^{max}_{re}=10^{15}$ GeV and that is called instantaneous reheating. Similarly, the minimum reheating temperature $T_{re}=T_{BBN}\sim 10$ MeV \cite{Ste1,Kawasaki:2000en,Kawasaki:1999na}, corresponds to a minimum allowed spectral index $n^{min}_s$ for a given the inflaton equation of state $w_\phi$. On the other hand, for $w_\phi>1/3$, one finds the opposite feature: maximum $T_{re}$ corresponds to the minimum spectral index $n^{min}_s$ and vice versa. When the reheating phase is dominated by purely gravitational interaction, the minimum possible reheating temperature fixes the maximum possible value of $n_s$ for the equation of state $ w_\phi\geq0.65$. 
For example, as shown in the Fig.\ref{ntre}, for $\omega_{\phi} = (0.82,0.99)$, we obtain $ T^{min}_{re}\simeq( 10^3, 10^6)$ GeV, respectively. From Fig.(\ref{ntre}), it is clear that thermal feedback to the decay rate does not affect the variation of reheating temperature with $n_s$. In Table \ref{boundBR}, we have given the possible bound on the inflationary parameters such as spectral index $n_s$ and the maximum inflationary e-folding number $N^{max}_k$ where $N^{max}_k$ is the maximum inflationary e-folding number corresponding to the maximum reheating temperature $T^{max}_{re}\sim 10^{15}$ GeV.
%%%%%%%%%%%%%%%%%%%%%%%%%%%%%%%%%%%%%%%%%%%%%
\begin{longtable}{|c|c|c|c|c|c|c|c|c|c|}
         \caption{Bosonic reheating: Bounds of the inflationary parameters}{\label{boundBR}}\\
               \hline
            \multirow{3}{*}{Parameters}& \multicolumn{5}{|c|}{
                 $\phi\rightarrow ss$
 }& \multicolumn{4}{|c|}{$\phi\phi\rightarrow ss$}
             \\              
   \cline{2-10}
         &  $w_\phi=0.0$ &$w_\phi=0.20$&$w_\phi=0.50$  &  $w_\phi=0.82$&$w_\phi=0.99$ & $w_\phi=0.23$&$w_\phi=0.50$&$w_\phi=0.82$&$w_\phi=0.99$\\
                \hline
           $n^{min}_s$  &0.95520&0.96220&0.96473&0.96455&0.96440&0.96294&0.96473&0.96455&0.96440\\\hline
                 $n^{max}_s$  &0.96540&0.96505&0.96722&0.96855&0.96809&0.96506&0.96722&0.96855&0.96809\\\hline
             $N^{max}_k$&55.69&55.78&55.71&55.83&56.04&55.78&55.71&55.83&56.04\\\hline       
               \end{longtable}
%%%%%%%%%%%%%%%%%%%%%%%%%%%%%%%%%%%%%%%%%%%
\begin{figure}[t] 
 	\begin{center} 		\includegraphics[width=15.0cm,height=11.0cm]{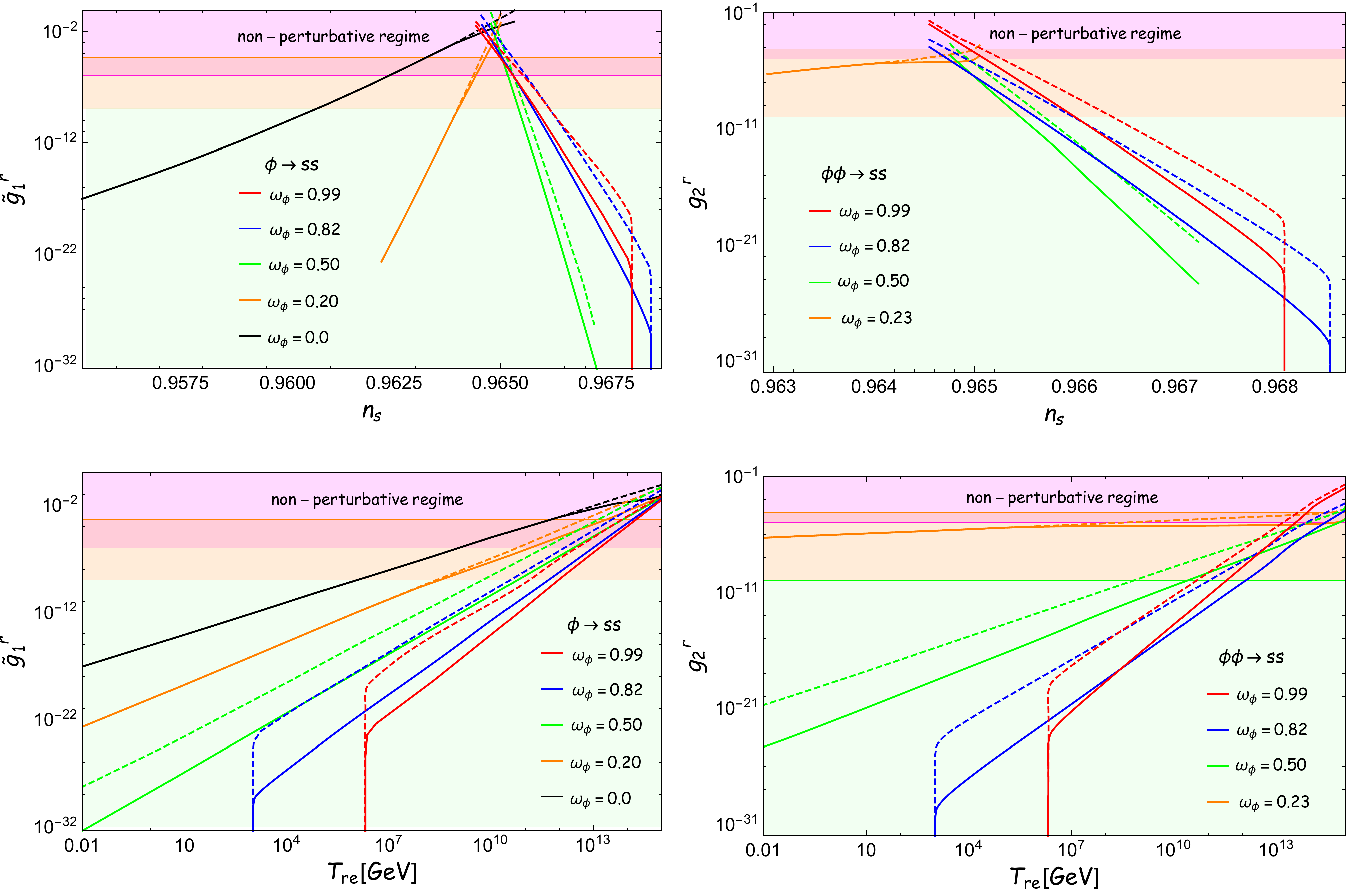}\quad		
 		\caption{\textbf{Upper panel :} Variation of dimensionless coupling parameters with respect to the spectral index $n_s$ for  $w_\phi=0.0$ (black), $0.20$ (orange)$, 0.50$ (green), $0.82$ (red), $0.99$ (blue). The dashed lines are for without the thermal feedback effect, and the solid lines are for with the thermal feedback effect. The purple-shaded region corresponds to the non-perturbative regime. The region below the green and orange lines corresponds to self-resonance dominated regions for $w_\phi=0.5$ and $0.2$ respectively. However, for other values of $w_\phi=0.0,0.82,0.99$, there is no self-resonance-dominated region.\textbf{Lower panel :} Variation of dimensionless coupling parameters as a function of reheating temperature $T_{re}$. The description of this plot is the same as the upper panel. }
 		\label{Fig5}
 	\end{center}
 \end{figure}
 \noindent
 (ii) \underline{Constraining inflaton couplings with bosonic radiation $(g^r_1,g^r_2)$}: 
One of the most important findings of our present analysis is illustrated in Fig.\ref{couplingbound}. The figure clearly depicts different regions in the parameter space of $(w_{\phi}, g^r_i)$, where the effect of different inflaton decay channels on the reheating process can be understood. At this point, let us reiterate different regions again: i) the light-cyan region is where reheating is entirely controlled by the gravity-mediated decay channel (gravitational reheating), ii) the yellow region is controlled by mostly inflaton-scalar coupling, iii) the light-red region is where successful reheating cannot be achieved as reheating temperature $T_{re} < T_{BBN}$, and iv) pink region where initial parametric resonance will be important which we ignored in this paper. Furthermore, in the right panel of Fig.\ref{couplingbound}, there is a light-gray region where reheating is not possible for the decay process $\phi\phi\to ss$. For this process, if the thermal effect is subdominant at the beginning ($m_\phi^{end}<T_{rad}^{max}$ limit),  the ratio between inflaton and radiation energy density varies as $\frac{\rho_\phi}{\rho_s^r}\propto A^{\frac{3}{2}(1-5w_\phi)}$ (see, for instance, Eqs.\ref{i1} and \ref{s1}), and hence if $w_{\phi} < 1/5$, the universe will always be inflaton dominated irrespective of the value of inflaton-scalar coupling $g^r_2$. In addition to that, if the thermal effect starts dominating from the beginning ($m_\phi^{end}>T_{rad}^{max}$ limit),  the ratio varies as $\frac{\rho_\phi}{\rho_s^r}\propto A^{(3-13w_\phi)}$  (see, for instance, Eqs.\ref{i1} and \ref{s1}) which implies that if $w_{\phi} < 3/13\sim0.23$ achieving radiation domination is not possible. However, for extremely large coupling, parametric resonance may have some effect. 

From Fig.\ref{couplingbound}, a generic feature can be  observed, and that is related to the monotonic decrease of $g^r_i$ with $w_{\phi}$ for a fixed reheating temperature $T_{re}$. The reason behind this behavior can be understood as follows: with increasing $w_\phi$, inflaton energy density dilutes faster, and hence to achieve the reheating condition $\rho_\phi=\rho^r_s$, one needs to lower the coupling. Furthermore, $m_\phi(t)$ decays faster with increasing $w_\phi$, and for both types of bosonic decay channels ($\phi\to ss$ and $\phi\phi\to ss$), the production rate goes as $\propto 1/m_\phi(t)$, which will boost up the production. As a result, to keep reheating temperature fixed, one needs to lower the value of $g^r_i$ again. 

Due to very nature of the bosonic particles the finite temperature correction in the decay width enhances the particle production rate from the inflaton condensate. As discussed, this physical fact is imprinted in the reheating dynamics and is further reflected in the parameter plot shown in Fig.\ref{Fig5}.
 Finite temperature correction naturally increases the effective decay width of the inflaton to scalar radiation, and consequently, one needs to lower the values of the dimensionless coupling parameters $g^r_1=\tilde{g^r_1} {m^{end}_\phi}$ and $g_2^r$ as compared to their zero temperature case to have successful reheating. This can be observed in Fig.\ref{Fig5} both with respect to reheating temperature ($T_{re}$) (lower two plots) and CMB spectral index $n_s$ (upper two plots).
 
 The maximum limiting value of the coupling parameters will naturally be set by the maximum possible reheating temperature $T^{max}_{re}\simeq10^{15}$ GeV, where all the lines converge (see Fig.\ref{Fig5}). 
 If the reheating dynamics are controlled directly by the inflaton-radiation coupling, the minimum possible value of the coupling will be set by the minimum reheating temperature. 
 However, such a limit on the inflaton coupling is observed to be dependent on the finite temperature correct, which will be discussed in detail. 
 When the radiation temperature $T_{rad} \gg m_\phi(t)$, the thermal effect significantly influences the radiation dynamics and consequently affects on the possible constraints on the coupling parameter as compared to the zero temperature case. It can be observed that higher the value of $\omega_{\phi}$, more will be the effect of finite temperature correction on the thermal bath.  
 For $w_\phi=0$, the effective mass of the inflaton $m_\phi(\sim 10^{13})$ remains constant; as a result, the thermal effect manifests (see left two plots of Fig.\ref{Fig5}) itself only very near and above the reheating temperature $\sim 10^{13}$ GeV (or for $n_s> 0.9645$). On the other hand, for $w_\phi > 0$, the rate of decrease of effective inflaton mass $m_\phi \propto \partial^2_{\phi} V$ increases with increasing $\omega_{\phi}$ such that the condition $T_{rad} >  m_\phi(t)$ becomes easier to satisfy even at a lower temperature. For example, for $w_\phi=0.2$, the above condition begins to satisfy (see left  Fig.\ref{Fig5}) when $n_s>0.9639 \,(T\sim 10^8)$ GeV, and accordingly, the finite temperature effect (solid line) manifest itself after $T_{r e} \sim10^8$ GeV. For $w_\phi=0.50,0.82,0.99$, the condition $T_{rad} > m_\phi(t)$ can be observed to satisfy throughout the whole range of reheating temperature. 
 
To this end, we would like to elaborate on the finite temperature effect for low reheating temperatures. The author in the reference \cite{3} claims that the thermal effect will be insignificant at low reheating temperatures. However, generically such an effect depends on the evolution of the ratio $m_{\phi}/T_{rad}$. And we found that the finite temperature effect can be significant at low reheating temperature for the higher equation of state $w_\phi=(0.20,\,0.50,\,0.82,\,0.99)$ (see solid and dotted lines in Fig.\ref{Fig5}), for which inflaton mass undergoes non-trivial evolution.
As can be seen from the second row of Fig.\ref{Fig5}, for the higher equation of state finite temperature effect becomes more prominent at lower reheating temperature mainly because inflaton mass can become significantly smaller during the course of reheating. Moreover, for higher equation state ($w=0.82,0.99$), the gravitational reheating has been observed to give a lower limit of the reheating temperature ($10^3,10^6$ GeV), which is again found to be directly corresponding to specific inflationary scalar spectral index $n_s$. The  spectral indices associated with those temperatures are $n_s=0.96855$ and $0.9681$ for $w=0.82$ and $0.99$, respectively. When $n_s$ reaches these values, the coupling parameter tends towards zero, i.e., gravitational scattering solely controls the reheating dynamics. 
%%%%%%%%%%%%%%%%%%%%%%%%%%%%%%%%%%%%%%%%%%%%%%%
\section{Fermionic Reheating: results and constraints} \label{bgdfer}
\subsection{Dynamics: probing different decay channels}
\begin{figure}
          \begin{center}
          \includegraphics[width=17.00cm]
         % {FDuniversalcoupling2.pdf}
         {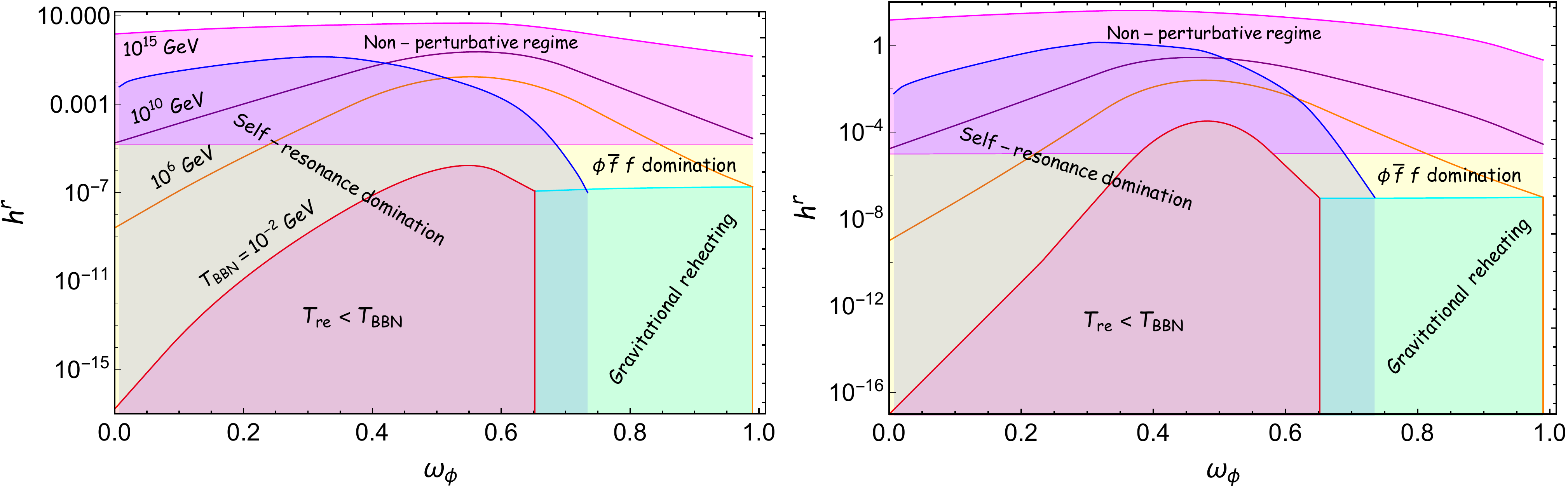}
          \caption{The description of this plot is the same as Fig.\ref{couplingbound}, the main difference is that we have shown the results in $(h^r,\,w_\phi)$ plane for  the fermionic decay channel $\phi\to \bar{f}f$. On the left panel, we have plotted for without thermal effect, whereas on the right, the results are for the with thermal effect. The pink-shaded region corresponds to the non-perturbative regime where bounds on coupling 
$h^r \geq \left({V^{1/2}_{end}(m^{end}_\phi)^3}/{\sqrt{2}M_p \phi^4_{end}}\right)^{1/4}$       
 are obtained from the resonance condition of the Mathieu equation for the fermion field \cite{Greene:2000ew,3,Greene:1998nh}.}
          \label{boundfermion}
          \end{center}
      \end{figure}

During reheating, if the dominant decay channel of the inflaton is through massless fermions, we call it fermionic reheating. For this purpose, we consider inflaton decaying only into massless Fermion though the standard Yukawa decay channel $\phi\to \bar{f}f$ along with gravitational scattering process $\phi\phi\rightarrow h_{\mu\nu}\rightarrow ss$. Instantaneous thermalization of those different components are assumed throughout. To study the evolution of the radiation energy density, we took the finite temperature effect arising due to Pauli blocking (see, for instance, Appendix-\ref{apfer} for details calculation). Similar to bosonic reheating, for the fermionic case, we identified distinct regions in $(w_{\phi}, h)$ plane depending upon different physical processes involved in controlling the reheating dynamics (see Fig.\ref{boundfermion}). For this case, we have plotted separately with and without the finite temperature effect and observed the quantitative change in parameter space due to the finite temperature effect where reheating would be successful.
The parameter space $(w_{\phi}, h)$ is again divided into four regions marked in different colors: i) Light-cyan region is where reheating is entirely controlled by the gravity-mediated decay channel (gravitational reheating), ii) Yellow region is controlled by mostly inflaton-Fermion coupling, iii) Light-red region is where successful reheating cannot be achieved as reheating temperature $T_{re} < T_{BBN}$, and iv) Pink region signifies initial parametric resonance domination which we ignored in this paper. As we discussed for bosonic reheating, based on whether gravitational or non-gravitational sectors dominate the dynamics, we have a fermionic critical coupling $\mathscr{H}_{c}$ which sets a boundary for two distinct scenarios:\\ 
1) {\bf Case-I:} The entire reheating dynamics is dominated by the Yukawa coupling. For this case, the fermionic coupling parameter is in the range $h^r>\mathscr{H}_{c}$. The critical coupling $\mathscr{H}_{c}$ is identified by equating the maximum energy densities from non-gravitational and gravitational sector $\rho_f^{r,\,max}=\rho_{gr}^{r,\,max}$. One can find the expression for the critical coupling for fermionic reheating as 
\bea
\mathscr{H}_c=\left[\frac{3(5-9w_\phi)H_{end}^2}{128M^2_p(1+15w_\phi)}\frac{\left[\left(\frac{9+15w_\phi}{8}\right)^{-\frac{8}{1+15w_\phi}}-\left(\frac{9+15w_\phi}{8}\right)^{-\frac{9+15w_\phi}{1+15w_\phi}}\right]}{\left[\left(\frac{8}{3(1+3w_\phi)}\right)^{-\frac{3(1+3w_\phi)}{5-9w_\phi}}-\left(\frac{8}{3(1+3w_\phi)}\right)^{-\frac{8}{5-9w_\phi}}\right]}\right]^{1/2}\,.
\eea
The maximum energy densities for both gravitational and non-gravitational sectors appear at the initial stage of reheating, and as the radiation bath associated with gravity mediated process dilutes faster than the non-gravitational one, for $h^r>\mathscr{H}_c$ reheating dynamics always have explicit fermionic coupling domination.\\2) {\bf Case-II:} For this case coupling parameter satisfies $h^r<\mathscr{H}_c$ and both sectors partially dominates the reheating phase. However, when EoS $w_\phi$ lies above $0.65$, gravity-mediated decay controls the reheating phase termed as gravitational reheating. In our succeeding discussion, we will discuss these two cases in detail:\\

\textbf{\bf Case-I: Coupling strength $\bf h^r>\mathscr{H}_c$:}
In this coupling regime, non-gravitational decay of inflaton into fermionic radiation controls the entire reheating process. In order to give analytical estimation and compare our result with the zero temperature scenario, we consider two separate regimes of the maximum radiation temperature (see  Fig.\ref{maxfer} for its depiction), and those are as follows: 

\underline{ When $ T_{rad}^{max}>m^{end}_\phi$}: If the maximum radiation temperature is greater than $m_{\phi}^{end}$, the coupling parameter associated with this region entirely lies in the pink region of Fig.\ref{boundfermion} (non-perturbative resonance-dominated region). Due to strong inflaton-Fermion coupling, 
the gravitational channel is naturally subdominant throughout the reheating, and the thermal effect is non-negligible from early stage of reheating. However, its effectiveness through out the reheating process will depend on how $T_{rad}/m_\phi$ evolves. Since $T_{rad}>m_\phi$ at the initial stage, decay width can be approximated as,
\begin{equation}{\label{mdr}}
   \Gamma_{\phi\rightarrow \bar ff}=\frac{(h^r)^2}{8\pi}\frac{m^2_\phi(t)}{4\,T_{rad}} \,.
\end{equation}
With the aforementioned decay width, one can estimate the behavior of radiation energy density,
\begin{equation}\label{radferwth}
    \rho_f^r(A)=\left[\frac{\zeta(w_\phi)\,(h^r)^2\epsilon^{1/4}M^2_P}{A^5}(m^{end}_\phi)^2H_{end}\left(A^{\frac{7-15w_\phi}{2}}-1\right)\right]^{4/5},~~\mbox{where}~~\zeta(w_\phi)=\frac{15(1+w_\phi)}{64\pi(7-15w_\phi)}.
\end{equation}
The above equation suggests that the evolution of the radiation component is entirely different in two different regimes
 \begin{itemize}
  \item \underline{ $ w_\phi>7/15$ :} Most of the production occurs at the initial reheating stage, and temperature decreases with the scale factor as $A^{-1}$. Since the ratio $T_{rad}/m_\phi$ behaves as $A^{3\,w_\phi-1}$, $T_{rad}$ always remains greater than $m_\phi$, and hence the finite temperature effect will be significant till the end of reheating. The reheating temperature for this case assumes the following form,
  \begin{equation}
    T_{re}=\left[\frac{\zeta(w_\phi)\,(h^r)^2M^2_P}{\epsilon}(m^{end}_\phi)^2H_{end}A^{-5}_{re}\right]^{1/5},\,A_{re}=\left[\frac{\zeta(w_\phi)\,h^2\epsilon^{1/4}M^2_p(m^{end}_\phi)^2H_{end}}{(3M^2_pH^2_{end})^{5/4}}\right]^{\frac{5(1-3w_\phi)}{4}}.
    \end{equation}
%%%%%%%%%%%%%%%%%%%%%%%%%%%%%%%%%%%%5
    \item \underline{ $0\leq w_\phi<7/15$ :} For this case, the ratio  ${T_{rad}}/m_\phi\propto A^{\frac{3}{10}(5w_\phi-1)}$ induces two different evolution history with regard to the finite temperature effect. It turns out that when EoS stays within $0\leq w_\phi<1/5$, due to initial high radiation temperature thermal correction will have a significant effect. As the reheating progresses, such effect diminishes with the complete takeover by the zero temperature dynamics at a certain value of scale factor, which depends on the inflaton equation of state as follows,
  \begin{equation}\label{fercross}
      A_{c}=\left(\frac{\zeta(w_\phi)\,(h^r)^2M^2_pH_{end}}{\epsilon(m^{end}_\phi)^3}\right)^{2/{(3-15w_\phi)}}\,.
  \end{equation}
  \begin{figure}
          \begin{center}
          \includegraphics[width=7.00cm]
          {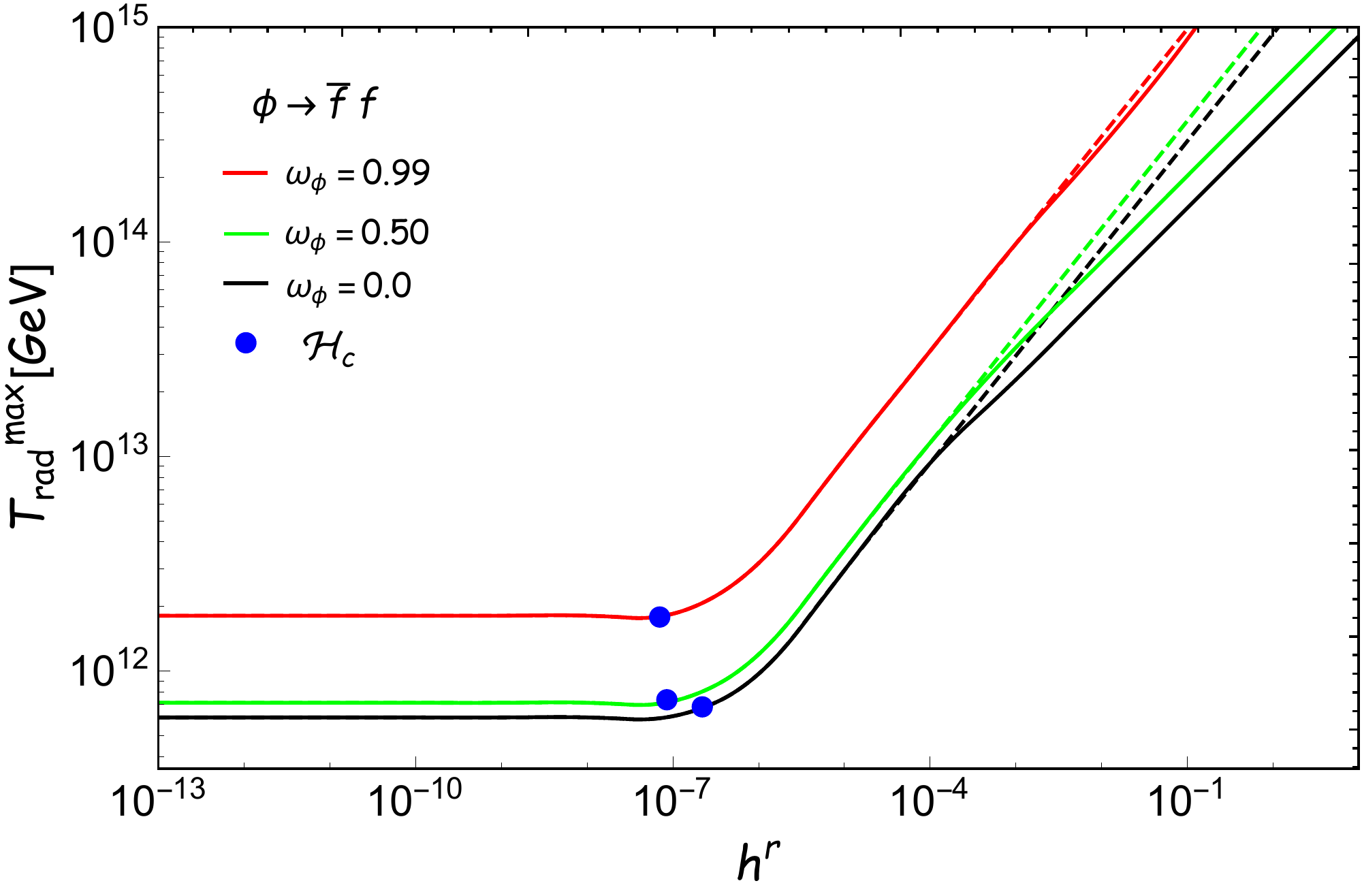}
          \caption{The description of this plot is the same as Fig.\ref{Tmax}, the main difference is that here we have shown results for the fermionic reheating.}
          \label{maxfer}
          \end{center}
      \end{figure}
  After this intermediate scale factor, the dynamics is governed by the zero temperature decay channel following the Eq.\ref{radevwth} (see without thermal effect section of Appendix-\ref{apfer} for details calculation) till the end of reheating and eventually equating $\rho_\phi=\rho_f^r$, we find the associated reheating temperature as,
\bea \label{trewthm}
T_{re}=\left(\frac{6M^2_p(1+w_\phi)m^{end}_\phi H_{end}}{8\pi\epsilon(5-9w_\phi)} (h^r)^2\right)^{1/4}A_{re}^{\frac{-3(1+3w_\phi}{8}},\,A_{re}=\left(\frac{8\pi(5-9w_\phi)H_{end}}{2(1+w_\phi)m^{end}_\phi}(h^r)^2\right)^{\frac{2}{3-3w_\phi}} .
\eea
On the other hand when $w_\phi$ is in between $1/5<w_\phi<7/15$, the thermal effect will be non-negligible throughout the entire reheating history, and we have the associated reheating temperature
\begin{equation} \label{trethm}
    T_{re}=\left(\frac{\zeta(w_\phi)\,(h^rM_P)^2}{\epsilon}(m^{end}_\phi)^2H_{end}\right)^{1/5}A^{\frac{-3-15w_\phi}{10}}_{re},A_{re}=\left[\frac{\zeta(w_\phi)\epsilon^{\frac{1}{4}}(h^rM_p m^{end}_\phi)^2H_{end}}{(3M^2_pH^2_{end})^{5/4}}\right]^{-\frac{4}{3(3+5w_\phi)}}.
\end{equation} 
\end{itemize}
\begin{figure}[t]\centering
            \includegraphics[width=13.50cm]
          {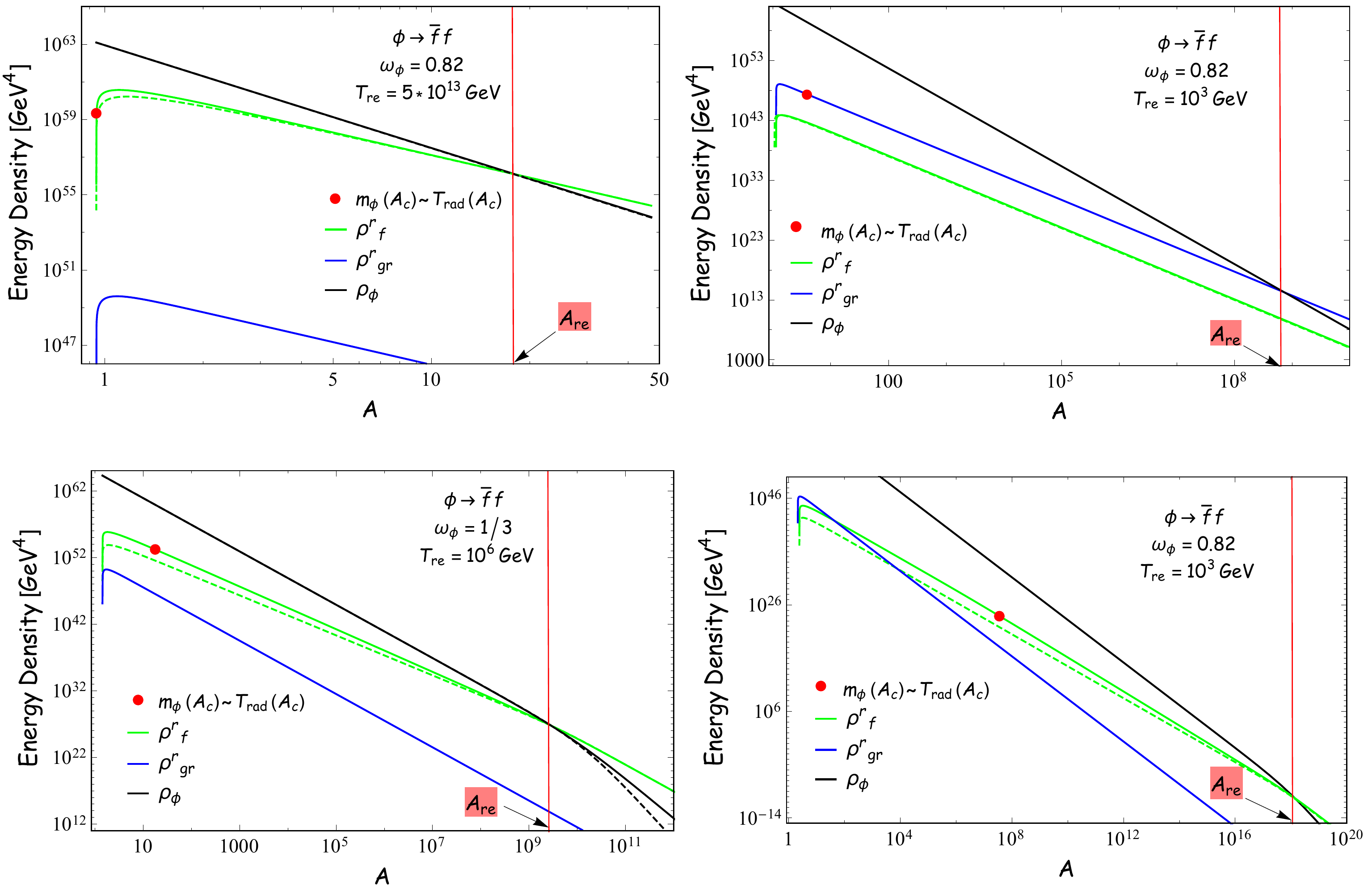}
          \caption{Evolution of inflaton and radiation energy density as a function of normalized scale factor $A$ for $\phi\rightarrow \bar{f}f$ decay channel with and without thermal effect (solid line for with thermal effect and dashed line for without thermal effect). \textbf{Left panel:} Coupling is in the range of  $h^r>\mathscr{H}_{c}$.  \textbf{Right panel:} Coupling is in the range of  $h^r<\mathscr{H}_{c}$. We have considered two distinct values of $w_\phi=(0.82,\,1/3)$ for these two cases.}
          \label{evfer}
          \end{figure}
\underline {When $ T_{gr}^{r,\,max}<T_{rad}^{max}<m^{end}_\phi$}: For this case, the coupling parameter mostly lies in the pink region of Fig.\ref{boundfermion}. Similar to the previous case, whole reheating dynamics are governed by non-gravitational coupling. Since $T_{rad}<m_\phi$ at the initial stage, the thermal effect is minimal. As reheating progresses, depending on the ratio $\frac{T_{rad}}{m_\phi}$, the thermal effect may state dominating the dynamics. Thus initially, the radiation component evolves as
\begin{equation}\label{couplingnth}
\rho_f^r(A)=\frac{6M^2_p(1+w_\phi)m^{end}_\phi H_{end}}{8\pi(5-9w_\phi)A^4} (h^r)^2(A^{\frac{5-9w_\phi}{2}}-1) \,.
\end{equation}
The aforementioned equation clearly suggests that the radiation component behaves differently for $w_\phi<5/9$ and $w_\phi>5/9$. Let us discuss these two cases,
\begin{itemize}
\item \underline{ $w_\phi>5/9$ :}
Maximum production happens initially, and the bath temperature  falls as $A^{-1}$. For this case, $T_{rad}/m_\phi \propto A^{3\,w_\phi-1}$ and hence finite temperature effect will be important near the final stage of reheating, and the reheating temperature can be expressed as,
\bea\label{trenthfermion}
T_{re}=\left(\frac{6M^2_p(1+w_\phi)m^{end}_\phi H_{end}}{8\pi\epsilon(9w_\phi-5)} (h^r)^2\right)^{1/4}A_{re}^{-1}\quad;\quad A_{re}=\left(\frac{8\pi(9w_\phi-5)H_{end}}{2h^2(1+w_\phi)m^{end}_\phi}\right)^{\frac{-1}{1-3w_\phi}}
\eea
\item \underline{ $0\leq w_\phi<5/9$ :} For this case, $T_{rad}/m_\phi \propto A^{-3/8(1-5w_\phi)}$ implies two different evolution histories depending on the value of equation state greater or less than $0.2$. For $0\leq w_\phi < 0.2$, the thermal correction will be subdominant, and the reheating temperature can be simply read off from Eq.\ref{trewthm}. Whereas for EoS $w_\phi>0.2$, the thermal effect will start to dominate at some intermediate time within the reheating phase, which we call the crossover point, 
\bea \label{Actwthcoupling}
A_c=\left(\frac{6M^2_p(1+w_\phi)m^{end}_\phi H_{end}}{8\pi\epsilon(5-9w_\phi)} (h^r)^2\right)^{\frac{-2}{3(1-5w_\phi)}},
\eea
and reheating temperature is given by Eqn.\ref{trethm}.
\end{itemize}
\begin{longtable}{|c|c|c|c|c|c|c|}
 \caption{The temperature evolution for fermionic reheating}\label{tempFR}\\
               \hline
            \multirow{3}{*}{\text{Channel}}& \multicolumn{2}{|c|}{
                 $T\ll m_\phi(t)$ (\text{Without thermal effect})
 }& \multicolumn{2}{|c|}{ $T\gg m_\phi(t)$ (\text{With thermal effect})}
        \\       
   \cline{2-5}
         & Non-gravitational & Gravitational  & Non-gravitational & Gravitational\\
               \hline
                 $\phi\rightarrow \bar ff$ &$A^{-\frac{3(1+3w_\phi)}{8}}$$(A^{-1})\,$\mbox{for} $w_\phi<5/9(>5/9)$&$A^{-1}$&$A^{-\frac{3(1+5w_\phi)}{10}}$$(A^{-1})$~\mbox{for} $w_\phi<\frac{7}{15}(>\frac{7}{15})$&$A^{-1}$\\\hline              
\end{longtable}

\textbf{Case-II: Coupling strength $\bf h^r< \mathscr{H}_{c}$ :} \\ 
In this coupling  regime, the maximum temperature is always controlled by the gravitational sector $T_{rad}^{max}=T^{r,\,max}_{gr}$. This condition generally satisfies within the entire allowed region shown in Fig.\ref{boundfermion} except the parametric resonance dominated region shaded in pink. Evolution of the different energy densities in two different regimes $h^r>\mathscr{H}_{c}$ and $h^r<\mathscr{H}_{c}$ with two distinct values of inflaton equation of state $w_\phi(1/3,\,0.82)$ are shown in Fig.\ref{evfer}. Depending on the inflaton equation of state, here also we have the following three different scenarios.
\begin{itemize}
    \item \underline{ $ w_\phi>0.65$ :} For this case, the gravitational sector governs the entire reheating phase, and we termed this as gravitational reheating. The parameter space where this condition is met is shaded in light cyan in both the Figs.\ref{boundfermion}. The reheating temperature can be followed from Eq.\ref{tgr}, which depends only on the reheating equation of state.

    \item \underline{ $5/9<w_\phi<0.65$ :} This case turned out to be within the light red shaded region in the $(h,\,w_\phi)$ plane of Fig.\ref{boundfermion}. As the figure suggests, reheating temperature evolved into below BBN temperature, which does not support the standard cosmological constraints. 

      \item \underline{ $0\leq w_\phi<5/9$ :} In this case, the competition between two sectors of production, along with the finite temperature effect, leads to two different physically distinguishable reheating dynamics.
      %A gravitation$\to$ non-gravitational sector crossover ($gr \to ngr$) occurs at any point during reheating, and another crossover point exists where the thermal effect starts affecting. 
      In the $(w_\phi, h)$ plane, the condition under consideration lies in the light yellow region of Fig.\ref{boundfermion}. Here we have two different possibilities depending on how the thermal effect plays its role during the reheating history. As discussed for the bosonic reheating case\\
      1) The thermal effect starts to influence the reheating dynamics in its early stage ($T_{rad}>m_\phi$) during the gravitational decay of inflaton. For this case, the behavior of the radiation component during non-gravitational sector domination is simply followed by Eq.\ref{radferwth}, and we have reheating temperature as in Eq.\ref{trethm}. \\
%      In this context, the $gr\to ngr$ cross-over point is identified as
 %     \begin{equation}
 %   A_{gr\rightarrow ngr}=\left(\frac{4\pi\epsilon(7-15w_\phi)(\mathcal{C}(w_\phi))^5}{3(1+w_\phi)H_{end}(M_pm^{end}_\phi h^r)^2}\right)^{\frac{2}{15(1-w_\phi)}}\,.
%\end{equation}\\
2) The thermal effect starts dominance during the later stage of reheating when it is governed by non-gravitational inflaton decay. The scale factor associated with the point where the thermal effect starts to influence the dynamics can be the same as the Eq.\ref{Actwthcoupling}, and reheating temperature is given by Eq.\ref{trethm}.
%the scale factor associated with this transition turns out as
%\begin{equation}
%     A_{gr\rightarrow ngr}=\left(\frac{8\pi\epsilon(5-9w_\phi)(\mathcal{C}(w_\phi))^4}{6(1+w_\phi)H_{end}m^{end}_\phi M^2_p(h^r)^2}\right)^{\frac{2}{(9w_\phi-5)}}\,.
%\end{equation}
\end{itemize}
 
%%%%%%%%%%%%%%%%%%%%%%%%%%%%%%%%%%%%%%%%%%%%%
\subsection{Inflaton phenomenology: Constraining reheating and fermionic decay parameters}
Similar to the bosonic reheating case, to illustrate our results in terms of inflationary parameter $n_s$ and to see how the coupling strength behaves as a function of reheating reheating temperature, we have taken five different sample values of $w_\phi=(0,\,0.2,\,0.5,\,0.82,\,0.99)$ and compare the results for with and without thermal effect.

i) \underline{Reheating temperature ($T_{re}$) in terms inflationary (CMB) parameter ($n_s$):}
The qualitative relation between reheating temperature and inflationary parameters remains similar to that of the bosonic case discussed earlier. Therefore, the possible bounds on inflationary parameters such as spectral index $n_s$, the maximum inflationary e-folding number $N^{max}_k$ do not depend on the details of the reheating dynamics but only the reheating temperature. As a result, bounds on the inflationary parameters remain the same as bosonic reheating (see, for instance, Table-\ref{boundBR}). This can easily be read off from the left plot of Fig.\ref{ntre}. In addition, Fig.(\ref{ntre}), further indicates that the thermal feedback on the decay process does not affect the reheating temperature variation with $n_s$.\\
%%%%%%%%%%%%%%%%%%%%%%%%%%%%%%%%%%%%%%%%%%%%%%            
\begin{figure}[t!] 
 	\begin{center}
 		\includegraphics[width=15.0cm,height=5.00cm]{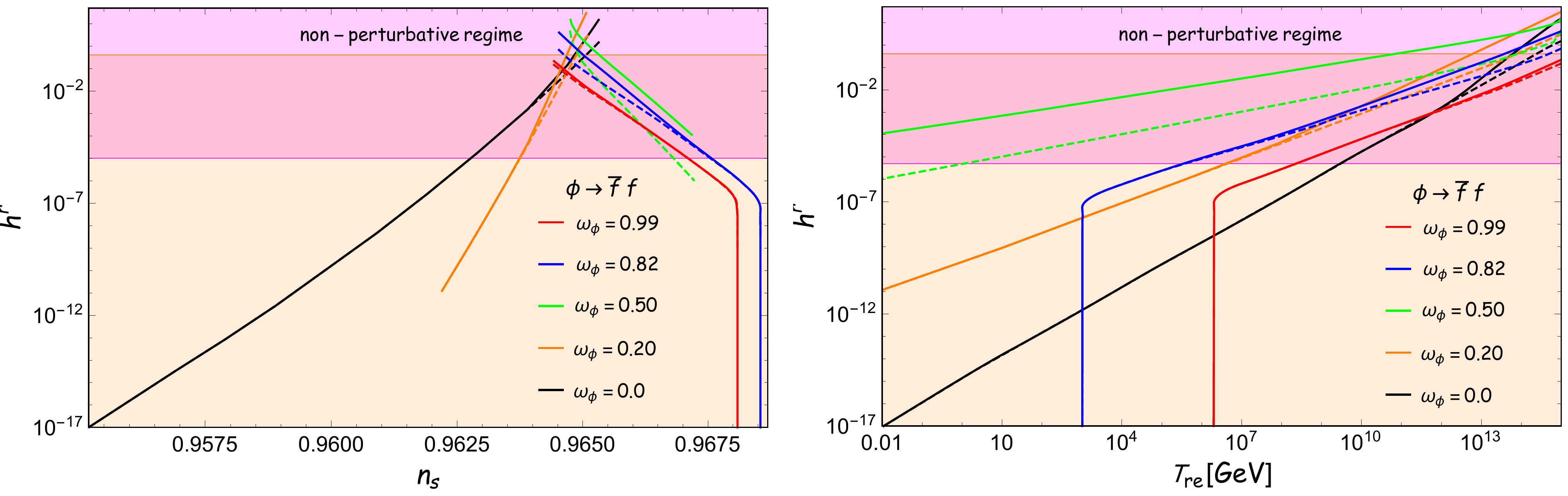}\quad
 		\caption{ The description of the plot is the same as Fig. (\ref{Fig5}), the only difference is that here we have plotted for fermionic reheating and the self-resonance dominated region is same for both $w_\phi=0.5$ and $0.2$ shaded by orange-shaded region.}
 		\label{Fig10}
 	\end{center}
 \end{figure}
 %%%%%%%%%%%%%%%%%%%%%%%%%%%%%%%%%%%%%%%%%%%%%%%%%%%%%%%
% \floatbarrier
 (ii) \underline{Constraining inflaton couplings with fermionic radiation $(h^r)$}:  
 For the fermionic reheating, the parameter space in Fig.\ref{boundfermion} illustrates different regions in $(w_{\phi}, h^r)$ plane, where the effect of different inflaton decay channels on the reheating process can be read off. We have given two plots with and without finite temperature effects for clear depiction.
  
 An interesting distinction can be observed as compared to the scalar reheating case is that for fermionic reheating inflaton-Fermion coupling $h^r$ does not vary monotonically with respect to $w_{\phi}$ given a fixed reheating temperature. There exists a critical value of $w_{\phi} \simeq 7/15 (5/9)$ for with (without) thermal effect, below which one requires a higher $h^r$ value for a higher equation state for a fixed reheating temperature. And this can be understood from the behavior of Fermion production rate $\propto \Gamma_{\phi\rightarrow \bar f f} \rho_{\phi} \propto (h^r)^2 m_\phi(t) \rho_{\phi}$. With increasing $w_{\phi}$, the effective mass of the inflaton decreases faster with time; hence, to achieve a fixed reheating temperature, $h^r$ needs to be enhanced. However, this simple physical argument is no longer tenable after $w_{\phi} \geq 7/15~(5/9)$ for with (without) thermal effect. For such cases, most of the production happens initially, and the radiation energy density simply dilutes as $A^{-4}$, which is slower than that of the inflaton energy density. In such a situation with increasing $w_\phi$ and fixed reheating temperature, we need a lower the value of $h^r$ to satisfy the reheating condition $\rho_\phi=\rho_f^r$.

Due to its intrinsic nature, the finite temperature Fermion bath diminishes its own production rate from the inflaton condensate. Consequently, for successful reheating one needs higher values of the dimensionless coupling parameters $h^r$ as compared to the zero temperature case. This can be clearly observed from Fig.\ref{Fig10}, and such behavior is opposite to that of the scalar reheating case. The qualitative behavior of the fermionic coupling in terms of the spectral index and reheating temperature are the same as that of the scalar reheating case. For example, for $w_\phi=0$ the coupling parameter $h^r$ with thermal effect begins to affect only at very high temperatures at around $\sim 10^{13}$ GeV, and in terms of the spectral index, the deviation is visible for $n_s>0.9645$. On the other hand, since the effective mass of the inflaton varies as $A^{-3w_\phi}$ (see, for instance, Eq.\ref{mphit}), for $w_\phi > 0$, $m_\phi (t)$ decrease faster with increasing $\omega_{\phi}$ and the $T_{rad} >  m_\phi(t)$ condition becomes easier to fulfill even at a lower radiation temperature. As a result, for $w_\phi=0.2$, the thermal effect starts dominating even at smaller radiation temperature $T_{rad}\geq 10^8$ GeV when $n_s>0.9639$. The situation is entirely different though for $w_\phi>5/9$. For EoS greater than $5/9$, the maximum radiation production happens initially, and hence thermal effect will be dominant from the beginning, which indicates maximum radiation temperature $T_{rad}^{max}>m_\phi^{end}$. From Fig.\ref{Fig10}, we can clearly see that for EoS $w_\phi=(0.82,\,0.99)$, the deviation between the results for with and without thermal effect start visible when $T^{max}_{rad}\sim m_\phi^{end}$.
%%%%%%%%%%%%%%%%%%%%%%%%%%%%%%%%%%%%%%%%
\section{\textbf{FIMPs and WIMPs during reheating and Observational constraints}}{\label{section5}} 
Discussion on DM will be considered in three parts. In the first part, we discuss the production of DM exclusively from the inflaton through non-gravitational and gravitational interaction. We mainly point out the constraints on the inflaton-DM coupling and DM mass from both theory and observation. Since it is produced solely from the inflton decay, we call these as FIMP like DM. 
\begin{figure}[t!] 
 	\begin{center}
 		   \includegraphics[width=16.0cm,height=2.0cm]{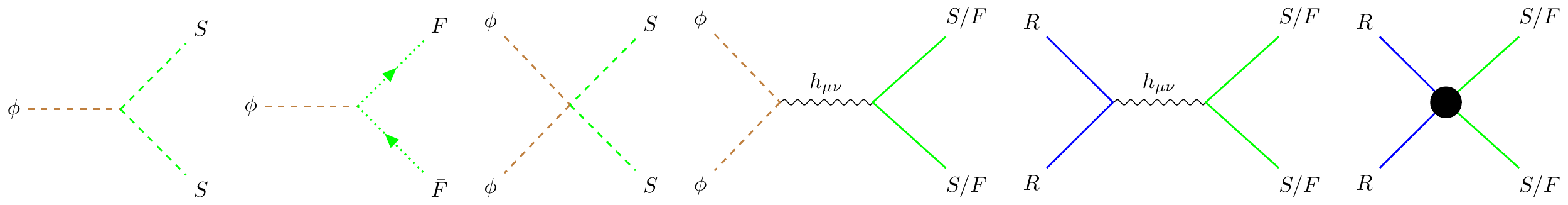}
  	\caption{Fynmann diagram for dark-matter (DM) production. The solid black circle corresponds to effective vertex representing $2\to2$ scattering process between the bath particle (R) and DM (S,F).}
 		\label{fynDM}
 	\end{center}
 \end{figure}
%%%%%%%%%%%%%%%%%%%%%%%%%%%%%%%%%%%%%
In the second part, we discuss the production from the thermal bath assuming an effective radiation to DM annihilation cross-section $\langle \sigma v\rangle$, added with the universal gravitational production discussed in the first section. For this case, we will have both the Freeze-in and Freeze-out production scenarios depending upon the strength of  $\langle \sigma v\rangle$. The DM produced due to thermal Freeze-out from the radiation bath will be generally called WIMPs. On the other hand, DM  produced by the decay process from the radiation bath via the Freeze-in mechanism will be called FIMPs.  
%%%%%%%%%%%%%%%%%%%%%%%%%%%%%%%%%%%%%%%
In the third part, we discuss about the experimental constraints on various reheating and DM scenarios.  
%%%%%%%%%%%%%%%%%%%%%%%%%%%%%%%%%%%%%%%
\subsection{Constraining light DM through $\Delta N_{\mbox{eff}}$ during BBN :} In this short section, we would like to point out the BBN bound on the additional light degrees of freedom. In our present paper, we use this bound to constrain the DM parameter space of both FIMPs and WIMPs scenarios that were relativistic at the time of BBN. If the DM is relativistic at the time of BBN, it would inevitably modify the  background expansion and may jeopardize the formation of the light elements, which is tightly constrained by the BBN observation. The total effective number of relativistic degrees of freedom is defined as $N_{\mbox{eff}}=N^{\mbox{SM}}_{\mbox{eff}}+\Delta N_{\mbox{eff}}$. At the time of BBN, if the active neutrinos are the only relativistic without any new particle, $N_{\mbox{eff}}=N^{\mbox{SM}}_{\mbox{eff}}=3.046$ ($\Delta N_{\mbox{eff}}=0$). BBN observation gives the bound of $\Delta N_{\mbox{eff}}\leq 0.5$ at $95\%$ \cite{Cyburt:2015mya,Knapen:2017xzo,Nollett:2014lwa,Paul:2018njm} confidence level. A general expression of $\Delta N_{\mbox{eff}}$ can be written as\cite{Jinno:2012xb}
\begin{equation}
    \Delta N_{\mbox{eff}}=\left(\frac{\mbox{extra radiation energy density}(\rho_{DM})}{\mbox{energy density of single SM neutrino species}(\rho_{\nu})}\right)_{T=T_{BBN}}
   =\left(\frac{43}{7}\right)\left(\frac{\rho_{DM}}{\rho_{rad}}\right)_{T=T_{BBN}}
\end{equation}
We will be discussing two production mechanisms. For FIMP like DM, we intend to separately discuss its production from the direct inflaton decay and radiation bath. For inflaton decaying into DM, production freezes during reheating or at the end of reheating, depending on the decay channels and DM mass. On the other hand, for DM from the thermal bath, its production rate crucially depends on the radiation production rate. Therefore, in this case, also freeze-in occurs mostly during or at the end of reheating, depending on DM mass. Overall, for the freeze-in mechanism, we, therefore, can always express
\begin{equation} \label{dneff}
    \Delta N_{\mbox{eff}}=\left(\frac{43}{7}\right)\left(\frac{\rho_{DM}}{\rho_{rad}}\right)_{T=T_{BBN}}=\left(\frac{43}{7}\right)\left(\frac{\rho_{DM}}{\rho_{rad}}\right)_{T=T_{re}} .
\end{equation}
If the DM is relativistic after Freeze-in, both radiation ($\rho_{rad}$) and relativistic DM energy density ($\rho_{DM}$) fall as $a^{-4}$. The ratio ${\rho_{DM}}/{\rho_{rad}}$, therefore, stays constant between reheating and BBN. Hence, the above equality Eq.\ref{dneff}, holds true generically for any FIMP scenario.\\
For WIMP, on the other hand, the situation becomes very different but simpler. For such cases, till it freezes out, DM remains in equilibrium with the thermal bath. Therefore, any relativistic DM being in the thermal bath at the time of BBN always behaves like an additional degree of freedom. Therefore, $\Delta N_{\mbox{eff}}$ naturally transforms into \cite{Wallisch:2018rzj,Jinno:2012xb}
 \begin{eqnarray}{\label{nef1}}
 &&  \Delta N_{\mbox{eff}}= \frac{4}{7}j_x = \left\{\begin{array}{ll}
        &0.571~~~\mbox{scalar DM}  \\
        & 1.14~~~\mbox{fermionic DM}
   \end{array}\right.
 \end{eqnarray}
 Where $j_x$ is the DM's intrinsic number of degrees of freedom, WIMPs mass lighter than $T_{BBN}\sim 10 $ MeV always behaves like dark radiation at the time of BBN.  Hence, Therefore, all the WIMPs of mass $m_x\leq 10$ MeV  violate the BBN bound of $\Delta N_{\mbox{eff}}$ (see Eq. (\ref{nef1})). 
%%%%%%%%%%%%%%%%%%%%%%%%%%%%%%%%%%%%%%%%%
\subsection{Freeze-in production of dark matter from inflaton decay and constraints}
Similar to massless radiation production, we will now discuss DM production during reheating, considering various decay channels for both scalar and Fermion DM. As discussed in the introduction, we considered two categories of production channels: (i) DM production from inflaton through gravitational scattering and (ii) production through explicit coupling-dependent decay channel.\\
The governing equation for the DM number density ($n_x$) produced from direct inflaton decay takes the following form
\begin{equation}{\label{dm1}}
%\dot{n}_x+3Hn_x=\frac{\Gamma^{\phi,c}_x\rho_\phi}{\langle E^c_x\rangle^\phi}+\frac{\Gamma^{\phi,g}_x\rho_\phi}{\langle E^g_x\rangle^\phi}
\dot{n}_x+3Hn_x=\frac{\Gamma^{\phi}_x\rho_\phi}{\langle E_x\rangle^\phi}
\end{equation} 
where $\Gamma^{\phi}_x$ is the inflaton decay width to DM and $\langle E_x\rangle^\phi$ the average energy of the DM particles.
\subsubsection{\bf freeze-in production via Gravitational interaction}
Gravitational freeze-in production of DM is universal in nature and hence will always be present in any inflationary scenario. In this work, we have considered scalar (S) and Fermion (F) DM (see Fig.\ref{fynDM} for relevant Feynman diagrams for gravitational scatting from inflaton). The decay rates associated with the gravitational production are\cite{Mambrini:2021zpp,Barman:2021ugy}
\begin{eqnarray}
&&\Gamma^\phi_x=\left\{
\begin{array}{ll}
     & \frac{\rho_\phi m_\phi}{1024\pi M^4_p}\left(1+\frac{m^2_x}{2m^2_\phi}\right)^2\sqrt{1-\frac{m^2_x}{m^2_\phi}}~~~~~~~\mbox{for}~~~~h_{\mu\nu}(T^{\mu\nu}_S +T^{\mu\nu}_\phi)\\
     & \frac{\rho_\phi m^2_f}{4096\pi M^4_pm_\phi}\left(1-\frac{m^2_x}{m^2_\phi}\right)^{3/2}~~~~~~~~~~~~~\mbox{for}~~~~h_{\mu\nu}(T^{\mu\nu}_F +T^{\mu\nu}_\phi) .
\end{array}\right.
\end{eqnarray}
Since, DM are feebly coupled with the radiation bath, the thermal correction to the decay width will be unimportant. Using the above equations in Eq.\ref{dm1}, we have obtained the following solutions for the number density,
\begin{eqnarray}
&&n^{g}_x(A)=\left\{
\begin{array}{ll}
     &\frac{3H^3_{end}}{512\pi(1+3w_\phi)A^3}\left(1-A^{-\frac{3(1+3w_\phi)}{2}}\right)~~~~~~~~~~~~~~~~~~~~~\mbox{for}~~~~h_{\mu\nu}(T^{\mu\nu}_S +T^{\mu\nu}_\phi)\\ 
     & \frac{3H^3_{end}}{2048\pi(1-w_\phi)A^3}\left(\frac{m_x}{m^{end}_\phi}\right)^2\left(1-A^{-\frac{3(1-w_\phi)}{2}}\right)~~~~~~~~\mbox{for}~~~~h_{\mu\nu}(T^{\mu\nu}_F +T^{\mu\nu}_\phi) .
\end{array}\right.
\end{eqnarray}
Therefore, the gravitational contribution to the DM abundance is calculated as\footnote{In the final DM abundance, we also include the gravitational production of DM from the thermal bath. But for scalar DMs, which are produced from the thermal bath scattering, it has no contribution to the final DM abundance as it is always subdominant compared to the production from inflaton scattering through gravitational interaction. However, for the fermionic DM, gravitational production from the radiation bath can be dominant when reheating temperature $T_{re}\geq 10^{13}$ GeV ( for the details calculation, see Ref.\cite{Haque:2021mab})} 
\begin{eqnarray}
&&\Omega^{g}_xh^2=\left\{
\begin{array}{ll}
     &\Omega_rh^2\frac{3m_xH^3_{end}}{512\pi\epsilon(1+3w_\phi) T_{now}}\left(\frac{\epsilon}{3M_p^2H^2_{end}}\right)^{1/{1+w_\phi}}T_{re}^{\frac{1-3w_\phi}{1+w_\phi}}~~~~~~~~~~~~~~\mbox{for}~~~~h_{\mu\nu}(T^{\mu\nu}_S +T^{\mu\nu}_\phi)\\ 
     & \Omega_rh^2\frac{3m^3_xH^3_{end}}{2048\pi\epsilon(1-w_\phi)\mathscr{M}_\phi^2 T_{now}}\left(\frac{\epsilon}{3M_p^2H^2_{end}}\right)^{\frac{1+2w_\phi}{1+w_\phi}}T_{re}^{\frac{1-3w_\phi}{1+w_\phi}}~~~~~~~~\mbox{for}~~~~h_{\mu\nu}(T^{\mu\nu}_F +T^{\mu\nu}_\phi) .
\end{array}\right.
\end{eqnarray}
Where the suffix "g" stands for production due to the gravitational scattering process. $\Omega_rh^2 \simeq 5\times 10^{-5}$ is the present value of the radiation abundance. It is clear from the expression above that gravitational production depends on $H_{end}$, $m^{end}_\phi$, and DM mass $m_x$. Given an inflation model, inflationary parameters such as $H_{end}$, $m^{end}_\phi$ are fixed by CMB observation. Therefore, $w_\phi$ and DM mass $m_x$ are the only free-parameter. Therefore, the present DM abundance can completely fix the DM mass once a particular inflaton equation of states $w_\phi$ is assumed. We will later observe that the aforesaid mass will 
set a maximum possible limit on the DM mass, which we symbolized as $m^{g,max}_{x}$, for a large range of coupling and the inflaton equation of state. To this end, let us reiterate that due to its universal nature, the gravitational contribution to DM must be added to all the additional production processes we take up in the following sections.
%%%%%%%%%%%%%%%%%%%%%%%%%%%%%%%%%%%%%%%%%
\subsubsection{\bf freeze-in production via inflaton direct decay}
We introduce different inflation coupling to DM. We consider three types of possible interaction: $g^x_1\phi S^2$,  $g^x_2\phi^2 S^2$, $h^x\phi\bar FF$ (see Fig.\ref{fynDM} for relevant Feynman diagrams), and corresponding decay widths are,
\begin{eqnarray}
&&\Gamma^{\phi,c}_x=\left\{
\begin{array}{ll}
&\frac{(g^x_1)^2}{8\pi m_\phi(t)}\sqrt{1-\frac{4m^2_x}{m^2_\phi}}~~~~~~~~~~~~~~~~~~~\mbox{for}~~~~g^x_1\phi S^2\\
&\frac{(g^x_2)^2}{8\pi}\frac{\rho_\phi(t)}{m^3_\phi(t)}\left(1-\frac{m^2_x}{m^2_\phi}\right)^{1/2}~~~~~~~~~~~\mbox{for}~~~~g^x_2\phi^2 S^2\\
&\frac{(h^x)^2}{8\pi}m_\phi(t)\left(1-\frac{4m^2_x}{m^2_\phi}\right)^{3/2}~~~~~~~~~\mbox{for}~~~h^x\phi\bar FF\\
\end{array}\right.
\end{eqnarray}
Using the above equation  in Eq. (\ref{dm1}), we have obtained the following solution of number density,
\begin{eqnarray}{\label{dmsoll}}
&&n_x(A)=\frac{M^2_pH_{end}}{2\pi}\left\{
\begin{array}{ll}
&\frac{(g^x_1)^2}{(1+3w_\phi)(m^{end}_\phi)^2A^3}\left(A^{\frac{3}{2}(1+3w_\phi)}-1\right)~~~~~~\mbox{for}~~~~g^x_1\phi^1 S^2\\
&\frac{3(g^x_2H_{end}M_p)^2}{2(5w_\phi-1)(m^{end}_\phi)^4A^3}\left(A^{-\frac{3}{2}(1-5w_\phi)}-1\right)~~~\mbox{for}~~~g^x_2\phi^2 S^2\\
&\frac{(h^x)^2}{(1-w_\phi)A^3}\left(A^{\frac{3}{2}(1-w_\phi)}-1\right)~~~~~~~~~~~~~~~~~\mbox{for}~~~~~~~h^x\phi\bar FF\\
\end{array}
\right.
\end{eqnarray}
\begin{figure}
          \begin{center}
       \includegraphics[width=005.30cm,height=003.95cm]
         {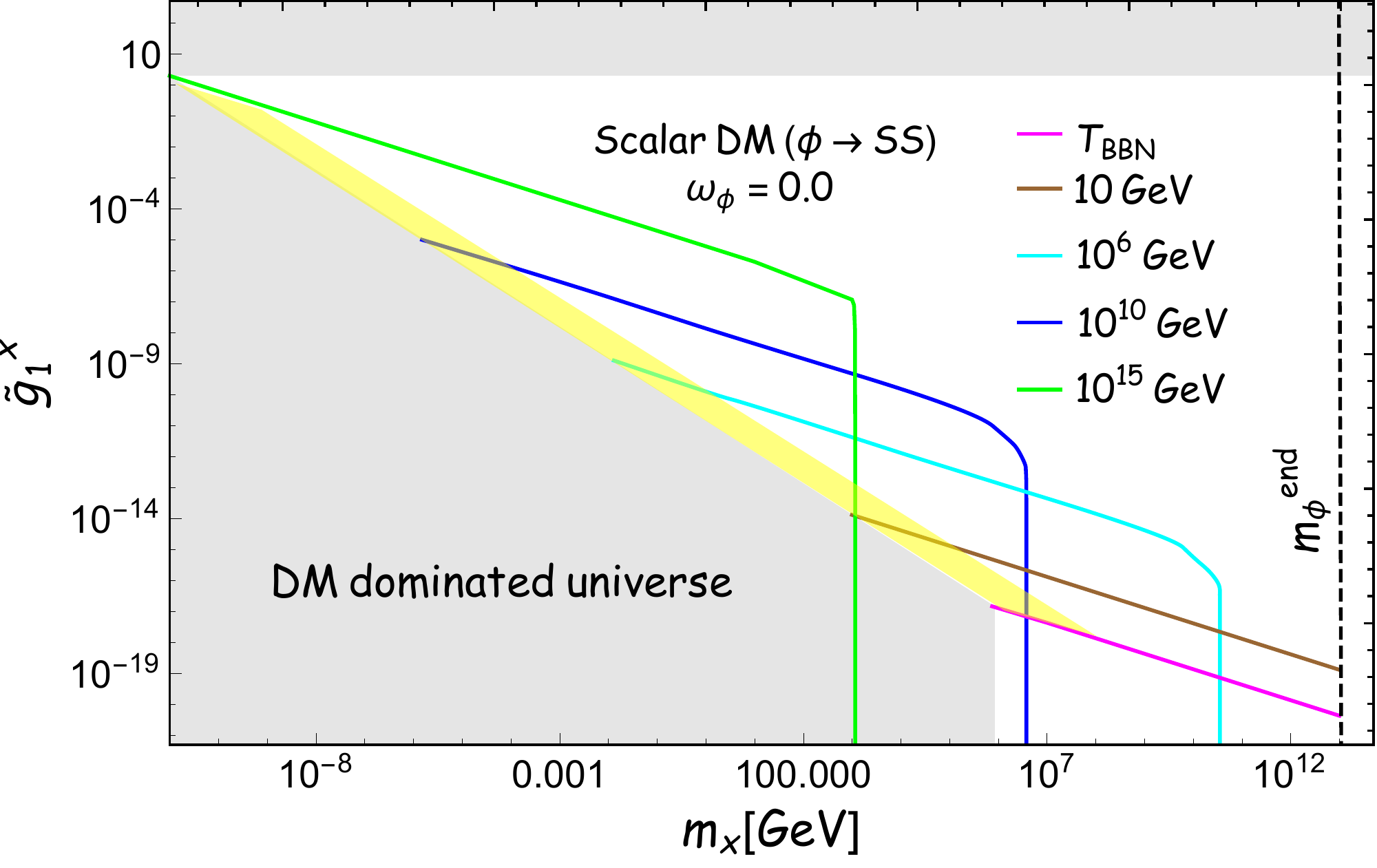}\quad
         \includegraphics[width=005.30cm,height=003.95cm]
          {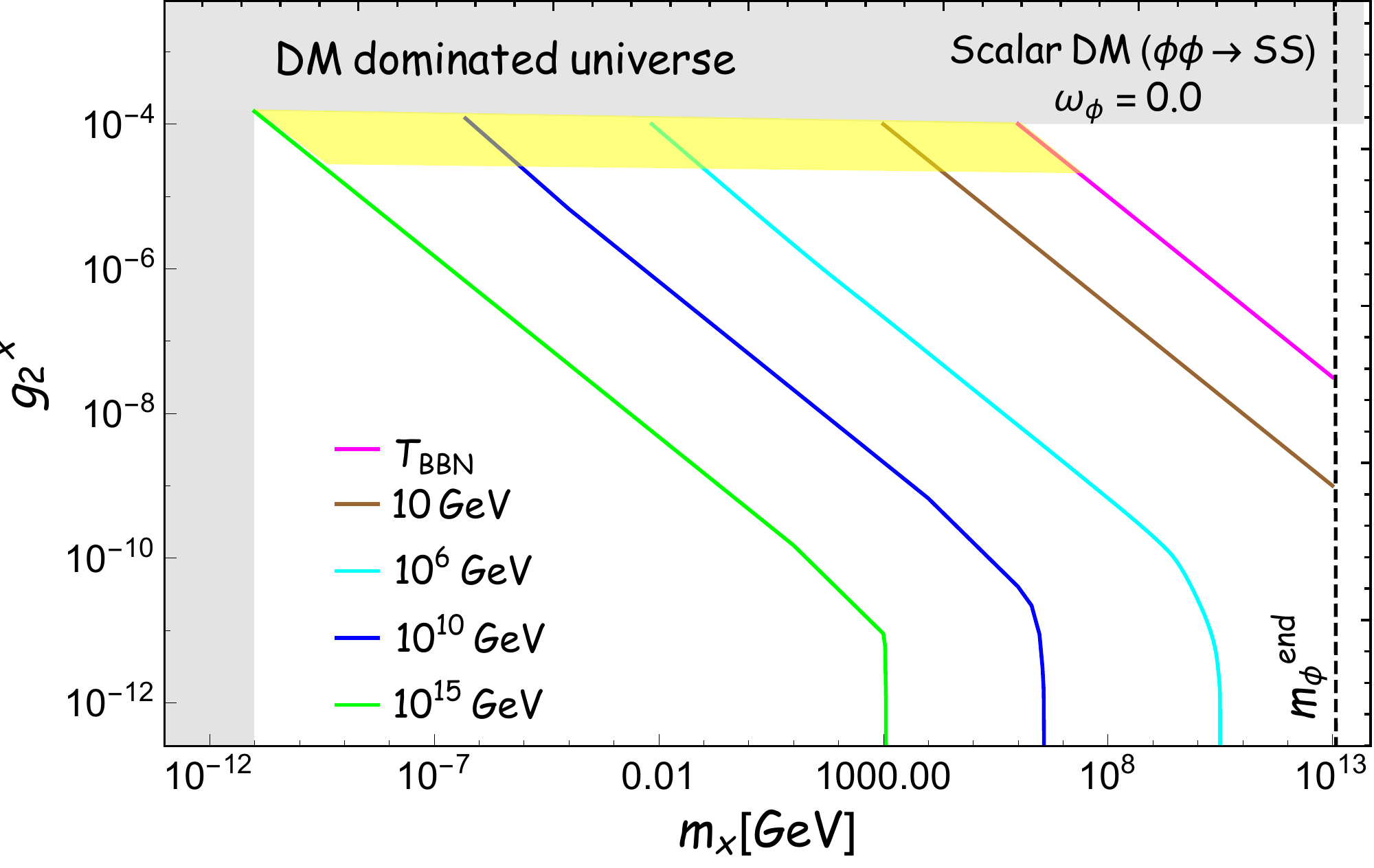}\quad
          \includegraphics[width=005.30cm,height=003.95cm]
          {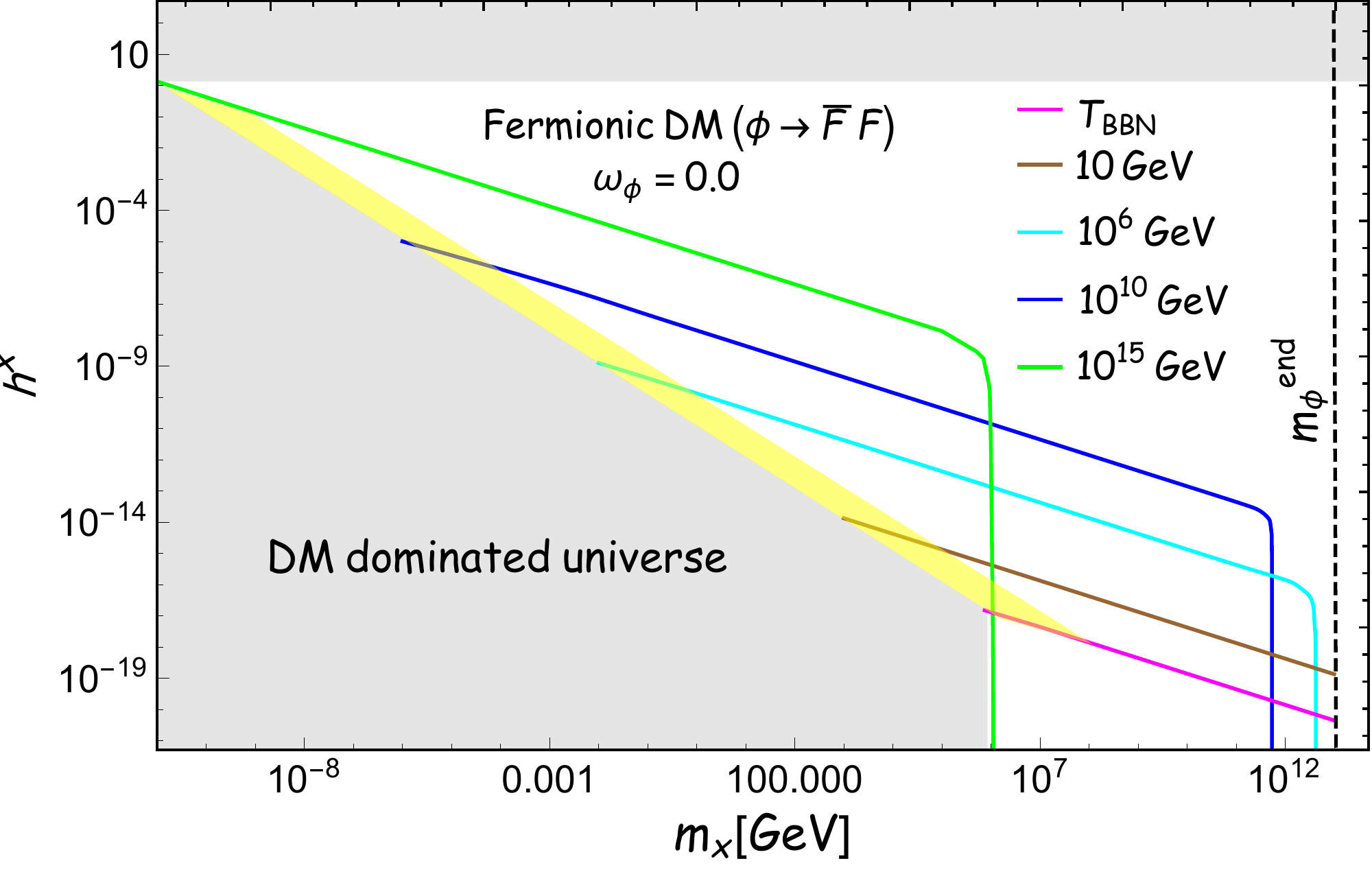}\quad
          	\includegraphics[width=005.30cm,height=003.95cm]
         {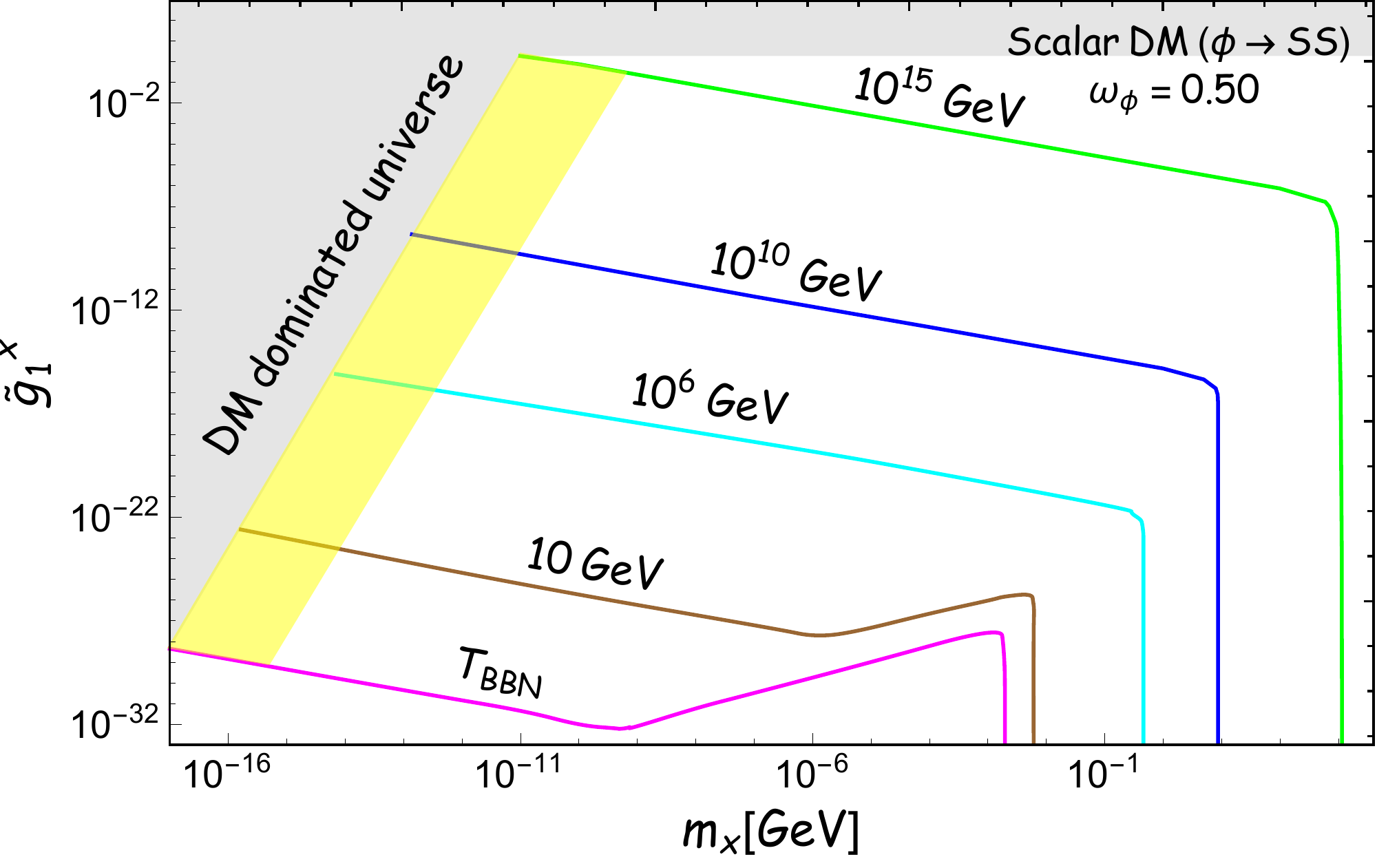}\quad
         \includegraphics[width=005.30cm,height=003.95cm]
          {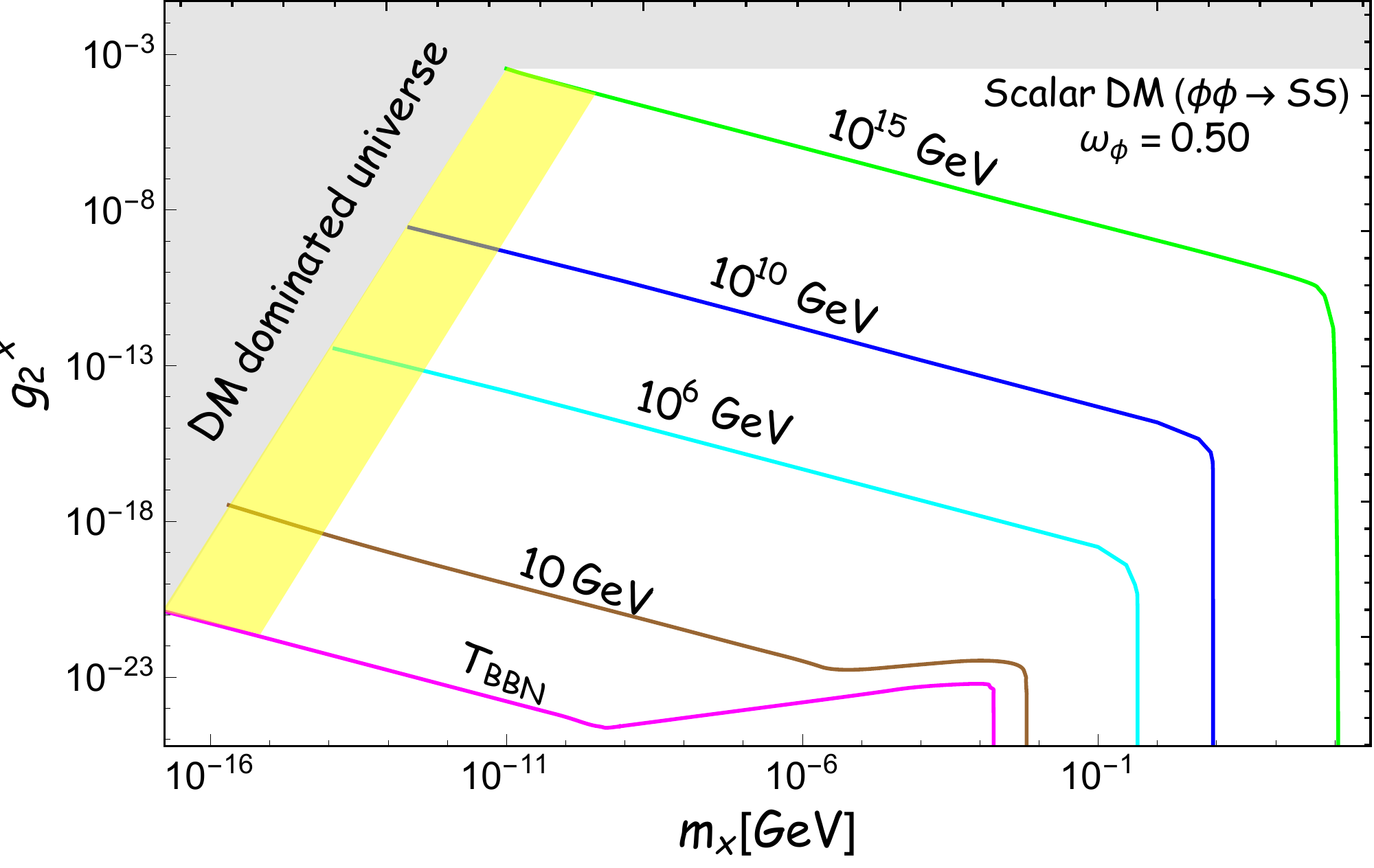}\quad
          \includegraphics[width=005.30cm,height=003.95cm]
          {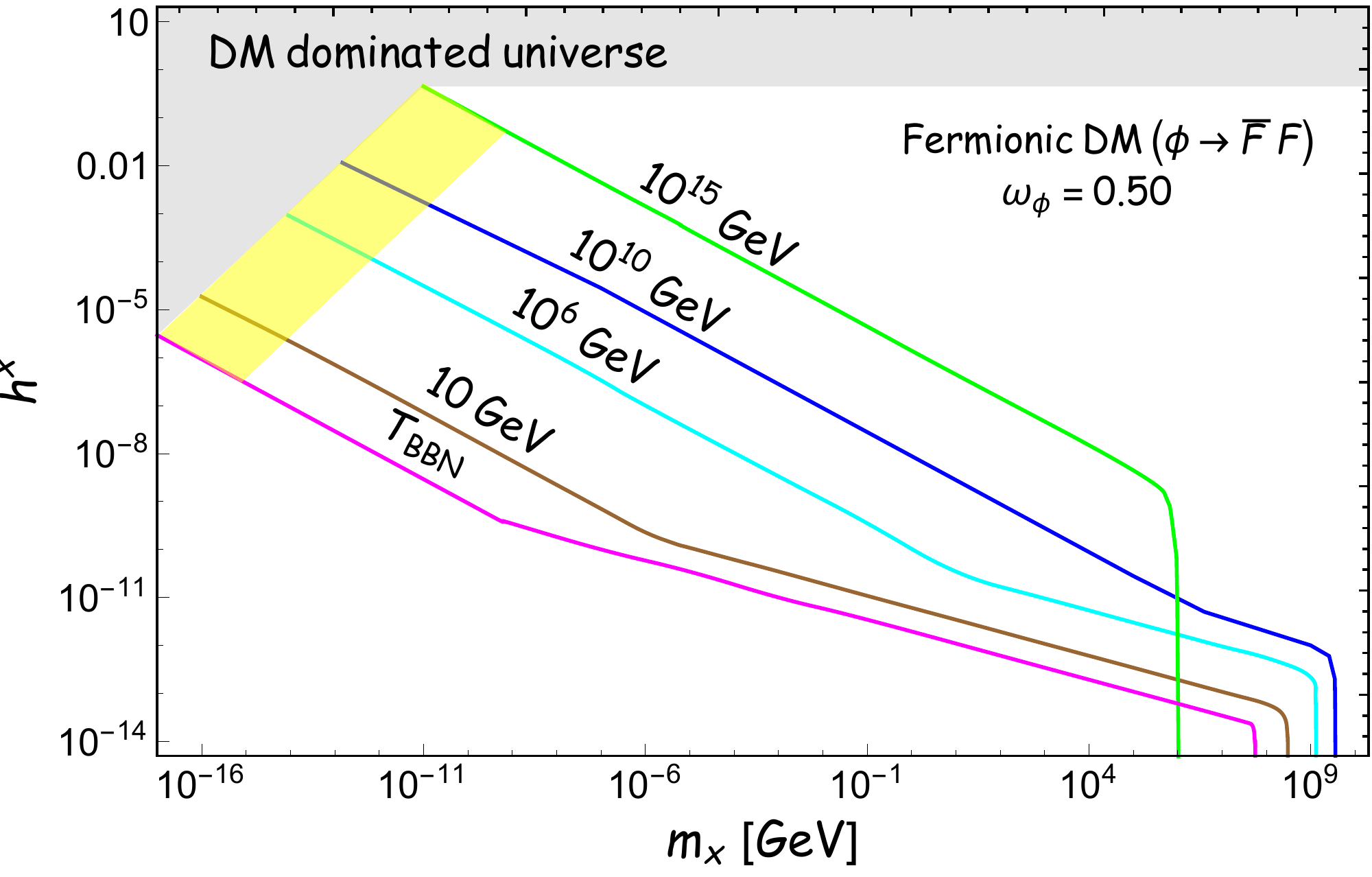}\quad
          \caption{Variation of the inflaton-DM matter coupling  against the DM mass for $w_\phi=0.0, 0.50$ for DM production processes $\phi\rightarrow SS$ (left), $\phi\phi\rightarrow SS$ (middle) and $\phi\rightarrow \bar FF$ (left). The different colour lines correspond to different reheating temperatures. The yellow-shaded region is ruled out by BBN bound of $\Delta N_{\mbox{eff}}$. The gray-shaded region corresponds to the no reheating region where the radiation domination era is impossible after inflaton domination. The vertical dashed lines (upper plot) correspond to the kinematically maximum allowed DM mass $m_\phi^{end}$. }
          \label{DM1}
          \end{center}
      \end{figure}
Unlike the previous gravitational production case, we now have additional  coupling parameters $g^{x}_{i},h^x$ in the problem. Therefore, the present DM abundance will provide the constraint equation between ($m_x, g^{x}_{i}/h^x$) once we fix a particular reheating history by fixing $(w_\phi, T_{re})$ and $g^{r}_{i}/h^r$. 

Depending upon the DM mass, we will have two different expressions for the DM abundance. If $m_x> m_{\phi}(A_{re})$, the DM freezes in before the end of reheating at some intermediate scale factor $A= A_{re}(m_x/{m_\phi(A_{re})})^{-1/{3w_\phi}}$ for $w_\phi\neq 0$, and, if  $m_x< m_{\phi}$, the DM freezes in after the end of reheating. The contribution to the DM abundance for different direct decay channels are calculated as ( $m_x< m_{\phi}(A_{re})$),   
 \begin{eqnarray}{\label{infdmw1}}
 \Omega^{dcay}_xh^2=\frac{ m_x n_x(A_{re})}{\epsilon T^3_{re}T_{now}}\Omega_rh^2= \Omega_rh^2\frac{M_p}{2\sqrt{3\epsilon}\pi T_{now}}\frac{m_x}{T_{re}}\left\{\begin{array}{ll}
       \frac{(g^x_1)^2}{(1+3w_\phi)\mathscr{M}^2_\phi}T^{\frac{-8w_\phi}{1+w_\phi}}_{re}~~~\mbox{for}~~~~g^x_1\phi S^2\\
       \frac{\epsilon (g^x_2)^2}{2(5w_\phi-1)\mathscr{M}^4_\phi} T^{\frac{4(1-3w_\phi)}{1+w_\phi}}_{re}~~~\mbox{for}~~~~g^x_2\phi^2 S^2~~\mbox{with}~~w_\phi>0.2\\
        \frac{\epsilon (g^x_2)^2T^{\frac{2(1-w_\phi)}{1+w_\phi}}_{re}}{2(1-5w_\phi)\mathscr{M}^4_\phi}\left(\frac{\epsilon}{3M_p^2H^2_{end}}\right)^{\frac{5w_\phi-1}{1+w_\phi}}~\mbox{for}~~g^x_2\phi^2 S^2~~\mbox{with}~~w_\phi<0.2\\
      \frac{(h^x)^2}{1-w_\phi}~~~~\mbox{for}~~~~h^x\phi\bar FF.
 \end{array}\right.
 \end{eqnarray}
 We introduce a new symbol, 
$\mathscr{M}_\phi=\sqrt{2n(2n-1)}\beta\Lambda^{2/n}\epsilon^{w/1+w}$ and $m_\phi =\mathscr{M}_\phi T_{re}^{4w_\phi/(1+w_\phi)}$ is the inflaton mass defined at the end of reheating. Note that for $\phi\phi\rightarrow SS$ (with $w_\phi<0.2$) DM production channel, most of the DM production happens at the initial phase of the reheating similar to the gravitational production. On the other hand if $m_x > m_{\phi}(A_{re})$, we have 
\begin{eqnarray}{\label{infdmw2}}
 \Omega^{dcay}_xh^2=\Omega_rh^2\frac{M_p}{2\sqrt{3\epsilon}\pi T_{now}}\frac{m_x}{T_{re}} \left\{\begin{array}{ll}
      & \frac{(g^x_1)^2}{(1+3w_\phi)\mathscr{M}^2_\phi}\left(\frac{m_x}{\mathscr{M}_\phi}\right)^{-\frac{1+3w_\phi}{2w_\phi}}T^{\frac{2(1-
     w_\phi)}{1+w_\phi}}_{re}~~~\mbox{for}~~~~g^x_1\phi S^2\\
     &\frac{\epsilon(g^x_2)}{2(5w_\phi-1)\mathscr{M}^4_\phi}\left(\frac{m_x}{\mathscr{M}_\phi}\right)^{\frac{1-5w_\phi}{2w_\phi}}T_{re}^{\frac{2(1-w_\phi)}{1+w_\phi}}~~~\mbox{for}~~~~g^x_2\phi^2 S^2~~\mbox{with}~~w_\phi>0.2\\
      &\frac{(h^x)^2}{1-w_\phi}\left(\frac{m_x}{\mathscr{M}_\phi}\right)^{\frac{w_\phi-1}{2w_\phi}}T_{re}^{\frac{2(1-w_\phi)}{1+w_\phi}}~~~~\mbox{for}~~~~h^x\phi\bar FF\\
 \end{array}\right.
 \end{eqnarray}
Once we obtain products from the direct decay, the total DM abundance can be expressed as 
\bea
\Omega_xh^2= \Omega^{g}_xh^2 +\Omega^{dcay}_xh^2 = 0.12 .
\eea
In order to constrain the coupling parameters, we consider the total DM abundance. 
 %%%%%%%%%%%%%%%%%%%%%%%%%%%%%%%%%%%
\subsubsection{\bf Constraining the inflaton-DM couplings $(g^x_i,h^x)$ and DM mass $(m_x)$}
Fig.(\ref{DM1}) depicts detailed allowed parameter space for which present DM abundance is satisfied. Let us emphasize again that while plotting the DM abundance, we take into account the contribution from both  the direct and the universal gravity-mediated decay of inflation. For the sake of presentation, we have considered two sample values of the inflation equation of states $w_\phi=(0.0,0.50)$ with reheating temperatures range between maximum and minimum values $T_{re}=(T_{BBN}, 10, 10^6, 10^{10}, 10^{15})$ GeV. 

In each plot for a fixed $(T_{re}, w_{\phi})$ we see the maximum limit on DM mass ($m^{g,max}_{x}$) \cite{Haque:2021mab,Clery:2021bwz} which is due to gravitational interaction as mentioned before.  However, since DMs are produced from the inflaton decay, purely kinematic constraints can also set the upper limit to be $m^{end}_\phi$ for some specific cases when $m^{g,max}_{x} > m^{end}_\phi$ which is observed for $w_{\phi} =0$ (see Fig.\ref{DM1}). It is natural to expect that for $m_x<m^{g,max}_{x}$, the DM from the decay channel solely controls the abundance. Upon $m_x$ approaching $m^{g,max}_{x}$ value, the gravitational contribution starts dominating the abundance till $m_x=m^{g,max}_{x}$. 

However, the lower bound on the DM mass will be fixed either from observation or theory. For example, the lower bound has been observed to be controlled by reheating temperature and the physical processes of reheating under consideration. In general, for this scenario lower the DM mass, the larger would be the inflaton-DM coupling to achieve the correct DM abundance. However, the inflaton-DM coupling should be bounded from above so that the universe should be radiation dominated from the end of reheating ($T_{re}$) to the matter-radiation equality ($T_{eq}\simeq10^{-9}$ GeV). Hence, there exists an upper limit of the DM coupling, above which we always get the DM-dominated universe after the inflaton domination, or in other words one never achieve the radiation-dominated universe. Since there is a one-to-one correspondence between the DM coupling and the DM mass, a lower limit of the DM mass corresponds to the upper limit of the DM coupling, and that can be obtained by equating the DM energy density and the radiation energy density at the time of matter-radiation equality. The upper limit of coupling are calculated as
\begin{equation}
    \begin{aligned}   &g^x_{1,\,c}=\left(\frac{9\pi\sqrt\epsilon(1+3w_\phi)\mathscr{M}_\phi}{\sqrt 3M_p}\right)^{1/2}T_{re}^{\frac{1+3w_\phi}{1+w_\phi}}~~~\mbox{for}~~g_1^x\phi S^2\\
    &g^x_{2,\,c}=\left(\frac{9\pi(5w_\phi-1)\mathscr{M}_\phi^3}{\sqrt 3\epsilon M_p}\right)^{1/2}T_{re}^{\frac{5w_\phi-1}{1+w_\phi}}~~~\mbox{for}~~g_2^x\phi^2 S^2~~\mbox{with}~~w_\phi>0.2\\
    &g^x_{2,\,c}=\left(\frac{12\pi(1-5w_\phi)\mathscr{M}_\phi^3}{\sqrt {3\epsilon} M_p}\right)^{1/2}\left(\frac{\epsilon}{3M_p^2H^2_{end}}\right)^{\frac{1-15w_\phi}{12(1+w_\phi)}}T_{re}^{\frac{2(3w_\phi-1)}{3(1+w_\phi)}}~~~\mbox{for}~~g_2^x\phi^2 S^2~~\mbox{with}~~w_\phi<0.2\\
 &h^x_c=\left(\frac{9\pi\sqrt\epsilon(1-w_\phi)}{\sqrt 3M_p\mathscr{M}_\phi}\right)^{1/2}T_{re}^{\frac{1-w_\phi}{1+w_\phi}}~~~\mbox{for}~~h^x\phi \bar FF .
    \end{aligned}
\end{equation}
which does not depend on the details of the reheating history but on the nature of DM, reheating temperature, and equation of states. The corresponding lower limit on the DM mass can similarly be calculated as (see Fig.\ref{DM1})
\begin{equation}
    m_{x,min}=4.3\times 10^{-10}\frac{m_\phi(T_{re})}{T_{re}}=4.3\times 10^{-10}\mathscr{M}_\phi T_{re}^{\frac{3w_\phi-1}{1+w_\phi}}
\end{equation}
However, such theoretical lower bound will be further constrained by the observational BBN bound on $\Delta N_{\mbox{eff}}\leq 0.50$ (yellow shaded region). Using this in Eq. (\ref{nef1}) one readily infers that the relativistic DM energy density should be less $8\%$ to the over all energy budget at the end of reheating. Upon incorporating such observational constraints we arrive at the following modified expression of lower limit of the DM mass as 
\begin{equation}
    m_{x,min}\simeq10^{-8}\mathscr{M}_\phi T_{re}^{\frac{3w_\phi-1}{1+w_\phi}}
\end{equation}
To this end, we would like to point out the fact that there exists Lyman$-\alpha$ bound on the DM mass $m^{Lyman}_x> 5\times 10^{-6}$ GeV. However, such bound on DM depends non-trivially on the details of its phase-space distribution and equation of state. Therefore, we defer this discussion in  detail for our future studies. \\
%%%%%%%%%%%%%%%%%%%%%%%%%%%%%%%%%%%%%%%
From our discussion so far, we have obtained two broad conditions on the DM mass, say $m_x>m_\phi (A_{re})$ and $m_x<m_\phi (A_{re})$. When the DM mass satisfies the condition $m_x<m_\phi (A_{re})$, its abundance decrease with increasing reheating temperature, as shown in
Eq.(\ref{infdmw1}) (see also Fig.(\ref{DM1})). As a result, in order to achieve correct abundance for a fixed DM mass, the inflaton-DM coupling must be increased for larger reheating temperature with an exception (see Fig. \ref{DM1}) for $\phi\phi\rightarrow SS$ production channel with $w_\phi < 1/3$. The reason is that for such a situation, the $\phi\phi\to SS$ channel produces DM only during the initial stage of reheating. On the other hand, when the DM satisfies $m_x>m_\phi (A_{re})$, the slope of the figure changes (see the bottom plot of Fig.\ref{DM1}), and the co-moving DM freezes in at any point during reheating due to kinematics reason where $m_x\sim m_\phi$. As a consequence, the mass dependency of the abundance $\Omega_xh^2$ also changes (see Eq. (\ref{infdmw2})).

%%%%%%%%%%%%%%%%%%%%%%%%%%%%%%%%%%%
 \subsection{Freeze-in and Freeze-out production of DM from radiation bath}
In this section, we will discuss DM production exclusively from the thermal bath. In addition, the universal gravitational production of DM will always be present, which can not be ignored.  The associated Boltzmann equations (see, for instance, Eqn.\ref{B3}) for the freeze-in production scenario are
\begin{eqnarray}
&&   \dot{\rho}^{r}_{tot}+4H\rho^r_{tot}-\Gamma^r_{\phi}(1+w_\phi)\rho_\phi-2\langle\sigma v\rangle\langle E_x\rangle^r\left[ (n^r_x)^2- (n^r_{x,eq})^2\right]=0, \label{darkr}\\
&&  \dot n^r_x+3Hn^r_x+\langle\sigma v\rangle\left[ (n^r_x)^2- (n^r_{x,eq})^2\right]=0
    \label{darkd} .
\end{eqnarray}
And for the freeze-out scenario, since the cross-section is strong enough, the gravitationally produced DM from inflaton and radiation occurs at the initial reheating stage and reaches thermal equilibrium within a short period.
%\textcolor{blue}{And for the freeze-out scenario, since the cross-section is strong enough, the produced DM from inflaton and radiation scattering through the exchange of graviton also be in equilibrium with the thermal bath}\footnote{\textcolor{red}{In our analysis, the re-annihilation of DM is not important because gravitational production happens just at the end of inflation and reaches thermal equilibrium for WIMP case. So in the final DM abundance, there is no contribution of gravitational DM.}}.
Thus, the Boltzmann equations associated with DM take the following form
\begin{eqnarray}
&&  \dot n^r_x+3Hn^r_x+\langle\sigma v\rangle\left[ (n^r_x)^2- (n^r_{x,eq})^2\right]-\frac{\Gamma_{\phi\phi\rightarrow SS/FF}\rho_\phi}{m_\phi(t)}-\frac{\gamma_{S/F}T^8_{rad}}{M^4_p}=0 .
    \label{darkd1}
\end{eqnarray}
 Where $\Gamma^r_{\phi}=\Gamma^{th}_{s/f}+\Gamma^{gr}_{\phi\phi\rightarrow RR}$, $\gamma_{S}=1.9\times 10^{-4}$ for scalar DM, $\gamma_F=1.9\times 10^{-3}$ for fermionic DM \cite{Barman:2021ugy}. In the above expression, the fourth and last terms are associated with the DM production through gravitational scattering from inflaton, and radiation bath respectively (see Fig.\ref{fynDM} for relevant Feynman diagrams for gravitational scattering from the thermal bath (R)). $\langle E_x\rangle^r=\sqrt{m^2_x+9\,T^2_{rad}}$ is the average energy per DM particle produced from the thermal bath \cite{Giudice:2000ex}. $\langle\sigma v\rangle $ be the thermally averaged cross section times velocity. $n^r_{x,eq}$ be the equilibrium number density of the DM, which can be expressed as 
\begin{equation}{\label{neq}}
    n^r_{x,eq}=\frac{j_x}{2\pi^2}\int_{m_x}^\infty\frac{\sqrt{E_x^2-m_x^2}}{e^{E_x/T_{rad}}+1}E_xdE_x=\frac{j_xT^3_{rad}}{2\pi^2}\left(\frac{m_x}{T_{rad}}\right)^2K_2\left(\frac{m_x}{T_{rad}}\right) ,
\end{equation}
where, $T_{rad}$ is the temperature of the radiation bath and $j_x$ be the internal degrees of freedom of DM and $K_2(x)$ is the modified Bessel function of the second kind. 
The expression of the DM relic abundance in terms of radiation abundance\cite{w1,w2} $\Omega_xh^2 = ({m_x N_x^r(A_F)\Omega_rh^2})/({\epsilon T^3_FA^3_F T_{now}})$, where $T_F$ be the temperature of the radiation bath at the very late times $A_F$, when both the radiation  and DM energy density became freezes, and $N^r_x=n^r_xA^3$ is the co-moving number density of DM. We constrain the DM parameter space ($m_x,\langle \sigma v\rangle$) in terms of ($w_\phi,T_{re}$). The population of the DM particles produced from the thermal bath strongly depends on the scattering cross-section $\langle\sigma v\rangle$. If the scattering cross-section is large enough, the produced DM particles reach thermal equilibrium, and when the bath temperature falls below the DM mass, the number density of DM freezes out-this mechanism is known as the freeze-out mechanism \cite{fo1,fo2,fo3,fo4,fo5,fo6,fo7,fo8}. On the other hand, if the scattering cross-section is small enough, the DM can never reach thermal equilibrium, and this mechanism is called the freeze-in mechanism \cite{Maity:2018dgy,Bernal:2020qyu,f1,f2,f3,f4,f5,f6}. In this paper, we will discuss both production mechanisms and analyze the parameter space needed to satisfy the correct relic.
%%%%%%%%%%%%%%%%%%%%%%%%%%%%%%%%%%%%%%%
\subsubsection{\bf Freeze-in from radiation bath: bosonic and fermionic reheating}
For the freeze-in from the thermal bath, the DMs will never be in thermal equilibrium, and hence DM number density generically satisfies $n^r_x\ll n^{r}_{x,eq}$. Dynamical equation Eq.\ref{darkd} then transformed into simplified form in terms of co-moving DM number density $N^r_x=n^r_xA^3$ as,
\begin{equation}\label{cnd1}
    \frac{d N^r_x(A)}{dA}=\frac{A^2}{H}\langle\sigma v\rangle \left(n_{x,eq}^r\right)^2.
\end{equation}
The above equation suggests that the production rate is simply proportional to the square of the equilibrium DM number density. In the $m_x<T_{rad}$ limit, it behaves as $\propto T_{rad}^6$. Thus freeze-in production from radiation bath naturally follows the way radiation temperature evolves up to the point $m_x\sim T_{rad}$. The production for the masses $m_x>T_{rad}$ will naturally be Boltzmann suppressed. As we have extensively discussed, the evolution of the bath temperature is conditioned non-trivially not only by the production process and its constituents, but also by the bath temperature itself. Therefore, details of the reheating history is expected to have interesting impact on DM evolution and its final abundance. Throughout our analysis, we will provide a detailed analysis of DM phenomenology and its dependence on the reheating history for different physical situations discussed before.\\

\noindent
$\clubsuit$ : \textbf{Freeze-in from the bosonic radiation bath} \\

Earlier, we discussed different possible bosonic reheating histories. In this section, we will quote our findings of the DM abundances for those different reheating histories. The detailed calculations are shown in the appendix-\ref{analytical exp DM}. As discussed for the bosonic reheating case, depending on the range of inflaton-scalar coupling, we have three different cases,   

\noindent
\textbf{\bf Case-I: Coupling strength $\bf g_i^{r}>\mathscr{G}^{1,\,th}_{ci}$:}
In this regime, direct decay of inflaton into radiation controls the entire reheating process, and as discussed we have the two broadly classified thermal histories,  

\underline {When $T^{max}_{rad}>m^{end}_\phi$:} For this case, the direct inflaton decay channel  controls the reheating dynamics and the thermal effect is effective throughout the reheating period for $w_\phi>w_\phi^c$, so the temperature evolves according to Eq.\ref{the1},\ref{the2}. With this reheating background, we now find the DM abundance for two different mass ranges. When $m_x < T_{re}$, the present-day DM abundance can be obtained as,
\begin{eqnarray}{\label{Wx1}}
\Omega_xh^2=\Omega_rh^2\frac{6M_p\langle\sigma v\rangle j_x^2}{(3\epsilon)^{3/2}\pi^4T_{now}}\left\{\begin{array}{ll}
     & \frac{m_xT_{re}}{1+7w_\phi}~~~\mbox{for}~~~ g_1^r\phi s^2 \\
     & \frac{m_xT_{re}}{11w_\phi-3}
     ~~~\mbox{for}~~~g_2^r\phi^2 s^2~~ \mbox{with}~~w_\phi>3/11\\
     & \frac{m_xT_{re}}{3-11w_\phi}\left(\frac{T_{re}}{T^{max}_{rad}}\right)^{\frac{3(11w_\phi-3)}{3-5w_\phi}}
    ~~~\mbox{for}~~~g_2^r\phi^2 s^2
    ~~ \mbox{with}~~w_\phi<3/11 .
\end{array}\right.
\end{eqnarray}
When the DM mass is lower than the reheating temperature, kinematically, DM production is expected to continue even after reheating until the point when $m_x\sim T_{rad}$. However, it is important to note that freeze-in production of DM from the radiation bath typically follows the evolution of the bath itself. In most cases, the comoving radiation energy density freezes at the end of reheating. Therefore, for analytical calculation, it is safe to take DM production up to the end of reheating even for $m_x<T_{re}$. On the other hand, there are some situations where radiation production happens initially, which is visible in most of the scenarios where $w_\phi<3/11$ for $\phi\phi\rightarrow ss$ reheating process. For such case, DM production similarly happens instantaneously at the end of inflation, and its number density turned out to be independent of mass but depends on the maximum radiation temperature $T_{rad}^{rad}$. The resulting expression for the  abundance is given in the last expression of Eq.\ref{Wx1}. Therefore, for this particular reheating, since maximum production happens initially, the above expression of the abundance will remain the same even for  $m_x>T_{re}$. However, this should not be true in general. 

Generically if one considers the mass $m_x>T_{re}$, the DM is naturally expected to be produced during reheating until $m_x\sim T_{rad}$, and the abundance for different decay channels are obtained as 
\begin{eqnarray}{\label{Wx2}}
\Omega_xh^2=\Omega_rh^2\frac{6M_p\langle\sigma v\rangle j_x^2}{(3\epsilon)^{3/2}\pi^4T_{now}}\left\{\begin{array}{ll}
     & \frac{m_xT_{re}}{1+7w_\phi}\left(\frac{T_{re}}{m_x}\right)^{\frac{3(1+7w_\phi)}{1-3w_\phi}}~~~\mbox{for}~~~ g_1^r\phi s^2\\
     & \frac{m_xT_{re}}{11w_\phi-3}\left(\frac{T_{re}}{m_x}\right)^{\frac{3(11w_\phi-3)}{3-5w_\phi}}~~~\mbox{for}~~~g_2^r\phi^2 s^2
     ~~ \mbox{with}~~w_\phi>3/11\\
   %  & \frac{m_xT_{re}}{3-11w_\phi}\left(\frac{T_{re}}{T_{max}}\right)^{\frac{3(11w_\phi-3)}{3-5w_\phi}}~~~\mbox{for}~~~\phi^2 S^2~~ \mbox{with}~~w_\phi<w_\phi^c
\end{array}\right.
\end{eqnarray}
At this point, we would like to elaborate the exceptional case for $w_\phi>w^t_\phi$, for which the reheating temperature equals the maximum radiation temperature (see, for instance, Eq.\ref{the1}). If the EoS satisfies $w_\phi>w_\phi^t$ and $m_x>T_{re}$, the DM production remains always suppressed and the dominating contribution comes from the initial stage of reheating. On the other hand, for $w_\phi<w_\phi^t$ where maximum radiation temperature appears at the initial phase of reheating and for $m_x>T_{re}$ DM production occurs till the point $m_x\sim T_{rad}$.\\

As discussed earlier, when $w_\phi < w^c_\phi$, we found an intermediate temperature scale $T_c$ (at the point $A_c$ defined in Eq.\ref{acc}) above which bath temperature dynamics is controlled by the thermally corrected decay width. Hence, for $m_x>T_{re}$, since the freeze-in occurs during the reheating epoch itself, two different possibilities arise. If $m_x> T_c >T_{re}$, the DM will freeze in during the early phase of reheating, where the evolution of bath temperature is controlled by thermally corrected production rate, and the abundance will take the following form,  
\begin{equation}{\label{wx4}}
\Omega_xh^2=\Omega_rh^2\frac{6M_p\langle\sigma v\rangle j^2}{(3\epsilon)^{3/2}\pi^4T_{now}}
\left\{\begin{array}{ll}
    & \frac{m_xT_{re}}{1+7w_\phi}\left(\frac{T_{c}}{m_x}\right)^{\frac{3(1+7w_\phi)}{1-3w_\phi}}\left(\frac{T_{re}}{T_{c}}\right)^{\frac{2(3+5w_\phi)}{1-w_\phi}}~~~\mbox{for}~~~g_1^r\phi s^2\,. \\
     &\frac{m_xT_{re}}{3-11w_\phi}\left(\frac{T_{re}}{T_{max}}\right)^{\frac{3(11w_\phi-3)}{3-5w_\phi}}~~~\mbox{for}~~~g_2^r\phi^2  s^2
\end{array}\right.
\end{equation}
Whereas for $T_c> m_x > T_{re}$, the DM will freeze in during the later part of the reheating phase when finite temperature effect is diminished, and the abundance assumes different form as,
  \begin{equation}{\label{wx5}}
\Omega_xh^2=\Omega_rh^2\frac{12M_p\langle\sigma v\rangle j^2}{(3\epsilon)^{3/2}\pi^4T_{now}}
\left\{\begin{array}{ll}
&\frac{m_xT_{re}}{3+5w_\phi}\left(\frac{T_{re}}{m_x}\right)^{\frac{2(3+5w_\phi)}{1-w_\phi}}~~~\mbox{for}~~~g_1^r\phi s^2\,.\\
 &\frac{m_xT_{re}}{3-11w_\phi}\left(\frac{T_{re}}{T_{max}}\right)^{\frac{2(11w_\phi-3)}{3(1-w_\phi)}}~~~\mbox{for}~~~g_2^r\phi^2 s^2%~~ \mbox{with}~~w_\phi<w_\phi^c\,.\\
\end{array}\right.
\end{equation}
And if $m_x < T_{re}$, the DM abundance at the present time can be written as
 \begin{equation}{\label{wx3}}
 \Omega_xh^2=\Omega_rh^2\frac{12M_p\langle\sigma v\rangle j^2}{(3\epsilon)^{3/2}\pi^4T_{now}}\left\{\begin{array}{ll}
 &\frac{m_xT_{re}}{3+5w_\phi}~~\mbox{for}~~g_1^r\phi s^2\\
    &\frac{m_xT_{re}}{3-11w_\phi}\left(\frac{T_{re}}{T_{max}}\right)^{\frac{2(11w_\phi-3)}{3(1-w_\phi)}}~~~\mbox{for}~~~g_2^r\phi^2 s^2%~~ \mbox{with}~~w_\phi<w_\phi^c\,.\\
    \end{array}\right.
 \end{equation}
\underline{ When $ T^{r,max}_{gr}<T^{max}_{rad}<m^{end}_\phi$:} For this case, direct inflaton decay channel controls the entire reheating dynamics. Similar to the discussion above for $m_x < T_{re}$, the abundance can be written as
\begin{eqnarray}\label{wx6}
\Omega_xh^2=\Omega_rh^2\frac{12M_p\langle\sigma v\rangle j^2}{(3\epsilon)^{3/2}\pi^4T_{now}}\left\{\begin{array}{ll}
&\frac{m_xT_{re}}{3+5w_\phi}~~~\mbox{for}~~~g_1^r\phi s^2~~\mbox{with}~~~ w_\phi<w^c_\phi\\
&\frac{m_xT_{re}}{1+7w_\phi}~~~\mbox{for}~~~g_1^r\phi s^2~~\mbox{with}~~~ w_\phi>w^c_\phi\\
&\frac{m_xT_{re}}{3-11w_\phi}\left(\frac{T_{re}}{T^{max}_{rad}}\right)^{\frac{3(11w_\phi-3)}{3-5w_\phi}}~~~\mbox{for}~~~g_2^r\phi^2 s^2~~ \mbox{with}~~w_\phi<w_\phi^c\\
 & \frac{m_xT_{re}}{11w_\phi-3}~~~\mbox{for}~~~g_2^r\phi^2 s^2~~ \mbox{with}~~w_\phi>w^c_\phi .
\end{array}\right.
\end{eqnarray}
However, for $m_x > T_{re}$ freeze-in will naturally occur during reheating, and for $w<w_\phi^c$ the abundance can be found to be the same as Eq.\ref{wx5}. 
On the other hand for $w>w_\phi^c$ the abundance will be same as Eq.(\ref{Wx2}) for $m_x < T_c$,  and for $m_x>T_c$ is
\begin{eqnarray}{\label{wx7}}
\Omega_xh^2=\Omega_rh^2\frac{12M_p\langle\sigma v\rangle j^2}{(3\epsilon)^{3/2}\pi^4T_{now}}\left\{\begin{array}{ll}
&\frac{m_xT_{re}}{3+5w_\phi}\left(\frac{T_c}{m_x}\right)^{\frac{2(3+5w_\phi)}{1-w}}\left(\frac{T_{re}}{T_c}\right)^{\frac{3(1+7w_\phi)}{1-3w_\phi}}~~~\mbox{for}~~~g_1^r\phi s^2~~\\
&\frac{m_xT_{re}}{11w_\phi-3}\left(\frac{T_c}{m_x}\right)^{\frac{2(3+5w_\phi)}{1-w}}\left(\frac{T_{re}}{T_c}\right)^{\frac{3(11w_\phi-3)}{3-5w_\phi}}~~~\mbox{for}~~~g_2^r\phi^2 s^2~~\,,\\
\end{array}\right.
\end{eqnarray}

\noindent
\textbf{Case-II: Coupling strength in between $\bf \mathscr{G}^{2,\,th}_{ci}<g_i^{r}<\mathscr{G}^{1,\,th}_{ci}$ :}  
As discussed earlier, for this coupling range, the gravitational interaction drives the dynamics of reheating at the initial stage. Therefore, the maximum temperature is always controlled by the gravitational sector $T^{r,\, max}_{gr}=T^{max}_{rad}$. In this coupling range, a cross-over temperature scale $T_s$ (at the point $A_{gr\rightarrow ngr}$-see Eq.\ref{gr2ngr}) exists across which gravitational decay dominated to non-gravitational decay-dominated reheating occurs. When $m_x<T_{re}$, the DM will be produced up to the end of reheating (except for $g^r_2\phi^2 s^2$ with $w_\phi<3/11$), and hence the maximum production occurs at the end of reheating, and the abundance follows Eqs.\ref{Wx1}, \ref{wx3}. However, if $T_s>m_x>T_{re}$ and freeze-in occurs during the decay channel-dominated phase, and the final abundance follows the same form as expressed in Eqs. (\ref{Wx2}), and (\ref{wx5}) with $T_c$ being replaced by $T_s$. However, if $m_x>T_s> T_{re}$ and freeze-in happens during the universal gravitational decay-dominated phase, and we have 
\begin{eqnarray}{\label{wx8}}
   \Omega_xh^2=\Omega_rh^2\frac{6M_p\langle\sigma v\rangle j^2}{(3\epsilon)^{3/2}\pi^4T_{now}}\frac{m_xT_{re}}{1-w_\phi}\left(\frac{T^{max}_{rad}}{T_{s}}\right)^{\frac{3(1-w_\phi)}{2}}\left\{\begin{array}{ll}
        &\left(\frac{T_{re}}{T_{s}}\right)^{\frac{3(1+7w_\phi)}{1-3w_\phi}}~~\mbox{for}~~ g_1^r\phi s^2~~\mbox{with}~~ w_\phi>w_\phi^c  \\
        & \left(\frac{T_{re}}{T_{s}}\right)^{\frac{3(11w_\phi-3)}{3-5w_\phi}}~~\mbox{for}~~ g_2^r\phi s^2~~\mbox{with}~~ w_\phi>w_\phi^c \,, \\
   \end{array}\right.
\end{eqnarray}
and
\begin{eqnarray}{\label{wx9}}
   \Omega_xh^2=\Omega_rh^2\frac{6M_p\langle\sigma v\rangle j^2}{(3\epsilon)^{3/2}\pi^4T_{now}}\frac{m_xT_{re}}{1-w_\phi}\left(\frac{T^{max}_{rad}}{T_{s}}\right)^{\frac{3(1-w_\phi)}{2}}\left\{\begin{array}{ll}
        &\left(\frac{T_{re}}{T_{s}}\right)^{\frac{2(3+5w_\phi)}{1-w_\phi}}~~\mbox{for}~~ g_1^r\phi s^2~~\mbox{with}~~ w_\phi<w_\phi^c  \\
       & \left(\frac{T_{re}}{T_{s}}\right)^{\frac{2(11w_\phi-3)}{1-w_\phi}}~~\mbox{for}~~ g_2^r\phi s^2~~\mbox{with}~~ w_\phi<w_\phi^c\,.  \\
   \end{array}\right.
\end{eqnarray}
\textbf{Case III: when $\bf g_i^{r} < \mathscr{G}^2_{ci}$:} For this case the gravity-mediated decay of inflaton controls the entire dynamics of reheating (termed as gravitational reheating). This particular phase is realized for $w_\phi>0.65$ (see the light-cyan region of Fig.\ref{couplingbound}).
Similar to the previous case, $T^{r,max}_{gr}=T^{max}_{rad}$ will always holds.  Since gravitational radiation production happens only at the beginning of reheating, most of the DM production is also expected to happen at the initial phase of the reheating, and the abundance for such a case is calculated to be,
\begin{equation}{\label{grdmw}}
    \Omega_xh^2=\Omega_rh^2\frac{6M_p\langle\sigma v\rangle j^2}{(3\epsilon)^{3/2}\pi^4T_{now}}\frac{m_xT_{re}}{(1-w_\phi)}(T_{re}/T_{max}^{rad})^{3(w_\phi-1)/2}\,.
\end{equation}

\noindent
$\clubsuit$ : \textbf{Freeze-in from the Fermionic radiation bath} \\
Details of the fermionic reheating has been discussed. In this section, we will quote our findings of the DM abundances for those different reheating histories. The detailed calculations are shown in the appendix-\ref{analytical exp DM}. The distinct behaviour of fermionic reheating, as opposed to bosonic one, mainly arises for the higher value of inflation equation of state, say $w_\phi>7/15 \,(5/9)$ for with (without) thermal effect. In addition to that, if $w_\phi>7/15 \,(5/9)$, most of the fermionic radiation production occurs during the initial stage of reheating due to its specific behaviour of production rate from the inflaton.
Similar to the bosonic one, for fermionic reheating depending on the range of inflaton-Fermion coupling we have two different possibilities,\\
%%%%%%%%%%%%%%%%%%%%%%%%%%%%%%%%%%%%%
\textbf{\bf Case-I: Coupling strength, $\bf h^r>\mathscr{H}_c$:} 
%%%%%%%%%%%%%%%%%%%%%%%%%%%%%%%%%
As discussed before, for this coupling regime, non-gravitational decay of inflaton into radiation controls the entire reheating process. In the following subsection consider different temperature regimes 

\underline{When $T^{max}_{rad}>m^{end}_\phi$}: For this case, the thermal correction in the decay rate will be dominant from the beginning. Depending upon the equation of state we have two different possibilities of evolution of the radiation component: (i) $1> w_\phi\geq 7/15$, the radiation production mainly takes place at the initial stage, (ii) $0\leq w_\phi<7/15$, radiation production happens throughout the reheating phase (see, for instance, Eqn.\ref{radferwth}).

\begin{itemize}
 \item {$\bf 1>w_\phi\geq 7/15$ :} As just pointed out, radiation production occurs at the initial stage and hence the comoving radiation density becomes constant early. Following the radiation evolution, the DMs are also produced at the beginning of reheating. As a result the abundance (see, Eqn.\ref{fdmw1}), naturally be controlled by the $T^{max}_{rad}$, as follows
\begin{equation}{\label{fdmw1}}
        \Omega_xh^2=\Omega_rh^2\frac{6M_p\langle\sigma v\rangle j_x^2}{(3\epsilon)^{3/2}\pi^4(1-w_\phi)T_{now}}m_xT_{re}\left(\frac{T_{re}}{T_{rad}^{max}}\right)^{\frac{3(w_\phi-1)}{2}}\,,
    \end{equation}
\end{itemize}
\begin{itemize} 
\item {$\bf 0\leq w_\phi< 7/15$ :} For this range of EoS, the ratio $T_{rad}/m_\phi$ varies as $A^{-\frac{3}{10}(1-5w_\phi)}$ which indicates that the thermal effect is dominant throughout the entire reheating phase when $7/15>w_\phi>1/5$. However, for
 $0\leq w_\phi<1/5$, there exists an intermediate temperature scale $T_c$ with the scale factor $A_c$ (see Eq.\ref{bgdfer}) above which the thermal effects drops down drastically. Let us discuss two possible scenarios in this context:

1.\underline{\it Dominant finite temperature effect during entire reheating period for EoS $1/5 < w_\phi < 7/15\,:$}  We found two different sub-possibilities depending on EoS.\\
a) When EoS is in the range of $9/25 <w_\phi <7/15$, comoving DM freezes at the initial stage of reheating, and due to that, there is an explicit maximum temperature dependence in the DM abundance. Henceforth, for both $m_x>T_{re}$ and $m_x<T_{re}$, we have the same DM abundance expression, and that is,
 \begin{eqnarray}{\label{fdmw2}}
     \Omega_xh^2=\Omega_rh^2\frac{30M_p\langle\sigma v\rangle j_x^2}{(3\epsilon)^{3/2}\pi^4T_{now}} \, \frac{m_xT_{re}}{(25w_\phi-9)}\left(\frac{T_{re}}{T_{rad}^{max}}\right)^{\frac{9-25w_\phi}{1+5w_\phi}}\,,
    \end{eqnarray}  
b) When EoS lies between $1/5<w_\phi<9/25$, the expression for DM abundance will be different for $m_x<T_{re}$ (comoving DM freezes at the end of reheating) and $m_x>T_{re}$ (comoving DM freezes at any intermediate point during reheating where $m_x \sim T_{rad}$).
For $m_x<T_{re}$, DM abundance is found to be,
     \begin{eqnarray}{\label{fdmw2}}
     \Omega_xh^2=\Omega_rh^2\frac{30M_p\langle\sigma v\rangle j_x^2}{(3\epsilon)^{3/2}\pi^4T_{now}}\,\frac{m_xT_{re}}{9-25w_\phi}\,,
    \end{eqnarray}  
 and for $m_x>T_{re}$ we have,
    \begin{equation}{\label{fdmw3}}
        \Omega_xh^2=\Omega_rh^2\frac{30M_p\langle\sigma v\rangle j_x^2}{(3\epsilon)^{3/2}\pi^4T_{now}}\frac{m_xT_{re}}{9-25w_\phi}\left(\frac{T_{re}}{m_x}\right)^{\frac{9-25w_\phi}{1+5w_\phi}}\,.
    \end{equation}  
    %%%%%%%%%%%%%%%%%%%%%%%%%%%%%%%%%%%%%%%%%%%%%%%
 2. \underline{\it Finite temperature effect will not be dominant during the entire reheating period for EoS $ 0 \leq w_\phi < 1/5\,$:} As already discussed earlier, for this range of EoS, there exits an intermediate temperature scale $T_c$ across which thermal effect drops down ( see discussion around Eq.\ref{fercross}). Therefore, for $m_x > T_{re}$ (comoving DM freezes in during reheating), we have two different possibilities:\\
i) When $m_x>T_c>T_{re}$, the DM freezes in before thermal to non-thermal domination crossover, and we get  
\begin{equation}{\label{fdmw4}}
    \Omega_xh^2=\Omega_rh^2\frac{30M_p\langle\sigma v\rangle j_x^2}{(3\epsilon)^{3/2}\pi^4T_{now}} \frac{m_xT_{re}}{9-25w_\phi}\left(\frac{T_c}{m_x}\right)^{\frac{(9-25w_\phi)}{1+5w_\phi}}\left(\frac{T_{re}}{T_{c}}\right)^{\frac{2(3-7w_\phi)}{1+3w_\phi}}\,.
\end{equation}
ii) When $T_{re}<m_x<T_c$, the comoving DM freezes after thermal to non-thermal domination crossover, and abundance assumes, 
\begin{equation}{\label{fdmw5}}
     \Omega_xh^2=\Omega_rh^2\frac{4M_p\langle\sigma v\rangle j_x^2}{(3\epsilon)^{3/2}\pi^4T_{now}}\frac{m_xT_{re}}{3-7w_\phi}\left(\frac{T_{re}}{m_x}\right)^{\frac{2(3-7w_\phi)}{1+3w_\phi}}\,.
\end{equation}
 when $m_x<T_{re}$, the Comoving DM freezes at the end of reheating, and the abundance can be expressed as 
 \begin{equation}{\label{fdmw31}}
     \Omega_xh^2=\Omega_rh^2\frac{4M_p\langle\sigma v\rangle j_x^2}{(3\epsilon)^{3/2}\pi^4T_{now}}\frac{m_xT_{re}}{3-7w_\phi}\,.
\end{equation}
\end{itemize}

\underline{$T^{r,\,max}_{gr}<T^{max}_{rad}<m^{end}_\phi$ :} In this case, as described earlier in detail (see, for instance, sec-\ref{bgdfer}) there is no thermal effect at the beginning of reheating. Depending on the reheating background, there are two different possibilities (see, for instance, Eq.\ref{couplingnth}): 
\begin{itemize}
    \item {$\bf 1>w_\phi\geq5/9$ :} The bath temperature always falls as $A^{-1}$ since the radiation component is frozen at the beginning of reheating. For the freeze-in mechanism from the thermal bath, it is expected that the DM component follows the radiation and freezes at the beginning, irrespective of its mass. Thus the expression for the abundance will be exactly the same as defined earlier in Eq.\ref{fdmw1}.
\end{itemize}
\begin{itemize}
    \item {$\bf 0\leq w_\phi<5/9$ :} In this scenario, the thermal effect is subdominant at the beginning, but there is a chance of the thermal effect being dominant at any intermediate scale where $T_{rad}\sim T_c$. However, depending upon the way the finite temperature effect made its presence on the DM evolution, the EoS range is divided into three sub-ranges, 

 1.\underline{\it Sub-dominant finite temperature effect during the reheating for EoS $0 \leq w_\phi < 1/5\,$:}
If the thermal correction is not applicable throughout reheating, then the abundance follows the Eq. (\ref{fdmw31}) for $m_x<T_{re}$ and Eq.\ref{fdmw5} for $m_x>T_{re}$.\\

2.\underline{ \it Dominant finite temperature effect at intermediate temperature $T_c$ for $1/5< w_\phi < 3/7\,$ :}  Here we have two different cases depending on different EoS regimes:\\
i) In the presence of the thermal effect when EoS lies within $9/25<w_\phi<3/7$, most of the DM production occurs before the intermediate temperature scale $T_c$. Hence, for any value of $m_x < T_{c}$, the DM always freezes-in its production near around the $T_c$, and DM abundance assumes the form, 
 \begin{equation}{}
     \Omega_xh^2=\Omega_rh^2\frac{4M_p\langle\sigma v\rangle j_x^2}{(3\epsilon)^{3/2}\pi^4T_{now}}\frac{m_xT_{re}}{3-7w_\phi}\left(\frac{T_{c}}{T_{re}}\right)^{\frac{-9+25w_\phi}{1+5w_\phi}}\,.
 \end{equation}
Whereas, for the same EoS range when $m_x>T_{c}>T_{re}$, the comoving DM will freeze in during the initial phase when the thermal effect would be subdominant, and the abundance takes the following form
 \begin{equation}{\label{fdmw6}}
      \Omega_xh^2=\Omega_rh^2\frac{4M_p\langle\sigma v\rangle j_x^2}{(3\epsilon)^{3/2}\pi^4T_{now}}\frac{m_xT_{re}}{3-7w_\phi}\left(\frac{T_{c}}{m_{x}}\right)^{\frac{2(3-7w_\phi}{1+3w_\phi}}\left(\frac{T_{re}}{T_c}\right)^{\frac{9-25w_\phi}{1+5w_\phi}} \,. 
 \end{equation}
 ii) For the range of EoS $1/5<w_\phi<9/25$, the DM production continues up to the end of reheating, and for $m_x<T_{re}$, the abundance follows the Eq.\ref{fdmw2}. Whereas, when $m_x>T_{re}$, the DM production continue up to $m_x\sim T_{rad}$. Therefore, for $T_c>m_x>T_{re}$, the DM abundance will follow the Eq.\ref{fdmw3} and for $m_x>T_c>T_{re}$, Eq.\ref{fdmw6} will be the abundance expression.

 3. \underline{For EoS $3/7 < w_\phi < 5/9$}: For this case, it is observed that the comoving DM freezes immediately after the reheating begins irrespective of DM mass, and the abundance of the DM takes the following form,
    \bea
      \Omega_xh^2=\Omega_rh^2\frac{4M_p\langle\sigma v\rangle j_x^2}{(3\epsilon)^{3/2}\pi^4T_{now}}\frac{m_xT_{re}}{7w_\phi-3}\left(\frac{T_{re}}{T^{max}_{rad}}\right)^{\frac{2(3-7w_\phi)}{1+3w_\phi}}\,.
\eea
 \end{itemize}
 %%%%%%%%%%%%%%%%%%%%%%%%%%%%%%%%%%%%%%%%%%%%%%%%
\textbf{\bf Case-II: Coupling strength, $\bf h^r < \mathscr{H}_c$:} 
%%%%%%%%%%%%%%%%%%%%%%%%%%%%%%%%%%%%%%%%%%%%%%%%
As discussed before, for this coupling regime, gravitational decay of inflaton into radiation controls the entire reheating process. Therefore, $T_{rad}^{max}=T_{gr}^{r,\,max}$ will always be the case. Depending on the different EoS, there are the following possibilities,
 \begin{itemize}
     \item {$\bf 1> w_\phi>0.65$}: Since for this case $h^r<\mathscr{H}_{c}$ and $w_\phi>0.65$, the gravitational sector is the governing reheating dynamics, termed as gravitational reheating. In gravitational reheating following the radiation component, the DM component will also freeze just at the beginning of reheating, irrespective of its mass, and the abundance will be of the same form as expressed in Eq.\ref{grdmw}.        
 \end{itemize}
 \begin{itemize}
      \item {$\bf 0\leq w_\phi<0.65$}: For this case, purely gravitational production will not be sufficient to reheat the universe. Hence, to have successful reheating, one needs to have non-gravitational production during the later stage of reheating, and the reheating temperature is defined by non-gravitational fermionic coupling. We found three interesting cases, which are as follows: 
      
      1.\underline{\it Reheating temperature $T_{re} < T_{BBN}$ for EoS, $5/9 < w_\phi < 0.65$ :} It is observed that if the EoS lies in this range, both gravitational and non-gravitational production happen very early in reheating phase. Due to subsequent expansion, the reheating temperature turns out to be always $< T_{BBN}$ (see  the red shaded region of Fig.\ref{boundfermion}). 
      However, from the Fig.\ref{boundfermion} it is clear that for EoS between $0\leq w_\phi<5/9$, we have non-trivial dynamics \\
      
      2.\underline{\it Dominant finite temperature effect at intermediate temperature $T_s$ for EoS, $1/5< w_\phi < 5/9\,$ :} In the range of EoS, the finite temperature effect start to dominate at some intermediate temperature scale $T_s$ during reheating. Now, if $m_x<T_{re}$, the DMs are expected to freeze after the end of reheating, and consequently, its abundance is calculated to be the same as given in Eq.\ref{fdmw2}. If $T_s>m_x>T_{re}$, the DMs freeze in during the later phase of reheating when non-gravitational decay dominates, and the abundance assumes the form of Eq.\ref{fdmw3}. Finally, if $m_x>T_s>T_{re}$, the DM will freeze during the initial part of the reheating phase when gravity-mediated decay controls the reheating, and for such case, the abundance has been calculated as
   \begin{equation}
        \Omega_xh^2=\Omega_rh^2\frac{2M_p\langle\sigma v\rangle j_x^2}{(3\epsilon)^{3/2}\pi^4(1-w_\phi)T_{now}}m_xT_{re}\left(\frac{T_{s}}{T^{max}_{rad}}\right)^{\frac{3(w_\phi-1)}{2}}\left(\frac{T_{re}}{T_{s}}\right)^{\frac{9-25w_\phi}{1+5w_\phi}}\,.
    \end{equation}
     3.\underline{For EoS $0 < w_\phi<1/5$ :} When $m_x<T_{re}$, the DM abundance is the same expression as Eq.\ref{fdmw31} and when $T_s<m_x<T_{re}$, the abundance is same as defined in Eq.\ref{fdmw5}. Again, if $m_x>T_s$, the DM will freeze in during the initial gravitational channel domination sector; in such a case, the DM abundance follows the below equation
    \begin{equation}
        \Omega_xh^2=\Omega_rh^2\frac{2M_p\langle\sigma v\rangle j_x^2}{(3\epsilon)^{3/2}\pi^4(1-w_\phi)T_{now}}m_xT_{re}\left(\frac{T_{s}}{T^{max}_{rad}}\right)^{\frac{3(w_\phi-1)}{2}}\left(\frac{T_{re}}{T_{s}}\right)^{\frac{2(3-7w_\phi)}{1+3w_\phi}}\,.
    \end{equation}
\end{itemize}
\subsubsection{\bf Freeze-out from the bosonic and fermionic radiation bath}
For freeze in mechanism, the interaction cross-section $\langle\sigma v\rangle$ is so small that DM can never reach thermal equilibrium. However, if $\langle\sigma v\rangle$ is large enough, the DM can strongly interact with the SM bath and be in thermal equilibrium. The background expansion eventually helps the DM freeze out from the bath at a certain temperature $T_f$. Conventionally these DMs are called WIMP. The freeze-out temperature $T_f$ is defined as,
\begin{equation}{\label{fo1}}
    \langle\sigma v\rangle n^{r}_{x,eq}(T_f)=H(T_f) .
\end{equation}
During reheating, inflaton decays into radiation; hence, entropy is not conserved. Due to this physical situation, one can find two distinct situations:

{\textbf{Freeze out after reheating}}: If the mass of the DM $m_x < T_{re}$, it will freeze out after reheating, i.e., during the radiation-dominated era. Moreover, after the freeze out, the co-moving number density $N_x^r=n_x^rA^3$ will be much larger than the comoving equilibrium number density $N_{x,\,eq}^r$. Thus after freeze-out happens, one can neglect $N_{x,\,eq}^r$ in comparison with $N_x^r$ and from the Eq.\ref{darkd}, one can find\footnote{Near the DM thermal  freezes-out temperature, the non-thermal gravitational production can become dominant as compared to thermal one. This is usually interpreted as the re-annihilation phase \cite{Chu:2011be}. 
We thank the anonymous referee for pointing out this important fact. In our analysis, we ignored this effect. However, it is to be noted that if freeze-out occurs after reheating, such an effect turns out to be subdominant. The reason for this is that during radiation domination, the inflation energy density is negligible compared to the radiation energy density implying $\frac{\Gamma_{\phi\phi\rightarrow SS/FF}\rho_\phi}{m_\phi}<< \langle\sigma v\rangle(n^r_{x,eq})^2$ at $T_f$. On the other hand, for $T_f > T_{re}$, such an effect can become important, particularly for the higher DM mass range which we discussed in the appendix-\ref{appren}.} 
\begin{equation}
    \frac{dN_x^r}{dA}=-\frac{\langle\sigma v\rangle}{H(A_{re})}(A/A_{re})^2(N_x^r)^2\,.
\end{equation}
Integrating from freeze-out point ($A=A_f$) to the present time ($A=A_{0}$), we get 
\begin{equation}
    N_x^r(A_0)=\frac{H(A_{re})}{\langle\sigma v\rangle}A_fA^2_{re} ~~\implies ~~\Omega_xh^2=\frac{1}{\sqrt{3\epsilon} M_p\langle\sigma v\rangle}(m_x/T_f)({\Omega_rh^2}/{T_0})
\end{equation}
$N_0$ is identified as the present day comoving number density leading to  $\Omega_xh^2 \propto 1/{\langle\sigma v\rangle}$. Hence, the abundance decreases with increasing $\langle\sigma v\rangle$.

{\textbf{Freeze out during reheating} :} Alternatively, DM freeze-out could occur during reheating if $m_x >T_{re}$. During reheating $H=H(A_{re})(A/A_{re})^{-\frac{3(1+w)}{2}}$, after utilizing this in Eq.\ref{darkd}\footnote{To find out the analytical expressions of the DM abundance, here we neglect the re-annihilation of DM.}
\begin{equation}
    \frac{dN_x^r}{dA}=-\frac{\langle\sigma v\rangle}{H(A_{re})A_{re}^4}(A/A_{re})^{\frac{3w_\phi-5}{2}}(N^r_x)^2
\end{equation}
Freeze-out occurs during reheating at some intermediate scale factor $A_f < A_{re}$. Therefore, integrating the above equation for the number density from $A_f$ to $A_{re}$, the comoving number density at the end of reheating is
\begin{eqnarray}{\label{nsol}}
N^x_r(A_{re})=\frac{3(1-w_\phi)}{2\langle\sigma v\rangle}H(A_{re})A^3_{re}\left(A_{re}/{A_{f}}\right)^{-3(1-w_\phi)/2}
\end{eqnarray}
Therefore, the current abundance can be written as,
 \begin{equation}{\label{omegafout}}
    \Omega_xh^2= {\bf \Omega}^{\dagger} h^2 \left(\frac{m_x}{T_{re}}\right) \left(\frac{A_f}{A_{re}}\right)^{\frac{3(1-w_\phi)}{2}} ~~;~~~{\bf \Omega}^{\dagger} = \frac{\sqrt3(1-w_\phi)}{2\sqrt\epsilon M_p\langle\sigma v\rangle}\frac{\Omega_r}{T_{now}}
 \end{equation}
We will evaluate the abundance for the aforementioned two cases for different reheating models discussed earlier in detail.

{\textbf{Freeze out temperature} :} 
The freeze-out temperature $T_f$ in general can be computed from Eq.\ref{fo1} by assuming $H(T_f) \propto T_f^k$ as,
\begin{equation}
    T_f^{3/2}e^{-m_x/T_f}=\mathcal{K}(T_{re},T_c)  T_f^k .
\end{equation}
The general solution of the above equation is expressed in terms of Lambert function $W_{-1}(q)$ of branch $-1$ with argument $q$,
\begin{equation}{\label{Tf}}
T_f=-\frac{2m_x}{2k-3}\frac{1}{W_{-1}(q)} ~~\mbox{with}~~q=-\frac{2m_x}{2k-3}\mathcal{K}^{\frac{2}{2k-3}} .
\end{equation}
We further assume the approximate $\langle\sigma v\rangle$ being independent of temperature in the above expression and throughout our paper. If freeze-out happens after the reheating, one will simply have  $\mathcal{K}=\mathcal{K}_0 = \frac{\sqrt\epsilon}{\sqrt3 M_p\langle\sigma v\rangle j_x}(2\pi/m_x)^{3/2}$, and $k=2$. Consequently, the DM parameter space $(m_x\mbox{vs}\langle\sigma v\rangle)$ turns out to be the same for all reheating temperatures and inflaton equations of states. However, if freeze-out happens during reheating, the expression of $(\mathcal{K},k)$ will differ. It is observed that generically we can express $\mathcal{K}(T_{re},T_{c/s})= \mathcal{K}_0 T_{re}^{\mu} T_{c/s}^{\nu}$, where $(\mu, \nu)$ will assume different values for different reheating history and that will be our subject of our subsequent discussion.\\

 \noindent
$\clubsuit$ : \textbf{Freeze-out from the bosonic radiation bath} \\
\begin{figure}
          \begin{center}
  \includegraphics[width=16.5cm,height=4.50cm]
          {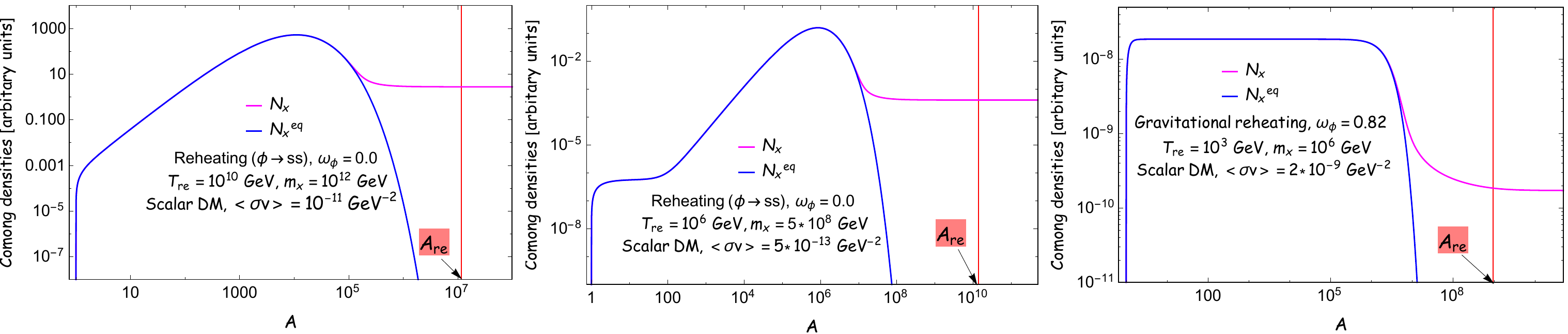}
          \caption{The evolution of the co-moving number density of the DM (WIMP) as a function of the normalized scale factor $A$ for bosonic reheating ($\phi\rightarrow ss$) for three different cases-I(left),II(middle),III(right).}
          \label{DM2}
          \end{center}
      \end{figure}
\noindent
\textbf{\bf Case-I: Coupling strength $\bf g_i^{r}>\mathscr{G}^{1,\,th}_{ci}$:} Similar to the freeze-in case, let us discuss two different  regimes.\\
\underline{ When $T^{max}_{rad}>m^{end}_\phi$ :} If the inflaton equation of state $w_{\phi} > w_{\phi}^c$, the abundance due to freeze-out can be calculated as, 
\begin{eqnarray}{\label{fzo1}}
\Omega_xh^2={\bf \Omega}^{\dagger} h^2 \left\{\begin{array}{ll}
      \left(\frac{m_x}{T_{re}}\right)\left(\frac{T_{re}}{T_f}\right)^{\frac{3(1-w_\phi)}{1-3w_\phi}}~~~\mbox{for}~~~g_1^r\phi s^2,~~\mbox{and}~~\nu = 0, \mu = 2- k, k=\frac{3(1+w_{\phi})}{1-3w_\phi}  \\
     \left(\frac{m_x}{T_{re}}\right)\left(\frac{T_{re}}{T_f}\right)^{\frac{3(1-w_\phi)}{3-5w_\phi}}~~~\mbox{for}~~~~g_2^r\phi^2 s^2,~~\mbox{and}~~\nu = 0, \mu = 2- k, k=\frac{3(1+w_{\phi})}{3-5w_\phi}\\
\end{array}\right.
\end{eqnarray}
where $T_f$ is the freeze-out temperature which we already defined in Eq. (\ref{Tf}). 

However, if $w_\phi<w^c_\phi$, as has already been discussed for the bosonic reheating, there exists an intermediate temperature scale $T_c$ at which the ratio $T_{rad}/m_{\phi}$ goes less than unity, and the thermal effect 
becomes subdominant at the later part of the reheating phase. Therefore, if $T_f > T_c$, DM freezes out during early reheating phase and the abundance is calculated to be
\begin{eqnarray}{\label{fzo2}}
\Omega_xh^2={\bf \Omega}^{\dagger} h^2\left\{\begin{array}{ll}
\left(\frac{m_x}{T_{re}}\right)\left(\frac{T_{re}}{T_c}\right)^4 \left(\frac{T_c}{T_{f}}\right)^{\frac{3(1-w_\phi)}{1-3w_\phi}} ~~~\mbox{for}~~~~g_1^r\phi s^2 ~~\mbox{and}~~\nu = \frac{4(1+w_\phi)}{1-w_\phi} -k, \mu = {\frac{-2(1+3w_\phi)}{1-w}}, k=\frac{3(1+w)}{1-3w_\phi}\\
\left(\frac{m_x}{T_{re}}\right)\left(\frac{T_{re}}{T_c}\right)^{\frac 4 3}\left(\frac{T_c}{T_{f}}\right)^{\frac{3(1-w_\phi)}{3-5w_\phi}}~~~\mbox{for}~~~~g_2^r\phi^2 s^2 ~~\mbox{and}~~\nu= {\frac{4(1+w_\phi)}{3(1-w_\phi)}-k}, \mu = {\frac{2(1-5w_\phi)}{3(1-w_\phi)}}, k=\frac{3(1+w_\phi)}{3-5w_\phi} \\
\end{array}\right.
\end{eqnarray}
However, if $T_c>T_f$, the freeze-out occurs during the later phase of reheating, and the abundance assumes a different form as
\begin{eqnarray}{\label{fzo3}}
\Omega_xh^2={\bf \Omega}^{\dagger} h^2 \left\{\begin{array}{ll}
\left(\frac{m_x}{T_{re}}\right)\left(\frac{T_{re}}{T_f}\right)^4 ~~~\mbox{for}~~~~~~g_1^r\phi s^2~~ \mbox{and}~~\nu = 0, \mu = 2- k, k=\frac{4(1+w_{\phi})}{1-w_\phi} \\
 \left(\frac{m_x}{T_{re}}\right)\left(\frac{T_{re}}{T_f}\right)^{4/3}~~~\mbox{for}~~~~g_2^r\phi^2 s^2~~~\mbox{and}~~\nu = 0, \mu = 2- k, k=\frac{4(1+w_{\phi})}{1-w_\phi} \\
\end{array}\right.
\end{eqnarray}
\underline{$T^{r,max}_{gr}<T^{max}_{rad}<m^{end}_\phi$ :} For this case, if the inflaton equation of state $w_\phi<w_\phi^c$, the effective mass of inflaton remains to be greater than the radiation temperature, and the finite temperature effect will be subdominant throughout. The DM abundance for such case will be same as Eq.\ref{fzo3}, and the evolution of the comoving number density is depicted in Fig.(\ref{DM2}). Whereas for $w_\phi>w_\phi^c$, a new temperature scale $T_c$ emerges as before. If DM mass happens to satisfy the condition $m_x < T_c$, DM freezes out with radiation temperature $T < T_c$, and the abundance assumes the same form as expressed in Eq.(\ref{fzo1}). On the other hand if $m_x > T_c$, freeze-out occurs during the initial non-thermal phase, and DM abundance becomes,
\begin{eqnarray}{\label{fzo4}}
\Omega_xh^2={\bf \Omega}^{\dagger} h^2 \left\{\begin{array}{ll}
& \left(\frac{m_x}{T_{re}}\right)\left(\frac{T_{c}}{T_f}\right)^4 \left(\frac{T_{re}}{T_c}\right)^{\frac{3(1-w_\phi)}{1-3w_\phi}} ~~~\mbox{for},~~~g_1^r\phi s^2,~~~~\nu = {\frac{3(1+w_\phi)}{1-3w_\phi}-k}, \mu = {\frac{-(1+9w_\phi)}{1-3w_\phi}}, k=\frac{4(1+w)}{1-w_\phi} \\
& \left(\frac{m_x}{T_{re}}\right)\left(\frac{T_{c}}{T_f}\right)^{4/3}\left(\frac{T_{re}}{T_c}\right)^{\frac{3(1-w_\phi)}{3-5w_\phi}}~~~\mbox{for}~~~g_2^r\phi^2 s^2,~~~~\nu ={\frac{3(1+w_\phi)}{3-5w_\phi}-k}, \mu = {\frac{3-13w_\phi}{3-5w_\phi)}}, k=\frac{4(1+w)}{3(1-w_\phi)}\\
\end{array}\right.
\end{eqnarray} 

\noindent
\textbf{Case-II: Coupling strength in between $\bf \mathscr{G}^{2,\,th}_{ci}<g_i^{r}<\mathscr{G}^{1,\,th}_{ci}$ :}This reheating history is described in the bosonic reheating section. The gravity-mediated decay channel controls the initial reheating dynamic up to $T=T_s$, and then the non-gravitational decay channel controls the reheating dynamics. As a result, initially, DM production is driven by the gravity-mediated decay channel (up to $T=T_s$) and then driven by the non-gravitational decay channel (see middle Fig.\ref{DM2}). For $w_\phi>w_\phi^c$, if the DM freezes out during the late non-gravitational decay channel domination phase, i.e., $T_f<T_s$, then the DM abundance follows the Eq.\ref{fzo1}, and if  the DM is frozen  out during gravity mediated reheating phase $T_f>T_s$, the DM has following abundance,
\begin{eqnarray}{\label{fzo4}}
\Omega_xh^2= {\bf \Omega}^{\dagger} h^2\left\{\begin{array}{ll}
& \left(\frac{m_x}{T_{re}}\right)\left(\frac{T_{s}}{T_f}\right)^{\frac{3(1-w_\phi)}{2}}\left(\frac{T_{re}}{T_s}\right)^{\frac{3(1-w_\phi)}{1-3w_\phi}}  ~~~\mbox{for},~~~g_1^r\phi s^2,~~\nu = {\frac{3(1+w_\phi)}{1-3w_\phi}-k}, \mu = {\frac{-(1+9w_\phi)}{1-3w_\phi}}, k=\frac{3(1+w)}{2} \\
& \left(\frac{m_x}{T_{re}}\right)\left(\frac{T_{s}}{T_f}\right)^{\frac{3(1-w_\phi)}{2}}\left(\frac{T_{re}}{T_s}\right)^{\frac{3(1-w_\phi)}{3-5w_\phi}}~~~\mbox{for}~~~g_2^r\phi^2 S^2,~~\nu ={\frac{3(1+w_\phi)}{3-5w_\phi}-k}, \mu = {\frac{3-13w_\phi}{3-5w_\phi}}, k=\frac{3(1+w)}{2}\\
\end{array}\right.
\end{eqnarray} 
 Again for $w_\phi<w_\phi^c$, if the DM is frozen out during the late decay channel domination phase, the abundance has the Eq. (\ref{fzo3}), and if the DM is frozen out during the initial gravity-mediated reheating phase, the DM abundance has the following equation
 \begin{eqnarray}{\label{fzo4}}
\Omega_xh^2= {\bf \Omega}^{\dagger} h^2\left\{\begin{array}{ll}
& \left(\frac{m_x}{T_{re}}\right)\left(\frac{T_{s}}{T_f}\right)^{\frac{3(1-w_\phi)}{2}}\left(\frac{T_{re}}{T_s}\right)^4  ~~~\mbox{for},~~~g_1^r\phi s^2,~~\nu = {\frac{4(1+w_\phi)}{1-w_\phi}-k}, \mu = {\frac{-2(1+3w_\phi)}{1-w_\phi}}, k=\frac{3(1+w)}{2} \\
& \left(\frac{m_x}{T_{re}}\right)\left(\frac{T_{s}}{T_f}\right)^{\frac{3(1-w_\phi)}{2}}\left(\frac{T_{re}}{T_s}\right)^{4/3}~~~\mbox{for}~~~g_2^r\phi^2 S^2,~~\nu = {\frac{4(1+w_\phi)}{3(1-w_\phi)}-k}, \mu = {\frac{2(1-5w_\phi)}{3(1-w_\phi)}}, k=\frac{3(1+w)}{2}\\
\end{array}\right.
\end{eqnarray} 
 \\
\textbf{Case-III :} For this reheating scenario, the gravitational sector is the governing reheating
dynamics termed as gravitational reheating (GR). The evolution of co-moving number density is shown in the left plot in Fig. (\ref{DM2}). The DM abundance is 
\begin{equation}
\Omega_xh^2={\bf \Omega}^{\dagger} h^2 (m_x/T_{re})(T_{re}/T_f)^{\frac{3(1-w_\phi)}{2}},~~\mbox{and}~~\nu=0,\mu=2-k,k=\frac{3}{2}(1+w_\phi)
\end{equation}
%where $\mathcal{K}=\mathcal{K}_0T_{re}^{2-k}$ and $k=\frac{3}{2}(1+w_\phi)$.\\

\noindent
$\clubsuit$ : \textbf{Freeze-out from the Fermionic radiation bath} \\

\begin{itemize}
    \item \underline{$w_\phi>5/9$} : For $w_\phi>5/9$, the bath temperature always behaves as $T_{rad}=T_{re}(A_{re}/A)$, and using this equation into Eq.(\ref{omegafout}), the abundance is,
     \begin{equation}{\label{fdout1}}
         \Omega_xh^2={\bf \Omega}^{\dagger} h^2 (m_x/T_{re})(T_{re}/T_f)^{\frac{3(1-w_\phi)}{2}},~~\mbox{and}~~\nu=0,\mu=2-k,k=\frac{3}{2}(1+w_\phi)
     \end{equation}
% where $\mathcal{K}=\mathcal{K}_0T_{re}^{2-k}$ and $k=\frac{3}{2}(1+w_\phi)$.  
\end{itemize}
\begin{itemize}
 \item \underline{$\bf 0 < w_\phi<5/9$}:  For this range of equation of state, we discuss three different possibilities as follows:
  \begin{figure}[t!] 
 	\begin{center}
 		\includegraphics[width=17.0cm,height=5.0cm]{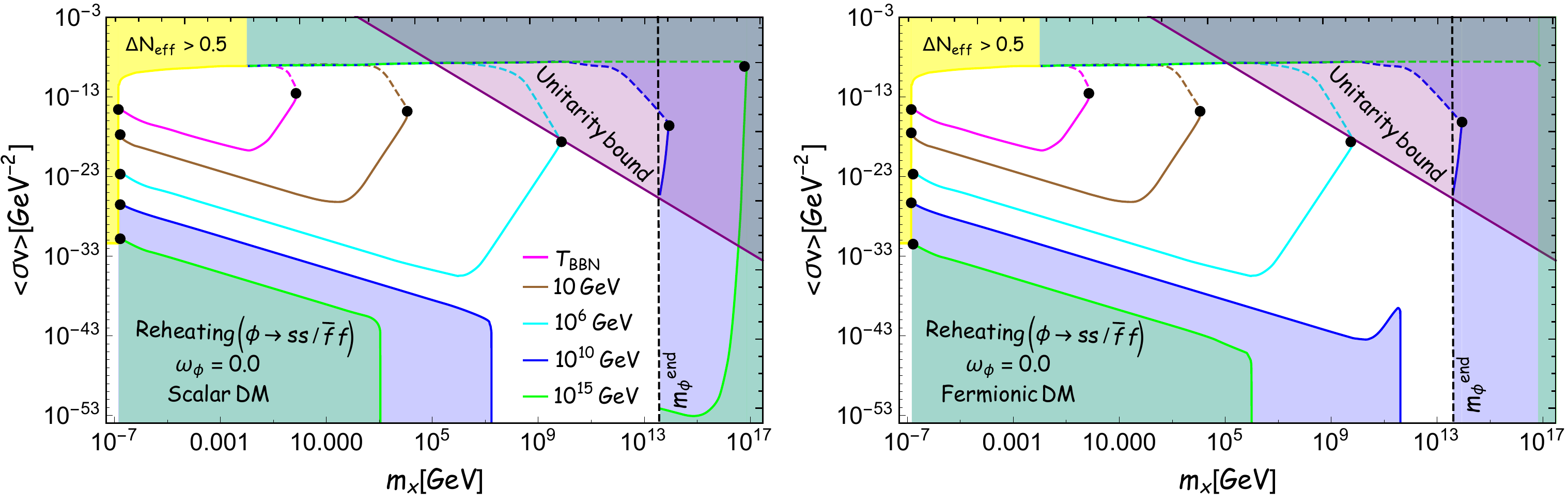}\quad
 		\caption{$\langle\sigma v\rangle$ vs $m_x$ for both freeze-in and freeze-out with five different reheating temperature $T_{re}=(T_{BBN},10, 10^6, 10^{10},10^{15}$ GeV for $w_\phi=0.0$. The left plot is for scalar DM, and the right plot fermionic DM. The solid (dashed) lines for freeze-in (freeze-out). The Small filled circle denotes the freeze-in and freeze-out coincidence point. The yellow-shaded region is ruled out by the $\Delta N_{\mbox{eff}}$ bound at BBN, and purple shaded region is ruled by the unitarity bound. }
 		\label{Fig.12}
 	\end{center}
 	\end{figure}
  
\noindent 
\textbf{\bf Case-I: Coupling strength $\bf h^{r}>\mathscr{H}_c$:}\\
$\underline{T_{rad}^{max} > m^{end}_\phi}$: 
 Depending upon the evolution of radiation, we will have different behavior of the DM abundance in terms of reheating temperature. For $7/15 < w_{\phi} < 5/9$ the radiation temperature behaves $A^{-1}$, and abundance follows the Eq.\ref{fdout1}. On the other hand if $1/5 < w_{\phi} < 7/15$, radiation temperature behaves $T_{rad}=T_{re}(A/A_{re})^{-\frac{3(1+5w_\phi)}{10}}$, and using this in Eq.\ref{omegafout}, we get
 \begin{equation}{\label{fdout2}}
         \Omega_xh^2={\bf \Omega}^{\dagger} h^2 (m_x/T_{re})(T_{re}/T_f)^{\frac{5(1-w_\phi)}{1+5w_\phi}}~~\mbox{and}~~\nu=0,\mu=2-k,k={5(1+w_\phi)}/{1+5w_\phi}
     \end{equation}
%where $\mathcal{K}=\mathcal{K}_0T_{re}^{2-k}$ and $k={5(1+w_\phi)}/{1+5w_\phi}$.
Finally, if the 
$0 \leq w_\phi < 1/5$ similar to scalar reheating case, the intermediate temperature scale $T_c$, leads to two different possibilities. If $m_x > T_c$, DM freezes out during the early reheating phase, and we have, 
      \begin{equation}{\label{fdout3}}
         \Omega_xh^2={\bf \Omega}^{\dagger} h^2 \left( \frac{m_x}{T_{re}}\right)\left(\frac{T_c}{T_f}\right)^{\frac{5(1-w_\phi)}{1+5w_\phi}}\left(\frac{T_{re}}{T_{c}}\right)^{\frac{4(1-w_\phi)}{1+3w_\phi}}~~~\mbox{and}~~\mu=\frac{2(w_\phi-1)}{1+3w_\phi},\nu=\frac{5(1+w_\phi)}{1+5w_\phi}-k,k=\frac{5(1+w_\phi)}{1+5w_\phi}
     \end{equation}
     where
On the other hand, if $m_x < T_c$, DM freezes out during the late reheating phase when the finite temperature effect is subdominant, and we have
 \begin{equation}{\label{fdout4}}
         \Omega_xh^2={\bf \Omega}^{\dagger} h^2  \left( \frac{m_x}{T_{re}}\right)\left(\frac{T_{re}}{T_f}\right)^{\frac{4(1-w_\phi)}{1+3w_\phi}}~~\mbox{and}~~\nu=0,\mu=2-k,k=\frac{4(1+w_\phi)}{1+3w_\phi}
     \end{equation}
 \underline{$T^{r, max}_{gr}<T^{max}_{rad}<m^{end}_\phi$:} For this case, finite temperature correction is subdominant at the beginning. In fact, for lower EoS, namely $0 < w_\phi<1/5$, such finite temperature effect will be subdominant throughout the reheating, and hence the abundance turned out to be same as Eq.(\ref{fdout4}). On the other hand, for $w_\phi>1/5$, the finite temperature effect will become dominant after an intermediate temperature scale $T_c$, and if the DM freeze-out temperature satisfies $T_f < T_c$, the abundance obeys the Eq.\ref{fdout2}. Conversely, if the DM freezes out during the early non-thermal  phase with $T_f > T_c$, one will have
    \begin{equation}
         \Omega_xh^2={\bf \Omega}^{\dagger} h^2 \left(\frac {m_x}{T_{re}}\right)\left(\frac{T_c}{T_f}\right)^{\frac{4(1-w_\phi)}{1+3w_\phi}}\left(\frac{T_{re}}{T_{c}}\right)^{\frac{5(1-w_\phi)}{1+5w_\phi}}~~\mbox{and}~~\nu =\frac {4(1+w_\phi)}{1+3w_\phi} -k,\mu =\frac{ -3+5w_\phi}{1+5w_\phi}, k =\frac{ 4(1+w_\phi}{1+3w_\phi}
     \end{equation}
\begin{figure}[t!] 
 	\begin{center}
 	    	\includegraphics[width=17.0cm,height=5.0cm]{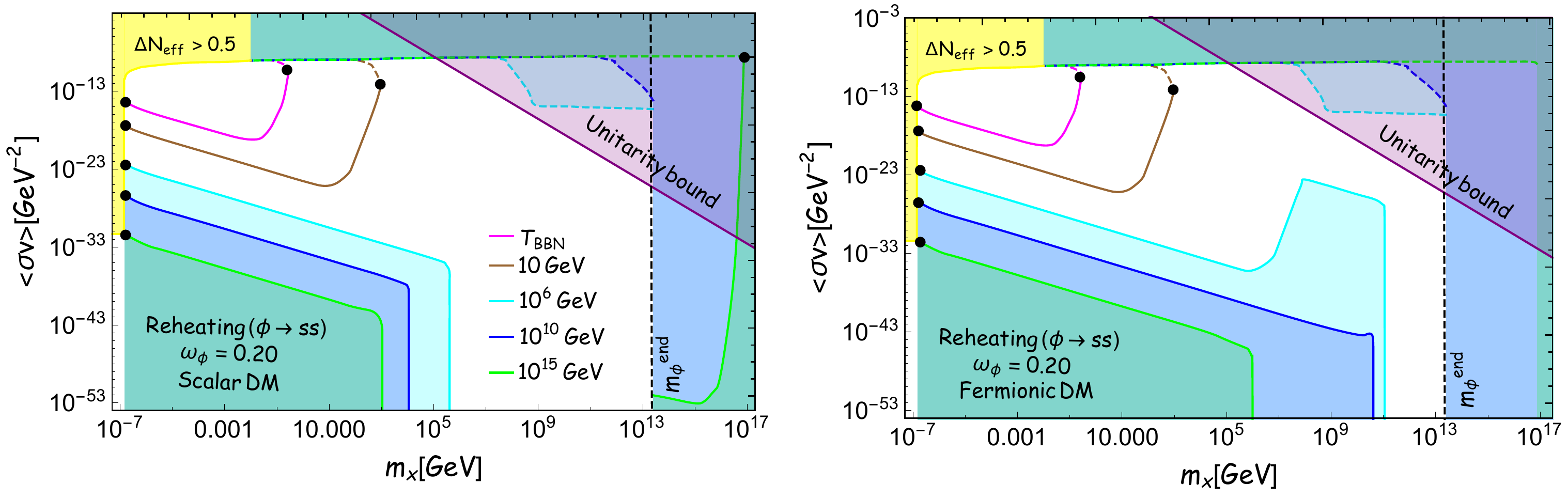}\quad
 				\includegraphics[width=17.0cm,height=5.0cm]{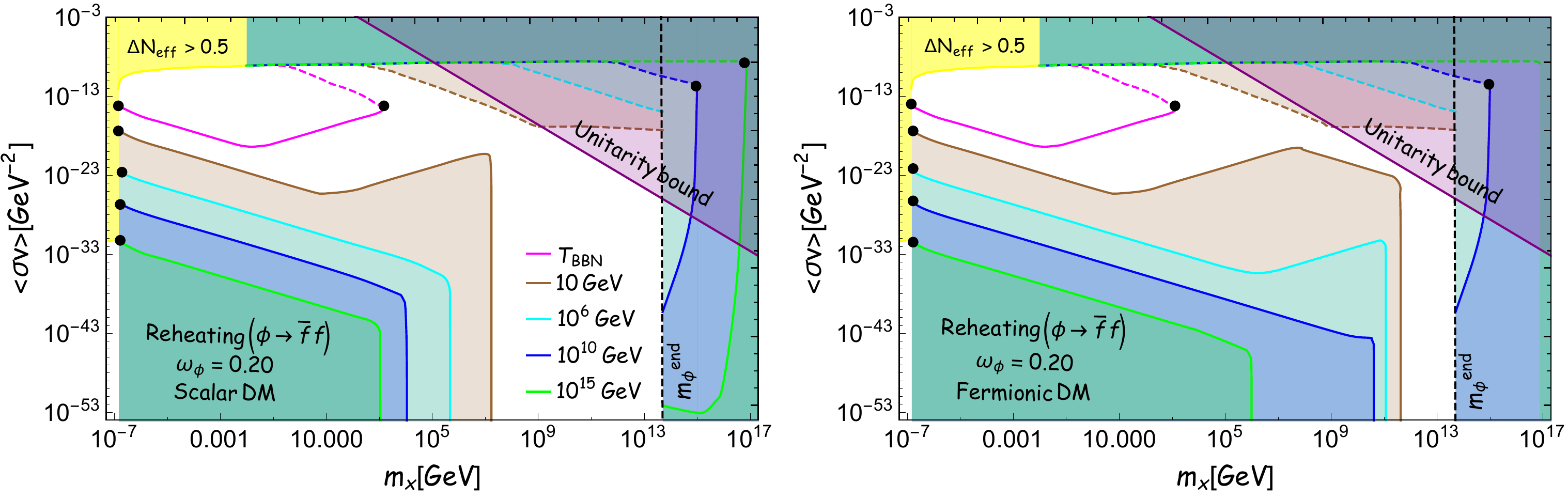}
     		\caption{$\langle\sigma v\rangle$ vs $m_x$ for both freeze-in (solid lines) and freeze-out (dashed lines) with five different reheating temperature $T_{re}={BBN,10,10^6,10^{10},10^{15}}$ GeV for $w_\phi=0.20$. The left plot is for scalar DM, and the right plot is for fermionic DM. The small filled circle is the freeze-in and freeze-out meet point. The yellow-shaded region is ruled out by the $\Delta N_{\mbox{eff}}$ bound at BBN, and purple shaded region is ruled by the unitarity bound.}
 		\label{Fig.13}
 	\end{center}
 \end{figure}
ii) \underline{\textbf{Coupling strength }$h^{r}<\mathscr{H}_{c}$:} For this case, initially, the reheating phase is governed by the gravity-mediated decay channel up to a point $A_{gr\rightarrow ngr}$, and the later part of the phase is dominated by direct inflaton decay. If the freeze-out happens during the later phase, the abundance will be the same as Eq.\ref{fdout2} for $w_\phi>1/5$ and Eq.\ref{fdout3} for $w_\phi<1/5$. If freeze-out occurs at the early phase of reheating, the abundance assumes,  
\begin{eqnarray}
         \Omega_xh^2={\bf \Omega}^{\dagger} h^2 \frac{m_x}{T_{re}}\left\{\begin{array}{l}
              \left(\frac{T_s}{T_f}\right)^{\frac{3(1-w_\phi)}{2}}\left(\frac{T_{re}}{T_{s}}\right)^{\frac{5(1-w_\phi)}{1+5w_\phi}}~~\mbox{for}~~w_\phi>1/5,~~\nu = {\frac{5(1+w_\phi)}{1+5w_\phi}-k},\mu = \frac{-3+5w_\phi}{1+5w_\phi}, k=\frac{3(1+w)}{2} \\
              \left(\frac{T_s}{T_f}\right)^{\frac{3(1-w_\phi)}{2}}\left(\frac{T_{re}}{T_{s}}\right)^{\frac{4(1-w_\phi)}{1+3w_\phi}}~~\mbox{for}~~w_\phi<1/5,~~\nu = {\frac{4(1+w_\phi)}{1-w_\phi}-k}, \mu = \frac{-2+2w_\phi}{1+3w_\phi}, k=\frac{3(1+w)}{2}
         \end{array}\right.
     \end{eqnarray}
\end{itemize}

\begin{figure}[t!] 
 	\begin{center}
     \includegraphics[width=15.0cm,height=15.0cm]{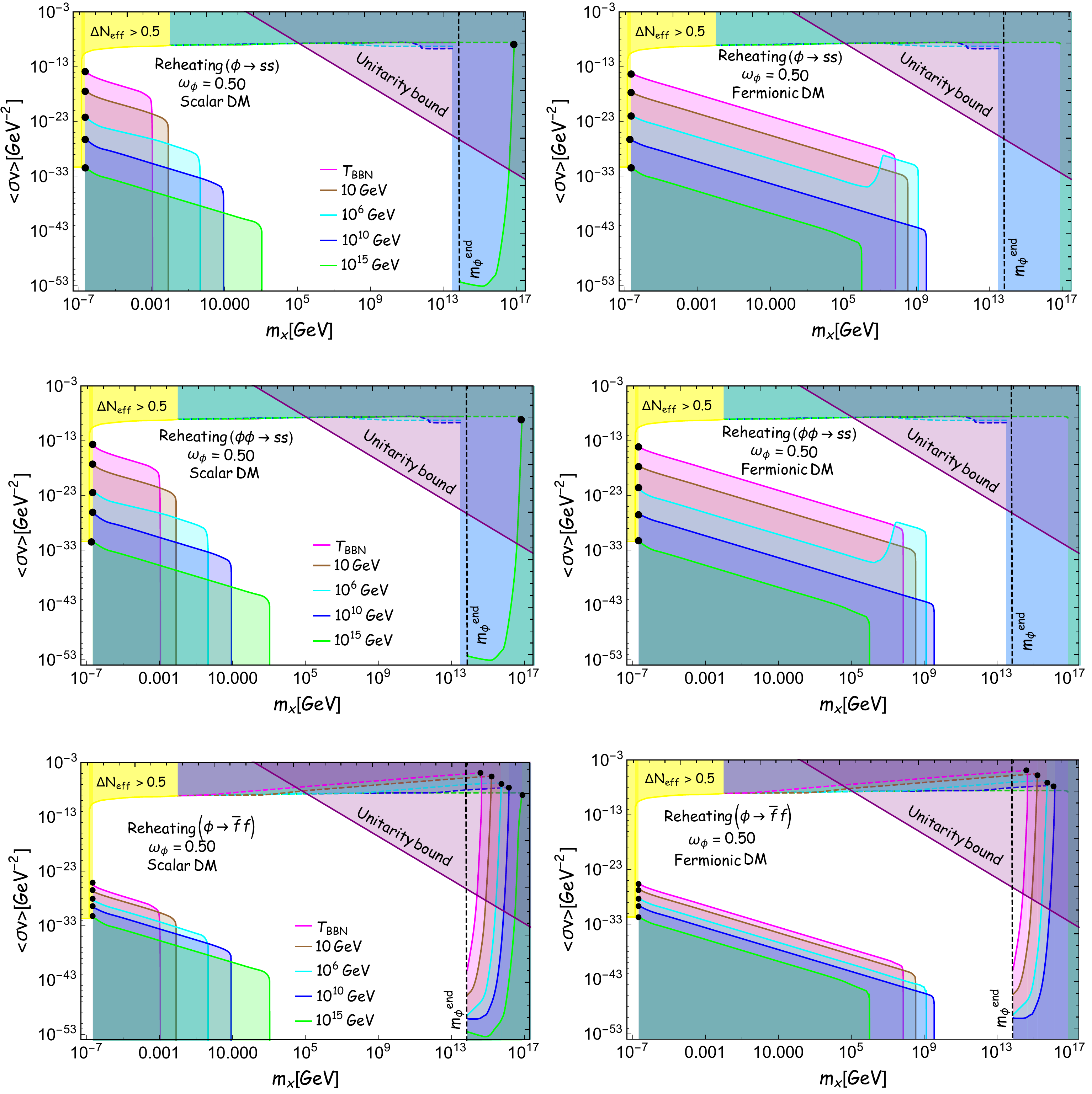}\quad
 		\caption{ Here we have plotted the $\langle\sigma v\rangle$ as a function of dark matter mass $m_x$ for $w_\phi=0.50$ for five different reheating temperature $T_{re}={BBN,10,10^6,10^{10},10^{15}}$ GeV. The plot is on the left side for scalar dark matter and the right side for fermionic dark matter. The solid lines for the freeze-in mechanism and the dotted lines for the freeze-out mechanism. The yellow-shaded region is ruled out by the $\Delta N_{\mbox{eff}}$ bound at BBN, and purple shaded region is ruled by the unitarity bound.}
 		\label{Fig.14}
 	\end{center}
 \end{figure}
\section{DM Parameter space, $(\langle \sigma v\rangle, m_x)$ for FIMPS and WIMPS from radiation bath}{\label{section6}}
In this section, we will discuss in detail the parameter region where DM production mechanisms are at play for different reheating histories. For the sake of completeness and understanding the DM parameter space, we consider three different EoS ($w_\phi=0.0,0.20,0.50$) with five different reheating temperature $(T_{BBN}, 10,10^6,10^{10}, 10^{15})$ GeV which includes both minimum and maximum reheating temperature. An important point we want to infer is that in the final contribution to DM abundance, we ignore any possible contribution from the non-gravitational inflaton-DM coupling but include  the universal gravitational production from both inflaton and radiation scattering. And as we pointed out before, such universal production has been observed to set a maximum limit on the DM mass $m_x^{g, max}$ ($\leq m_\phi^{end}$ of course) specifically for the freeze-in production.
%for higher reheating temperature \textcolor{blue}{for Eos $w_{\phi}=0,0,0.20$} (see Fig.\ref{Fig.12}, \ref{Fig.13}), and for any arbitrary reheating temperature \textcolor{blue}{EoS $w_{\phi}=0.5$} (see Fig.\ref{Fig.14}). 
Interestingly for a given $w_{\phi}$, we found a universal feature of lowest possible DM mass within $m_x^{min} \simeq 150-300$ eV (see Eq. (\ref{critmass})) irrespective of its nature and the reheating histories for which freeze-in and freeze-out mechanism coincides during the radiation-dominated era. Below this mass scale, all the DM has under abundance today. However, the critical cross-section  $\langle\sigma v\rangle_{crit}$ (for analytical expressions, see Appendix-\ref{critcal}) at which this coincidence occurs depends on the reheating temperature, which is represented by filled black circles in Figs.\ref{Fig.12}-\ref{Fig.14}. Moreover, another critical cross-section exists for higher DM mass $(m_x^{max})$ where the freeze-in and freeze-out mechanism coincides during reheating. 
%to occur at reheating temperature lower than around $\sim 10^7$ GeV with $w_{\phi} < 1/3$.
Unlike the lower DM mass bound, $m_x^{max}$ has been found to have non-trivial dependence on the reheating temperature but also depends on the EOS $w_\phi$, reheating background. All these features are clearly observed in Figs.\ref{Fig.12}, \ref{Fig.13}, \ref{Fig.14}. In some reheating temperatures, there is no meet point of freeze-in and freeze-out; it is due to the gravitational DM, which gives the overabundance for the freeze-in mechanism where we expect the meet point occurs.
%For $T_{re}=10^{15}$ GeV, fermionic DM has no freeze-in and freeze-out meet point, it is because, for the freeze-in process, the DM has always overabundance when $m_x>10^6$ GeV.
When $m_x^{max}<m_x^{g, max}$ condition is satisfied, then the abundance condition $\Omega_xh^2 = 0.12$ gives rise to a closed contour in $(m_x, \langle \sigma v\rangle)$ plane due to aforementioned two critical cross-sections for two different masses $(m_x^{min}, m_x^{max})$, where the smooth transition happens from freeze-in to freeze-out or vice versa. For example, see Fig. (\ref{Fig.12}) where we have found  close contours for $T_{re}=(10^{-2}, 10, 10^6)$ GeV and in  Fig. (\ref{Fig.13}) for $T_{re}=(10^{-2}, 10)$ GeV with bosonic reheating (for fermionic reheating we only have closed contour for $T_{re}=10$ GeV). On the other hand, if $m_\phi^{end}>m_x^{max}>m_x^{g, max}$, the critical cross-section disappears for higher mass values; for such cases, the maximum allowed DM mass is turned out to be $m^{g, max}_x$ ($m_{x}^{max}$) for freeze-in (freeze-out) mechanism. Again, when $m_x^{max}>m_\phi^{end}$ (it happens for higher reheating temperatures), one can find the freeze-in and freeze-out meet point, but the close contour is not formed. To this end, we would also like to point out that DM annihilation is dominant but under-abundant in the region outside the closed contours, whereas shaded regions are under-abundant for open contours. The minimum and the maximum DM masses up to which DM can give present abundance are $10^{-7}$ and $7\times 10^{16}$ GeV, respectively. But, further, some parameter space is ruled out by the $\Delta N_{\mbox{eff}}$ at BBN (yellow shaded region) and by unitarity bound of $\langle\sigma v\rangle\leq {8\pi}/{m_x^2}$\cite{Griest:1989wd,Giudice:2000ex,Bhatia:2020itt}. When WIMP DM has a mass scale in the order of $T_{BBN}$ or below, it violates the $\Delta N_{eff}$ bound at BBN. Again for WIMP DM, freeze-out happens during RD ($m_x<T_{re}$) if the mass scale lies above $10^5$ GeV, it violates the unitarity bound of $\langle\sigma v\rangle$, which is shown by the light purple color region. Using two bounds, for the freeze-in mechanism, the maximum and the minimum allowed scalar (fermionic) DM masses are $10^{-7}(10^{-7})$ GeV, $4\times 10^{16}(10^{15})$ GeV respectively. 

When $m_x < T_{re}$, we obtained some generic behavior of the DM abundance specifically for freeze-in production. Moreover, in the freeze-in production mechanism, if DM freezes during the radiation-dominated era, $\langle\sigma v\rangle$ behaves as $\propto 1/m_x$ (see Eqs.\ref{Wx1}, \ref{wx3}, \ref{wx6}). On the other hand, when $m_x < T_{re}$, for the freeze-out scenario, even though we do not have such simple relation, it turned out that $\langle\sigma v\rangle$ generically increases slowly with increasing $m_x$.
\section{WIMPs, experimental bounds and constraints on reheating}
\begin{figure}[t!] 
 	\begin{center}
 		\includegraphics[width=17.0cm,height=5.50cm]{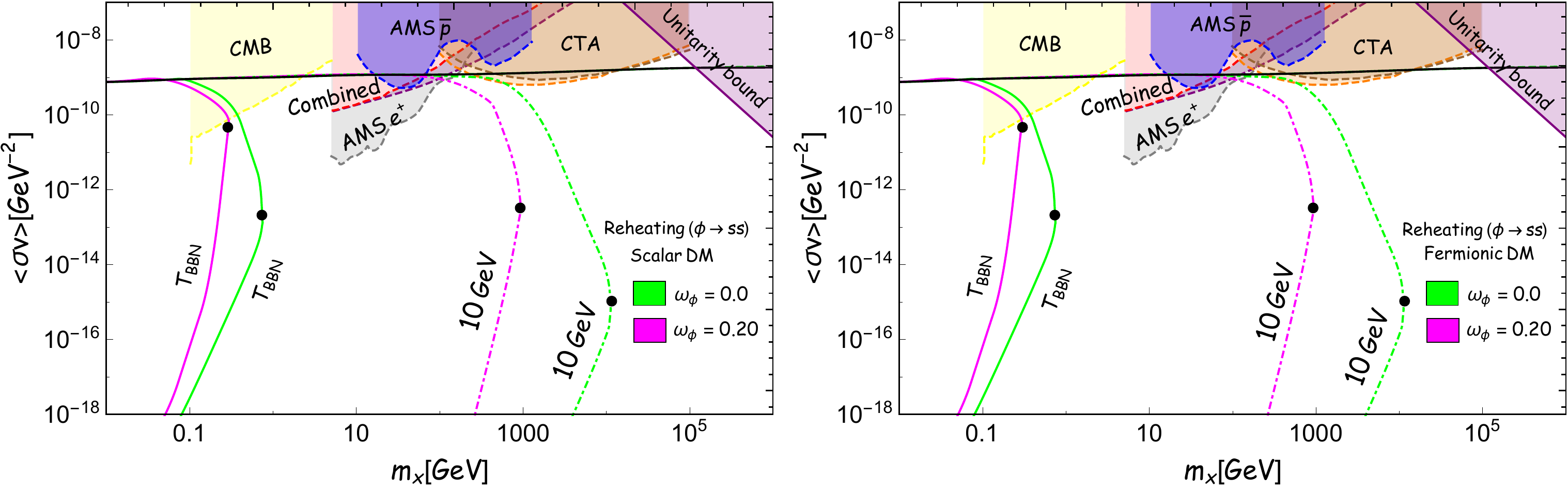}\quad
 		\includegraphics[width=17.0cm,height=5.50cm]{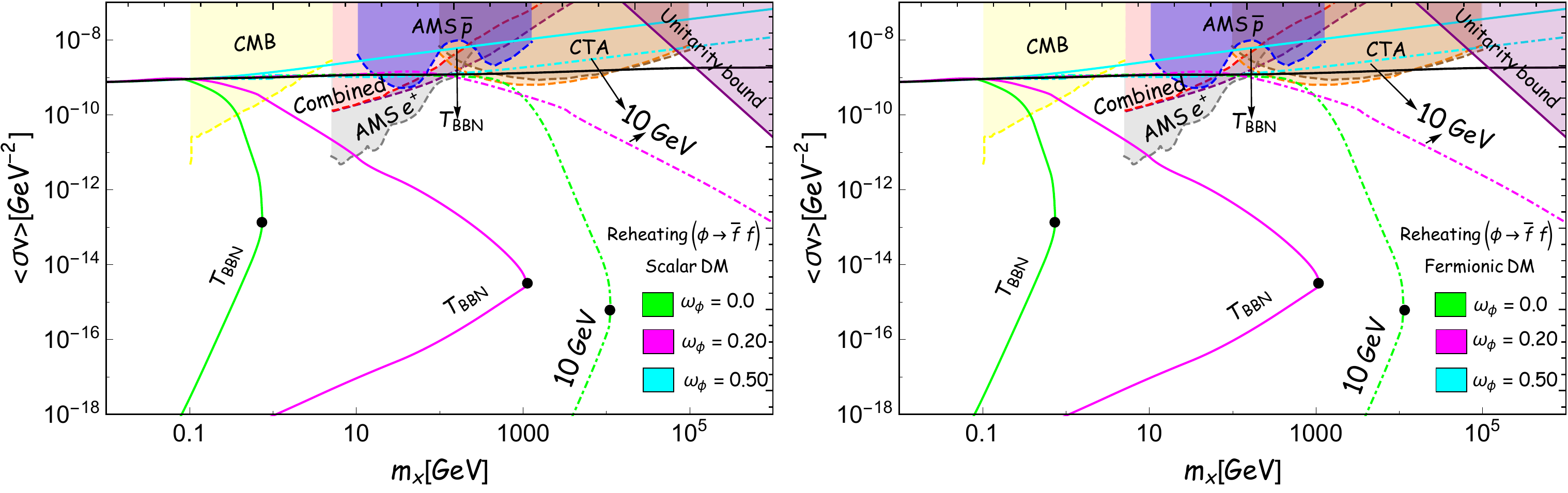}
 		\caption{$\langle\sigma v\rangle$ Vs $m_x$ for WIMP with experimental bound and future sensitivity for three different EOS $w_\phi=0.0(\mbox{green}),0.20(\mbox{magenta}),0.50(\mbox{cyan})$. The upper plot for the bosonic reheating $(\phi\rightarrow ss)$ and the left plot for the fermionic reheating $\phi\rightarrow \bar ff$. The left (right) plot for the scalar (fermionic) DM. The solid (dot-dashed) lines for $T_{re}=T_{BBN}(10)$ GeV and the small filled circle corresponding to the freeze-in and freeze-out meet point.}
 		\label{Expbound}
 	\end{center}
 \end{figure}

{\label{section7}}
In this section, we would like to discuss our results from the perspective of some indirect experimental constraints with their future projected sensitivity limit on the DM cross-section and its mass plane. BBN constraints on the effective number of relativistic degrees of freedom $\Delta N_{eff}$ put direct constraints on the possible lower limit on the DM mass. In all the $\langle\sigma v\rangle$ Vs $m_x$ plots, the yellow shaded regions depict the forbidden region coming from BBN bound. For WIMP-like DM, the condition of the $\Delta N_{eff}$ at BBN sets the approximate lowest possible DM mass $m_x \sim T_{BBN}$. In addition, for $2 \to 2$ scattering processes, the unitarity constraints further provide a bound on the maximum possible mass $m_x \sim 10^5$ GeV for thermally produced DMs which freeze out, particularly during the RD era ($T_{re}> m_x$). However, if freeze-out occurs during reheating era ($T_{re}< m_x$), DM mass can be large, which is observed from the pink dotted curve of the third and fourth plot of Fig.\ref{Expbound}. Therefore, reheating dynamics plays a significant role in setting the possible values of maximum DM mass, and it depends non-trivially on the inflationary parameter such as inflaton equation of state.  In Fig.(\ref{Expbound}), we have shown where our WIMP DM parameter space lies with the available sensitivity curve for indirect experiments within the mass range ($0.01-10^5$) GeV such as (i) PLANCK CMB measurement \cite{Leane:2018kjk} labeled as CMB shown in the light yellow region in the mass range of 0.1 to 10 GeV. (ii) Alpha magnetic spectrometer (AMS)-02 experiment  \cite{Bergstrom:2013jra,Calore:2022stf} provided cosmic ray positron ($e^+$) and antiproton ($\bar p$) data with unprecedented precision in the mass range $5\to10^3$ GeV. (iii) Combined bounds (labeled as Combined) from Fermi-LAT, HAWC, HESS, MAGIC, and VERITAS experiment \cite{Hess:2021cdp} provided a upper limit on the DM annihilation cross-section in the case of two annihilation channels $\bar bb$ (purple dashed) and $\tau^+\tau^{-}$ (red dashed). (iv) CTA experiment \cite{CTA:2020qlo} is a  ground-based gamma-ray detector that could be capable of searching the WIMP DM at the TeV scale with large sensitivity. The brown (orange) dashed line shows its expected sensitivity for WIMP annihilation to $\bar bb$ ($W^+W^-$). In the near future, CTA might be able to probe a significant portion of TeV scale WIMP DM that freeze-out during either RD or reheating era.\\
In order to illustrate our result, we choose $w_\phi=(0.0,0.20,0.50)$ with the background of both bosonic (upper plot) and fermionic reheating (lower plot). The solid and dot-dashed lines are for $T_{re}=T_{BBN}, 10$ GeV, respectively. The black solid lines correspond to those DMs which freeze out during the RD era and for which one requires $\langle\sigma v\rangle\sim \mathcal{O}(10^{-9})$ GeV$^{-2}$ to get correct present-day DM relic. Moreover, most of the parts of this black line are ruled out by the experimental bound (except CTA since it is a future experiment) as it lies inside the regime of the sensitivity curves. For bosonic reheating, if $w_\phi=0.5$, for both $T_{re}=T_{BBN}$ and $T_{re}=10$ GeV, the WIMPs follow the black line (see, for instance, first and the second plot of Fig. \ref{Expbound}) to achieve
the correct relic. For lower EOS, $w_\phi=0.0,0.20$ with low reheating temperature, WIMP-like DM freezes out during reheating, and one requires the lower value of $\langle\sigma v\rangle$ to obtain present-day DM relic. Therefore, the DM parameter space for freeze-out during reheating is still very much allowed. Thus in order to detect those DMs in the near future, we need to increase the sensitivity of the experiments.

\section{Conclusions and Outlook}
In this paper, we performed a detailed analysis of the physics of reheating after the end of inflation. The physics after the end of inflation is expected to play a significant role in every aspect of the late time universe. Apart from the well-established correspondence between the physics of CMB and the early inflationary era, the intriguing effects of reheating on our present universe can not be avoided. Unlike inflation, broadly reheating is similar to the usual phases of the standard big-bang model, except for the fact that it occurs right after the end of inflation. Therefore, experimentally it is challenging to look for its direct signatures. Further, it is generically argued that decoding any physics information is challenging because of non-linear thermalization processes during this phase. 
Over the years, however, endeavor toward understanding this phase has gained significant interest due to its rich new physics contents. Reheating is the phase that naturally encodes the physics of inflaton itself. The way inflaton decays into different fundamental fields is expected to be imprinted into different cosmological observables such as CMB anisotropy, DM, and gravitational waves in terms of new physics. 
Therefore, reheating could be an interesting  playground for phenomenological studies of DM and inflation in a unified framework that has not been studied extensively in the literature, and this is the main objective of our present paper. We have addressed two main topics:

{\bf Inflaton phenomenology}: In the first part, we have studied in detail the dynamics of reheating separately through inflaton decaying into scalar field (bosonic reheating) and decaying into fermions (fermionic reheating). Apart from the direct decay term, we further include two important effects, namely, the universal gravity-mediated decay of inflaton, and the finite temperature correction on different decay channels. The effects of gravitational decay have already been analyzed before \cite{Haque:2022kez}. It has been shown that purely gravitational reheating sets a lower bound on the reheating temperature when the inflaton equation of state satisfies $w_\phi>0.65$. Such a lower limit on the reheating temperature further sets constraints on the inflaton direct coupling parameter below which gravity-mediated decay will be the dominant channel for reheating process (see the cyan shaded region in Fig.\ref{couplingbound}, \ref{boundfermion}). The cyan zone of these plots is the important outcome of our present analysis. There is an equation of state-dependent critical values of the coupling constant below which reheating dynamics become completely independent of direct inflaton coupling. Combining this gravitational decay with the finite temperature correction in the direct decay, we observe intriguing interplay among those different physical effects during reheating. The thermal effect is expected to influence the dynamics when radiation temperature satisfies the condition $T_{rad}>m_\phi(t)$. Depending on the strength of the inflaton-radiation coupling, we observed several interesting reheating histories which have not been reported earlier. Let us outline the main interesting outcomes in the following discussion\\
$\spadesuit$ We particularly observed a new reheating history for which reheating temperature $(T_{re})$ and maximum radiation temperature $(T_{rad}^{max})$ coincide when the equation of state $w_\phi > (1/3, 3/5)$ for two different bosonic decay channels $\phi\rightarrow ss$ and $\phi\phi\rightarrow ss$ (see, Fig.\ref{Casemplot} for depiction) respectively.\\
$\spadesuit$ Another interesting observation in the case of fermionic reheating is that when EoS lies above $5/9 (7/15)$, maximum radiation production takes place at the initial stage of reheating. In this respect, therefore, fermionic reheating turns out to be qualitatively similar to gravitational reheating. The bosonic reheating through $\phi\phi\to ss$ decay process deserves special mention with regard to the fact that, if $w_{\phi} < 3/13 (1/5)$ with (without) thermal effect, successful reheating can not be achieved (see the grey shaded region of Fig.\ref{couplingbound}).\\
$\spadesuit$ The phenomenological constraints on the inflaton coupling are shown in Fig.\ref{couplingbound} for bosonic reheating and in Fig.\ref{boundfermion} for fermionic reheating. The plots depict an interesting connection between the inflaton coupling parameters ($g^r_i$,$h^r$) and inflationary parameter $n$ (power of the inflaton potential at its minimum), which is translated into effective EoS $w_\phi$ through the relation $(n-2)/(n+2)$ during reheating. Contrary to  the earlier claim \cite{3}, we 
observe that the thermal effect appeared to be most significant at low reheating temperatures for non-zero inflaton equation state, $w_\phi=(0.20,\,0.50,\,0.82,\,0.99)$ (see solid and dotted lines in Fig.\ref{Fig5}, \ref{Fig10}), for which inflaton mass undergoes non-trivial evolution.\\
$\spadesuit$ Thermal correction in the production rate significantly modifies the production rate (enhances for bosonic channels and diminishes for fermionic channel), which is imprinted in the radiation temperature evolution. Better visualization of how the thermal correction affects the evolution of the radiation temperature with respect to scale factor is shown in Tables-\ref{tempevboson}, \ref{tempFR}.\\
$\spadesuit$ For higher equation state ($w=0.82,0.99$), the gravitational reheating has been observed to give a lower limit on the reheating temperature ($10^3,10^6$) GeV, which is associated with the fixed scalar spectral index $n_s=0.9685$ and $0.9681$, respectively. When $n_s$ reaches these values, the coupling parameter tends towards zero, i.e., gravitational scattering solely controls the reheating dynamics.

{\bf DM phenomenology}: In the second half of the paper, we have extensively analyzed DM phenomenology in the context of production during reheating. We have discussed all known DM production mechanisms, namely: (i) gravitational production from both the inflaton and radiation scattering, (ii) production from direct inflaton decay through various decay channels, (iii) production from the thermal bath through effective average cross-section times velocity $\langle\sigma v\rangle$ in both freeze-in and freeze-out mechanism with  assuming $\langle\sigma v\rangle$ as a free parameter. The universal gravitational production of DM is always incorporated throughout our analysis. In the following, we summarize some of the main results and important outcomes of our analysis:\\
$\spadesuit$ When DMs are produced from direct decay of inflaton (see Fig.\ref{DM1}), the final abundance appears to depend only on the coupling strength, DM mass ($m_x$), and reheating temperature ($T_{re}$). We call this FIMP-like dark matter. For a fixed inflation EoS and reheating temperature, the minimum DM mass is fixed by the $\Delta N_{\mbox{eff}}$ bound at BBN, and the maximum allowed DM mass is fixed by either gravitational decay or the inflaton mass.\\
$\spadesuit$ For the production of DM from the thermal bath, we have considered both freeze-in (FIMP) and freeze-out (WIMP) mechanisms. Since the DM is produced from the thermal bath, the DM production strictly depends on the evolution of the bath temperature. Due to the thermal effect, the  bath temperature evolves differently during reheating, and the thermal effect directly influences DM production and its final abundance. \\
$\spadesuit$ In the context of both bosonic and fermionic reheating, we have discussed the DM production (both FIMP and WIMP) and derived the analytical expressions of the DM abundance. In Section-\ref {section6}, we have shown in detail the DM parameter space ($\langle\sigma v\rangle$ Vs. $m_x$) for both FIMP and WIMP production mechanisms. Interestingly, the lowest possible DM mass for both the mechanisms assumes a universal theoretical value $m_x\sim 10^{-7}$ GeV, which is independent of $T_{re}, w_\phi$ and the nature of DM. However, constraints from different physical considerations such as lyman-$\alpha$, $\Delta N_{eff}$ bounds may not satisfy this universal bound. For WIMPs, the minimum allowed value of DM mass is $T_{BBN}$ due to restriction from the $\Delta N_{eff}$. Depending on the reheating parameters such as reheating temperature, inflation EOS $w_\phi$, and the background reheating dynamics, a maximum allowed DM mass also exists. In most of the scenarios, the maximum allowed mass is given by a particular mass scale where there is a transition happening from freeze-in to Freeze-out mechanism or vice versa (see, for instance, black circle of Figs. \ref{Fig.12}-\ref{Fig.14}).\\
$\spadesuit$ For the DM production from the radiation bath, if one incorporates the unitarity bound, the DM mass lies above $10^5$ GeV is ruled out in most of the cases. However, for both fermionic and bosonic reheating, we still have some parameter space even for $m_x>10^5$ GeV (see, for instance, the third and fourth plot of Fig. \ref{Expbound}), which are consistent with the unitarity bound and satisfies the correct relic. The reason behind this is that if freeze-out happens during reheating ($m_x>T_{re}$), the freeze-out temperature shifts towards the maximum radiation temperature with the increasing mass, and we need to lower the cross-section to meet the correct relic. Such lowering of the cross-section with higher DM mass ($m_x > 10^5$ GeV), therefore, naturally evades the aforementioned unitarity bound.   \\
$\spadesuit$ Important distinction of fermionic decay width from that of the bosonic one is its proportional behavior in terms of inflaton mass, which dilutes faster with increasing $w_{\phi}$. 
Due to this, one finds a critical $w_{\phi} \simeq 7/15$ considering the thermal effect from the beginning of reheating, below which the inflaton decay process occurs throughout and above which it occurs at the beginning of the reheating. 
These two different reheating histories have had their distinct effects on the DM production, which is reflected on $\langle \sigma v\rangle~Vs~m_x$ parameter plane (see, for instance, third and forth plot of Figs.\ref{Fig.13}, and fifth and sixth plots of \ref{Fig.14}). 
 This phenomenon can also be explained from the Eqs.\ref{fdout1}-\ref{fdout4}. For $w_\phi=0.5 > 7/15$, a fixed value of $T_{re} < m_x$ and $T_f\sim m_f$, the relic abundance behaves as $ \Omega_xh^2\propto m_x^{1/4}/\langle \sigma v\rangle$ (see Eqn.\ref{fdout1}), which implies that with increasing $m_x$, $\langle \sigma v\rangle$ increases (see, for instance, fifth and sixth plots of Fig.\ref{Fig.14}). However, for $w_\phi=0 < 7/15$, for a fixed $T_{re}$ if freeze-out happens during reheating one has following relation $m_x\propto\langle \sigma v\rangle^{-3}$ (see, for instance, Eqn.\ref{fdout4}). This clearly suggests that with increasing DM mass above reheating energy scale, $\langle \sigma v\rangle$ decreases (see, for instance, Fig.\ref{Fig.12}).\\
$\spadesuit$ The oscillating inflaton condensate may be fragmented due to self-resonance (see, for instance, Refs. \cite{Lozanov:2016hid,Lozanov:2017hjm}), which is expected to cause reheating dynamics non-trivial. It is observed that such an effect is not significant for $n=1 \,(w_\phi=0)$. However, for higher $n$, self-resonance effects turned out to be important and may lead to radiation dominated (RD) universe depending on both $\alpha$ and $n$ values. When $\alpha<<1/6$, self-resonance is efficient, and the RD universe is established within less than an e-fold of expansion after inflation end. Whereas for $\alpha > 1/6$ and $n>2$, the self-resonance effect increasingly becomes inefficient, and the inflaton condensate remains intact for a long time without any other decay channel. It is in this parlance our present study is important. In our study, we considered $\alpha=1$ and further checked that the final results, namely the estimates of reheating temperature for different inflaton couplings and the whole DM parameters space, are not much sensitive with the increase of $\alpha$ within the limit set by recent Planck and BICEP/$keck$ data \cite{Ellis:2021kad,Chakraborty:2023ocr}. 
For higher $\alpha$ values, even though the self-resonance effect would be inefficient (as pointed out in \cite{Lozanov:2016hid,Lozanov:2017hjm}), it may have a non-trivial effect on the reheating parameter space.\\
%Considering such an effect in the present context may be important, and we leave it for our future study.}\\
Another important way forward of our present work would be to construct particle physics-motivated UV complete models that can  be directly probed through laboratory and cosmological experiments. One may further study the gravitational wave dynamics in this context and explore the potential signature of reheating through the present-day gravitational wave spectrum. Finally, the analysis of cosmological perturbation will be interesting to work on, which will be a good probe of early universe physics. We defer all these topics for our future work.   

\acknowledgments
M. R.H wish to acknowledge support from the Science and Engineering 
Research Board~(SERB), Government of India~(GoI), for the SERB National Post-Doctoral fellowship, File Number: PDF/2022/002988. DM wishes to acknowledge support from the Science and Engineering 
Research Board~(SERB), Department of Science and Technology~(DST), 
Government of India~(GoI), through the Core Research Grant CRG/2020/003664.
DM and RM wish to thank Nayan Das for the helpful discussions.
 We want to thank our Gravity and High Energy Physics groups at IIT Guwahati for illuminating discussions. We thank the anonymous referee for  his/her valuable comments, which helps us to improve our manuscript.

\appendix
%%%%%%%%%%%%%%%%%%%%%%%%%%%%%%%%%%
\section{Bosonic Reheating: analytical studies}\label{bosonanawth}
\subsection{Without thermal effect}  During reheating, the exponential decay term associated with the evolution of the inflaton energy density is always sub-dominant compared to the dilution due to the expansion. Thus, we can safely approximate the solution of Eq.(\ref{B1}) as
\begin{equation}{\label{i1}}
\rho_{\phi}=\rho^{end}_{\phi}A^{-3(1+w_{\phi})}e^{-(\Gamma_{s}+\Gamma_{\phi\phi\to RR})(t-t_{end})}\simeq\rho^{end}_{\phi}A^{-3(1+w_{\phi})}
\end{equation}
where $\rho^{end}_{\phi}$ and $t_{end}$ be the inflaton energy density and time at the inflation end. The evolution equation of radiation energy density
%for non-gravitational bosonic coupling
can be written as (see, for instance, Eqn.\ref{B2})
\begin{equation}{\label{A1}}
 d(\rho_s^rA^4)=\Gamma_{s}\,\rho_\phi\,(1+w_\phi)\,A^4\frac{da}{AH}
\end{equation}
 In the limit of $T_{rad}\ll m_\phi(t)$, finite temperature correction to the decay width \cite{Garcia:2020wiy} can be safely ignored (see, Eqn.\ref{t1})
\begin{eqnarray}{\label{a1}}
  \Gamma_{s}=&\left\{\begin{array}{ll}
  &\Gamma_{\phi \to ss}\simeq \frac{(g^r_1)^2}{8\pi m_\phi(t)}~~~\mbox{for}~~ g^r_1 \phi s^2 \\
       &\Gamma_{\phi\phi \to ss}\simeq\frac{(g^r_2)^2\rho_\phi(t)}{8\pi m^3_\phi(t)}~~~\mbox{for}~~ g^r_2 \phi^2 s^2 .
  \end{array}\right.
\end{eqnarray}
Where $m_\phi(t)$ be the time-dependent inflaton mass which is defined as $ m^2_\phi(t)=V^{''}(\phi_0(t))$. Since reheating happens near the minimum of the potential, we can expand the inflaton potential (Eq.\ref{pot1}) in the limit $\phi\ll M_p$ as
\begin{equation}
 V(\phi)\simeq\Lambda^4\beta^{2n}\phi^{2n}
\end{equation}
where $\beta=\sqrt{\frac{2}{3\alpha}}\frac{1}{M_p}$. The field-dependent mass becomes, 
\begin{equation}{\label{m1}}
 m^2_{\phi}=V^{''}(\phi_0)\simeq2n(2n-1)\Lambda^4\beta^2\left(\frac{{}V(\phi_0)}{\Lambda^4}\right)^{1-\frac{1}{n}} .
\end{equation}
Using the envelope approximation at any instant of time, the envelope value of the field $\phi_0$ must satisfy $V(\phi_0)\simeq \rho_\phi(t)$. Under this approximation, the inflaton mass at the inflation end is given by
\begin{equation}{\label{end}}
( m^{end}_{\phi})^2\simeq2n(2n-1)\Lambda^4\beta^2\left(\frac{{}\rho_\phi^{end}}{\Lambda^4}\right)^{1-\frac{1}{n}} .
\end{equation}
After utilizing Eqns.(\ref{m1}), (\ref{end}) and (\ref{i1}), one can find the evolution of the inflaton mass as 
\begin{equation}\label{mphit}
 m^2_{\phi}(t)=( m^{end}_{\phi})^2\left(\frac{\rho_\phi(t)}{\rho^{end}_{\phi}}\right)^{1-\frac{1}{n}}\simeq( m^{end}_{\phi})^2A^{-6w_\phi}\,.
\end{equation}
Inserting the above equation in Eqn.(\ref{a1}), we can re-write decay width in terms of the scale factor as
\begin{equation}{\label{A2}}
    \begin{aligned}
\Gamma_{\phi\rightarrow ss}=\frac{(g^r_1)^2}{8\pi m^{end}_\phi }A^{3w_\phi},\,\Gamma_{\phi\phi\rightarrow ss}=\frac{(g^r_2)^2\rho^{end}_\phi}{8\pi (m^{end}_\phi)^3 }A^{-3(1-2w_\phi)} .
    \end{aligned}
\end{equation}
Upon substitution of the above decay width (Eqn.\ref{A2}) in Eqn.\ref{A1}, we have the evolution of the energy density associated with the bosonic non-gravitational channel as
\begin{eqnarray}{\label{A3}}
    d(\rho^r_sA^4)&=\left\{\begin{array}{ll}
         &\frac{3M^2_p(1+w_\phi)H_{end}}{8\pi m^{end}_\phi}(g^r_1)^2A^{\frac{3}{2}(1+w_\phi)}\,dA~~~\mbox{for}~~ g^r_1 \phi s^2\,, \\
         & \frac{9M^4_p(1+w_\phi)H^3_{end}}{8\pi (m^{end}_\phi)^3}(g^r_2)^2A^{\frac{3}{2}(3w_\phi-1)}\,dA~~~\mbox{for}~~ g^r_2 \phi^2 s^2 \,,\\
    \end{array}\right.
\end{eqnarray}
where, $H_{end}= \sqrt{\rho^{end}_\phi/3 M^2_p}$. Since during reheating, the background dynamics is mainly governed by the inflaton field, we approximate $H\simeq \sqrt{{\rho_\phi}/{3 M^2_p}} = H_{end} (a/a_{end})^{-(3+3w_\phi)/2}$, which is used to determine the above equation. After straightforward integration of Eq.(\ref{A3}), we obtain,
\begin{eqnarray}{\label{s1}}
&& \rho^r_s(A) = \left\{\begin{array}{ll} 
&\frac{6 M^2_p(1+w_\phi)H_{end}(g^r_1)^2}{8\pi(5+3w_\phi) m^{end}_\phi A^4} \left(A^{\frac{5+3w_\phi}{2}}-1\right) ~~~~~~~\mbox{for}~~ g^r_1 \phi s^2\,,\\
&\frac{9 M^4_p(1+w_\phi)H^3_{end}(g^r_2)^2}{4\pi(9w_\phi-1)( m^{end}_\phi)^3 A^4} \left(A^{\frac{9w_\phi-1}{2}}-1\right) ~~~~~~~\mbox{for}~~ g^r_2 \phi^2 s^2\,.
\end{array} \right. \\\nonumber
\end{eqnarray}
We will follow the similar procedure to find the expression of gravitational radiation energy density, and the production width associated with the gravitational sector can be expressed as
\begin{equation}{\label{Gdw}}
\Gamma_{\phi\phi\rightarrow RR}=\frac{3M^2_pH^2_{end}m^{end}_\phi}{1024\pi M^4_p}A^{-3-6w_\phi}
\end{equation}
Combining Eqns.(\ref{B3}) and (\ref{Gdw}), we have the following solution for the radiation energy density associated with the gravitational sector
\begin{equation}{\label{gr1}}
 \rho_{gr}^{r}(A)=\frac{9(1+w_\phi)H^3_{end}m^{end}_\phi}{512\pi(1+15w_\phi)A^4 }\left(1-A^{-\frac{1+15w_\phi}{2}}\right)\,.
\end{equation}
The total energy density of the radiation bath is simply sum of the contribution from both gravitational and non-gravitational sector $\rho^{r}_{tot}(A)=\rho^r_s(A)+\rho_{gr}^r(A)$ and the corresponding radiation temperature  $T_{rad}=\left({30 \rho^r_{tot}}/({{\pi^2}g_\star)}\right)^{1/4}$. Where $g_\star$ is the total relativistic degrees of freedom associated with the thermal bath. Now let us find the maximum radiation energy density for both gravitational and non-gravitational sectors separately.\\
%\begin{equation}
% \rho^{Rad}_{tot}(A)=\frac{6 M^2_p(1+w_\phi)H_{end}}{8\pi m^{end}_\phi A^4} g^2\left(\frac{A^{\frac{5+3w_\phi}{2}}-1}{5+3w_\phi}\right)+\frac{9(1+w_\phi)(1+\delta)H^3_{end}m^{end}_\phi}{512\pi(1+15w_\phi)A^4 }\left(1-A^{-\frac{1+15w_\phi}{2}}\right)
%\end{equation}
%\begin{equation}{\label{gmb}}
%\begin{aligned}
 %T_{rad}(A)&=\left[\frac{6 M^2_p(1+w_\phi)H_{end}}{8\pi\epsilon(5+3w_\phi) m^{end}_\phi A^4} (g^r_1)^2\left(A^{\frac{5+3w_\phi}{2}}-1\right)\right.\\
 %&~~~~~~~~~~~~\left.+\frac{9(1+w_\phi)(1+\gamma)H^3_{end}m^{end}_\phi}{512\pi\epsilon(1+15w_\phi)A^4 }\left(1-A^{-\frac{1+15w_\phi}{2}}\right)\right]^{1/4}\mbox{for}~~ g^r_1 \phi s^2\\\\
 %\end{aligned}
%\end{equation}
%\begin{equation}{\label{gmb1}}
%\begin{aligned}
% T_{rad}(A)&=\left[\frac{9 M^4_p(1+w_\phi)H^3_{end}(g^r_2)^2}{4\pi\epsilon(9w_\phi-1)( m^{end}_\phi)^3 A^4} \left(A^{\frac{9w_\phi-1}{2}}-1\right)\right.\\
% &~~~~~~~~~~~~\left.+\frac{9(1+w_\phi)(1+\gamma)H^3_{end}m^{end}_\phi}{512\pi\epsilon(1+15w_\phi)A^4 }\left(1-A^{-\frac{1+15w_\phi}{2}}\right)\right]^{1/4}\mbox{for}~~ g^r_2 \phi^2 s^2\\\\
% \end{aligned}
%\end{equation}
%where $\epsilon=\frac{\pi^2}{30}g_\star$ and $g_\star$ is the total degrees of freedom of the SM bath.
At the beginning of the reheating,  the individual components of radiation (both non-gravitational and gravitational) set to be $\rho_s^r(A=1)=\rho_{gr}^r(A=1)=0$. Within a very short time, during the initial stage of reheating, both the components attain a maximum value  when $\frac{d\,\rho_{s/gr}^r}{d\,A}=0$, which is associated with a specific value of the scale factor 
 \begin{eqnarray}\label{amaxnthboson}
&& A_{max} = \left\{\begin{array}{ll}
&\left[\frac{8}{3(1-w_\phi)}\right]^{\frac{2}{5+3w_\phi}}~~~~\mbox{for}~~ g^r_1 \phi s^2\,,\\
&\left[\frac{9(1-w_\phi)}{8}\right]^{\frac{2}{1-9w_\phi}}~~~~\mbox{for}~~ g^r_2 \phi^2 s^2\,,\\
&\left[\frac{9+15w_\phi}{8}\right]^{\frac{2}{1+15w_\phi}}~~~~\mbox{for gravitational reheating \,.} 
\end{array}\right.
\end{eqnarray}
That, in turn, fixes the maximum value of the  radiation component as
\begin{eqnarray}{\label{mm2}}
 &&\rho^{r,\,max}_{s}=\left\{
 \begin{array}{ll}
 &\frac{6M^2_p(1+w_\phi)H_{end}}{8\pi m^{end}_\phi(5+3w_\phi)}(g^r_1)^2\left[\left(\frac{8}{3(1-w_\phi)}\right)^{\frac{3(w_\phi-1)}{(5+3w_\phi)}}-\left(\frac{8}{3(1-w_\phi)}\right)^{-\frac{8}{5+3w_\phi}}\right]~\mbox{for}~~ g^r_2 \phi^2 s^2\,,\\
 &\frac{9 M^4_p(1+w_\phi)H^3_{end}(g^r_2)^2}{4\pi(9w_\phi-1)( m^{end}_\phi)^3 A^4} \left[\left(\frac{9(1-w_\phi)}{8}\right)^{\frac{9(w_\phi-1)}{1-9w_\phi}}-\left(\frac{9(1-w_\phi)}{8}\right)^{\frac{-8}{1-9w_\phi}}\right] ~~~~\mbox{for}~~ g^r_2 \phi^2 s^2\,,\\
 \end{array}\right.
\end{eqnarray}

\begin{equation}{\label{gm}}
 \rho^{r,\,max}_{gr}=\frac{9(1+w_\phi)H^3_{end}m^{end}_\phi}{512\pi(1+15w_\phi)}\left[\left(\frac{9+15w_\phi}{8}\right)^{-\frac{8}{1+15w_\phi}}-\left(\frac{9+15w_\phi}{8}\right)^{-\frac{9+15w_\phi}{1+15w_\phi}}\right]\,.
 \end{equation}
 Since the gravitational scattering process is a kind of irreducible background, which will be present regardless of the coupling-dependent decay channel. However, a critical coupling exists $\mathscr{G}^1_{ci}$, above which the non-gravitational decay channel always controls the reheating dynamics. The maximum gravitational production happens initially, and the temperature associated with it falls off faster than the non-gravitational one. Thus, the critical coupling $\mathscr{G}^1_{ci}$, above which value the non-gravitational sector dominates throughout reheating, is defined from the condition $\rho_s^{r,\, max}=\rho_{gr}^{r,\,max}$. Since in the gravitational sector domination, the maximum radiation temperature comes out to be $10^{11}\to10^{12}$ GeV, which is less than the inflaton mass $m_\phi (A_{max})$; thermal effect not in working situation. Therefore the expression for critical coupling for with and without thermal effect be the same $\mathscr{G}^{1,\,th}_{ci}=\mathscr{G}^1_{ci}$ (see, for instance, Eqn.\ref{gcr1} ). In the limit of $g_{ci}>\mathscr{G}^1_{ci}$, the maximum radiation temperature is followed by Eqn.\ref{mm2}, whereas, for $g_{ci}<\mathscr{G}^1_{ci}$ gravitational sector determines the maximum radiation temperature (see, Eqn.\ref{gm}).\\
If the bosonic coupling strength is less than $\mathscr{G}^1_{ci}$, the gravitational sector can not reheat the universe before BBN energy scale for the equation of state $w_\phi<0.65$. Therefore those $(g_i^r,\,w_\phi)$ parameter space are forbidden from BBN constraints (see, for instance, light red region of Fig.\ref{couplingbound}). However, for $w_\phi>0.65$, the gravitational sector can successfully reheat the universe and sets the lower bound of reheating temperature that is defined in Eq.(\ref{tgr}). There exists another critical coupling parameter $\mathscr{G}^2_{ci}$, indicating below which only the gravitational sector defines the reheating temperature (see, for instance, light cyan region of Fig.\ref{couplingbound}). We have the expression of $\mathscr{G}^2_{ci}$ as follows
\begin{eqnarray}
&&\mathscr{G}^2_{ci}=\left\{\begin{array}{ll}
&\left[\left(\frac{3(5+3w_\phi)(H_{end}m^{end}_\phi)^2}{128M^2_p(1+15w_\phi)}\right)\left(\frac{512\pi M^2_p(1+15w_\phi)}{3H_{end}m^{end}_\phi(1+w_\phi)}\right)^{\frac{5+3w_\phi}{2(3w_\phi-1)}}\right]^{1/2}~\mbox{ for $g^r_1\phi s^2$ and $w_\phi\geq 0.65$}\,,\\
&\left[\left(\frac{(9w_\phi-1)(m^{end}_\phi)^4}{64M^4_p(1+15w_\phi)}\right)\left(\frac{512\pi M^2_p(1+15w_\phi)}{3H_{end}m^{end}_\phi(1+w_\phi)}\right)^{\frac{1-9w_\phi}{2(3w_\phi-1)}}\right]^{1/2}~\mbox{ for $g^r_2\phi^2 s^2$ and $w_\phi\geq 0.65$}\,.
\end{array}\right.
\end{eqnarray}
and when the gravitational sector controls the reheating temperature, the reheating temperature can be expressed as
\begin{equation}{\label{tgr}}
 T^{gr}_{re}=\left(\frac{9H^3_{end}m^{end}_\phi(1+w_\phi)}{512\epsilon\pi(1+15w_\phi)(A^{gr}_{re})^4}\right)^{1/4},A^{gr}_{re}=\left(\frac{512\pi M^2_p(1+15w_\phi)}{3H_{end}m^{end}_\phi(1+w_\phi)}\right)^{\frac{1}{3w_\phi-1}}\,,
\end{equation}
where $A^{gr}_{re}$ is the normalized scale factor at the end of gravitational reheating. When the coupling parameter $g^r_i>\mathscr{G}^2_{ci}$, the decay channel controls the reheating temperature and that can be defined as
\begin{eqnarray}
&&T_{re}=\left\{\begin{array}{ll}
&\left(\frac{6M^2_p(1+w_\phi)H_{end}}{8\pi\epsilon(5+3w_\phi) m^{end}_\phi}(g^r_1)^2\right)^{1/4}A^{-\frac{3}{8}(1-w_\phi)}_{re}~~~~~\mbox{for } g^r_1\phi s^2\,,\\
&\left(\frac{9 M^4_p(1+w_\phi)H^3_{end}}{4\pi\epsilon(9w_\phi-1)( m^{end}_\phi)^3}(g^r_2)^2\right)^{1/4} A^{\frac{9(w_\phi-1)}{8}}_{re}~~~~~\mbox{for } g^r_2\phi^2 s^2\,,\\
\end{array}\right.
\end{eqnarray} 
where
\begin{eqnarray}
&&A_{re}=\left\{\begin{array}{ll}
&\left(\frac{4\pi(5+3w_\phi)H_{end}m^{end}_\phi}{(1+w_\phi)(g^r_1)^2}\right)^{\frac{2}{3+9w_\phi}}~~~~~~~~\mbox{for } g^r_1\phi s^2\,,\\
&\left(\frac{4\pi(9w_\phi-1)(m^{end}_\phi)^3}{3(1+w_\phi)M^2_pH_{end}(g^r_2)^2}\right)^{\frac{2}{3(5w_\phi-1)}}~~~~~\mbox{for } g^r_2\phi^2 s^2\,.
\end{array}\right.
\end{eqnarray}\\
\subsection{With thermal effect} If the radiation temperature is much smaller than the inflaton mass scale, we can safely use the zero-temperature decay width. But, for the cases where radiation temperature is much higher than the inflaton mass scale, the thermal effect becomes very efficient. In the limit $T_{rad}>>m_\phi(t)$, the inflaton decay rate for different decay channels can be expressed as
\begin{equation}
    \begin{aligned}
             &\Gamma_{\phi\rightarrow ss}=\frac{(g^r_1)^2}{8\pi m_\phi(t) }\frac{4T_{rad}}{m_\phi(t)}\,,\\
       &\Gamma_{\phi\phi\rightarrow ss}=\frac{(g^r_2)^2\rho_\phi(t)}{8\pi (m_\phi(t))^3 }\frac{2T_{rad}}{m_\phi(t)}\,.\\
    \end{aligned}
\end{equation}
When the bosonic non-gravitational channel dominates over the gravitational one, we can approximate radiation temperature  as $T_{rad}=\left(\frac{\rho^{r}_{tot}}{\epsilon}\right)^{\frac{1}{4}}\simeq\left(\frac{\rho^{r}_{s}}{\epsilon}\right)^{\frac{1}{4}}$,where $\epsilon=\frac{\pi^2}{30}g_*(T_{rad})$. And under this consideration utilizing Eq.(\ref{mphit}), one can find the decay width as
\begin{equation}{\label{tb1}}
\begin{aligned}
       &\Gamma_{\phi\rightarrow ss}=\frac{4(g^r_1)^2}{8\pi\epsilon^{1/4} (m^{end}_\phi)^2 }\frac{\left(\rho_{s}^{r}(A)A^4\right)^{1/4}}{A^{1-6w_\phi}}\\
        &\Gamma_{\phi\phi\rightarrow ss}=\frac{2(g^r_2)^2}{8\pi\epsilon^{1/4} (m^{end}_\phi)^4 }\frac{\left(\rho_{s}^{r}(A)A^4\right)^{1/4}}{A^{3-9w_\phi}}
    \end{aligned}
    \end{equation}
    
Further Combining Eq.(\ref{A1}) and Eq. (\ref{tb1}), we have   
\begin{eqnarray}
  d(\rho_{s}^{r}(A)A^4)=&&\left\{\begin{array}{cc}
       & \frac{12(g^r_1)^2M^2_p(1+w_\phi)H_{end}}{8\pi \epsilon^{1/4}(m^{end}_\phi)^2}A^{\frac{1+9w_\phi}{2}}\left(\rho_{s}^{r}(A)A^4\right)^{1/4}dA\,, \\
       &\frac{18(g^r_2)M^4_p(1+w_\phi)H^3_{end}}{8\pi \epsilon^{1/4}(m^{end}_\phi)^4}A^{\frac{5(3w_\phi-1)}{2}}\left(\rho_{s}^{r}(A)A^4\right)^{1/4}dA\,.
  \end{array}\right.
\end{eqnarray}
Straightforward integration of the above equation, radiation energy density takes the following form 
\begin{equation}{\label{the1}}
\begin{aligned}
\rho_{s}^{r}(A)&=\left[\frac{3M^2_p(1+w_\phi)H_{end}}{4\pi\epsilon^{1/4}(1+3w_\phi)(m^{end}_\phi)^2A^3}{(g^r_1)}^2\left(A^{\frac{3(1+3w_\phi)}{2}}-1\right)
%&~~~~~~~~~~+\left.\left.\frac{3(1+\gamma)H^2_{end}(m^{end}_\phi)^2}{256M^2_p}(A^{-3}-A^{\frac{-9(1+w_\phi)}{2}})\right)
\right]^{4/3}~~\mbox{for}~~ g^r_1\phi s^2\,,\\
%&\simeq\left[\frac{3M^2_p(1+w_\phi)H_{end}}{4\pi\epsilon^{1/4}(1+3w_\phi)(m^{end}_\phi)^2}{(g^r_1)}^2A^{\frac{-3(1-3w_\phi)}{2}}\right]^{4/3}~~\mbox{when}~~A\gg 1 \,,
\end{aligned}
\end{equation}
\begin{equation}{\label{the2}}
\begin{aligned}
\rho_{s}^{r}(A)&=\left[\frac{9M^4_p(1+w_\phi)H^3_{end}}{8\pi\epsilon^{1/4}(5w_\phi-1)(m^{end}_\phi)^4A^3}{(g^r_2)}^2\left(A^{\frac{3(-1+5w_\phi)}{2}}-1\right)
%&~~~~~~~~~~+\left.\left.\frac{(1+\gamma)(5w_\phi-1)(m^{end}_\phi)^4}{128(1+3w_\phi) M^4_p}(A^{-3}-A^{\frac{-9(1+w_\phi)}{2}})\right)
\right]^{4/3}~~\mbox{for}~~ g^r_2\phi^2 s^2\,.\\
%&\simeq\left[\frac{9M^4_p(1+w_\phi)H^3_{end}}{8\pi\epsilon^{1/4}(5w_\phi-1)(m^{end}_\phi)^4}{(g^r_2)}^2A^{\frac{-3(3-5w_\phi)}{2}}\right]^{4/3}~~\mbox{when}~~A\gg 1 \,,
\end{aligned}
\end{equation}
These aforementioned equations represent the modified expression of the bosonic radiation energy density when the thermal effect is functional ($T_{rad}>>m_\phi(t)$). The maximum value of this radiation component associated with the scale factor $A_{max}$ at which the $d\rho_s^r/dA=0$. We have the maximum radiation energy density
 \bea{\label{a2cth}}
\rho_{s}^{r,\,max}\simeq =\left\{
\begin{array}{ll}
	&\left[\frac{3M^2_p(1+w_\phi)H_{end}}{4\pi\epsilon^{1/4}(1+3w_\phi)(m^{end}_\phi)^2}(g^r_1)^2\bigg\{\left(\frac{2}{1-3w_\phi}\right)^{\frac{3w_\phi-1}{3+9w_\phi}}-\left(\frac{2}{1-3w_\phi}\right)^{\frac{-2}{3+9w_\phi}}\bigg\} \right]^{4/3}~~~~~\mbox{for}~~~g^r_1\phi s^2\,,\\
&\left[\frac{9M^4_p(1+w_\phi)H^3_{end}}{8\pi\epsilon^{1/4}(5w_\phi-1)(m^{end}_\phi)^4}{(g^r_2)}^2\bigg\{\left(\frac{2}{3-5w_\phi}\right)^{\frac{5w_\phi-3}{5w_\phi-1}}-\left(\frac{2}{3-5w_\phi}\right)^{\frac{-2}{5w_\phi-1}}\bigg\}\right]^{4/3}~~~~~\mbox{for}~~~g^r_2\phi^2 s^2\,,\\
\end{array} \right.
\eea
and the scale factor $A_{max}$ associated with the maximum radiation energy density can be expressed as
\begin{eqnarray}{\label{amaxth}}
 && A_{max}=\left\{
 \begin{array}{ll}
 &\left(\frac{2}{1-3w_\phi}\right)^{\frac{2}{3(1+3w_\phi)}}~~~~~\mbox{for}~~~g^r_1\phi s^2\,,\\
 &\left(\frac{2}{3-5w_\phi}\right)^{\frac{2}{3(5w_\phi-1)}}~~~~\mbox{for}~~g^r_2\phi^2 s^2\,.
 \end{array}\right.
 \end{eqnarray}
 Now let us define reheating temperature in this context. For the coupling $g^r_1\phi s^2$, we have
 \begin{equation}
\begin{aligned}
T_{re}=\left[\frac{3M^2_p(1+w_\phi)H_{end}(g^r_1)^2}{4\pi\epsilon(1+3w_\phi)(m^{end}_\phi)^2} A_{re}^{-\frac{3}{2}(1-3w_\phi)}\right]^{1/3},\,A_{re}=\left(\frac{4\pi(1+3w_\phi)(m^{end}_\phi)^2}{(1+w_\phi)(g^r_1)^2(\frac{\epsilon}{3})^{1/4}}\left(\frac{H_{end}}{M_p}\right)^{1/2}\right)^{\frac{4}{3(1+9w_\phi)}}\,.
\end{aligned}
\end{equation}
Whereas for the coupling $g_2^r\phi^2 s^2$
\begin{equation}
\begin{aligned}
T_{re}&=\left[\frac{9M^4_p(1+w_\phi)H^3_{end}\,(g^r_2)}{8\pi\epsilon(5w_\phi-1)(m^{end}_\phi)^4}^2A^{\frac{-3(3-5w_\phi)}{2}}_{re}\right]^{1/3},\,A_{re}=\left(\frac{8\pi(27\epsilon)^{1/4}(5w_\phi-1)}{9(1+w_\phi)M^{5/2}_pH^{3/2}_{end}}(m^{end}_\phi)^4\right)^{\frac{4}{3(13w_\phi-3)}}\,.
\end{aligned}
\end{equation}
 \section{Fermionic Reheating: analytical studies}\label{apfer}
\subsection{Without thermal effect} 
 In the absence of thermal effect, the decay width for $\phi\to\bar{f}f$ can be written as \cite{Garcia:2020wiy}
 \begin{equation}
  \Gamma_{\phi\rightarrow\bar ff}=\frac{({h^r})^2}{8\pi}m_\phi(t)\simeq\frac{({h^r})^2}{8\pi}m^{end}_\phi A^{-3w_\phi} .
 \end{equation}
To derive the above equation, we utilized Eqn.\ref{mphit}. Upon substitution of the above decay width in Eqn.\ref{B2}, the evolution equation for the radiation energy density arising from the fermionic non-gravitational coupling can be expressed as
\begin{equation}
 d \left(\rho_f^r(A)\,A^4\right)=\frac{3M^2_p(1+w_\phi)m^{end}_\phi H_{end}}{8\pi} (h^r)^2A^{\frac{3}{2}(1-3w_\phi)}\,dA\,.
\end{equation} 
After integrating the above equation, we have 
\begin{equation}\label{radevwth}
    \rho_f^r(A)=\frac{6M^2_p(1+w_\phi)m^{end}_\phi H_{end}}{8\pi(5-9w_\phi)A^4} (h^r)^2(A^{\frac{5-9w_\phi}{2}}-1)\,.
\end{equation}
Similarly, we can find $\rho_{gr}^r$, the energy density associated with the gravitational sector (see, for instance, Eqn.\ref{gr1}). We can follow the Bosonic reheating section in Appendix-\ref{bosonanawth} for detailed derivation. The total energy density of the radiation bath is simply sum of the radiation components originated from both gravitational and non-gravitational sector,  $\rho^{r}_{tot}(A)=\rho_f^r(A)+\rho_{gr}^r(A)$ 
\begin{equation}
 \rho^{r}_{tot}(A)=\frac{6M^2_p(1+w_\phi)m^{end}_\phi H_{end}}{8\pi(5-9w_\phi)A^4} (h^r)^2(A^{\frac{5-9w_\phi}{2}}-1)+\frac{9(1+w_\phi)(1+\gamma)H^3_{end}m^{end}_\phi}{512\pi(1+15w_\phi)A^4 }\left(1-A^{-\frac{1+15w_\phi}{2}}\right)\,,
\end{equation}
and the radiation temperature $T_{rad}=\left(\rho_{tot}^r/\epsilon\right)^{1/4}$. After the end of the inflation, the energy budget associated with the thermal bath begins to increase from zero initial value and attain a maximum value when $d\rho_{f/gr}^r/dA=0$, then starts to fall off. We obtain the maximum radiation energy density corresponding to both gravitational and non-gravitational sector as
\begin{eqnarray}{\label{mm1}}
 &&\rho^{r,\,max}_{f/gr}=\left\{
 \begin{array}{ll}
 &\left(\frac{6M^2_p(1+w_\phi)m^{end}_\phi H_{end}}{8\pi(5-9w_\phi)} (h^r)^2\right)\left[\left(\frac{8}{3+9w_\phi}\right)^{\frac{-3(1+3w_\phi}{5-9w_\phi}}-\left(\frac{8}{3+9w_\phi}\right)^{\frac{-8}{5-9w_\phi}}\right]~\mbox{for}~~ h^r \phi\bar{f}f\,,\\
 &\frac{9(1+w_\phi)H^3_{end}m^{end}_\phi}{512\pi(1+15w_\phi)}\left[\left(\frac{9+15w_\phi}{8}\right)^{-\frac{8}{1+15w_\phi}}-\left(\frac{9+15w_\phi}{8}\right)^{-\frac{9+15w_\phi}{1+15w_\phi}}\right] ~\mbox{for}~\mbox{gravitational scattering}\,.\\
 \end{array}\right.
\end{eqnarray}
And the scale factor correlated with these maximum energy densities turns out as
 \begin{eqnarray}\label{amaxnth}
&& A_{max} = \left\{\begin{array}{ll}
&(\frac{8}{3+9w_\phi})^{2/{5-9w_\phi}}~~~~\mbox{for}~~ h^r \phi\bar{f}f\,,\\
&\left[\frac{9+15w_\phi}{8}\right]^{\frac{2}{1+15w_\phi}}~~~~\mbox{for gravitational scattering } 
\end{array}\right.\,.
\end{eqnarray}
Since the gravitational scattering process is always present, there exists a critical value of coupling below which the gravitational sector always defines maximum radiation temperature $T_{rad}^{max}$. Equating $\rho_f^{r,\,max}=\rho_{gr}^{r,\,max}$, one can find the expression for the critical coupling for fermionic reheating as 
\bea
\mathscr{H}_c=\left[\frac{3(5-9w_\phi)H_{end}^2}{128M^2_p(1+15w_\phi)}\frac{\left[\left(\frac{9+15w_\phi}{8}\right)^{-\frac{8}{1+15w_\phi}}-\left(\frac{9+15w_\phi}{8}\right)^{-\frac{9+15w_\phi}{1+15w_\phi}}\right]}{\left[\left(\frac{8}{3(1+3w_\phi)}\right)^{-\frac{3(1+3w_\phi)}{5-9w_\phi}}-\left(\frac{8}{3(1+3w_\phi)}\right)^{-\frac{8}{5-9w_\phi}}\right]}\right]^{1/2}\quad\textit{$w_\phi\neq 5/9$}\,.
\eea
Therefore, in the limit of $h^r>\mathscr{H}_c$, $T_{rad}^{max}$ can be approximated as $T_{rad}^{max}\simeq \left(\rho_f^{r,\,max}/\epsilon \right)^{1/4}$, whereas for $h^r<\mathscr{H}_c$, $T_{rad}^{max}\simeq \left(\rho_{gr}^{r,\,max}/\epsilon \right)^{1/4}$.\\ Another interesting fact is that the behavior of the radiation component associated with the non-gravitational sector is different for $w_\phi<5/9$ and $w_\phi>5/9$ (see, for instance, Eq.\ref{radferwth}). For $w_\phi>5/9$, the radiation component decays as $A^{-4}$ as the gravitational sector. Thus, if your Yukawa coupling strength $h^r<\mathscr{H}_c$ and $w_\phi>0.65$, then reheating dynamics is governed by the gravitational sector, which is termed as gravitational reheating. Moreover, the reheating temperature is followed by Eq.\ref{tgr}. Otherwise, in all ($h^r,\,w_\phi$) parameter space reheating end is defined via explicit coupling.  In such cases, $T_{re}$ and $A_{re}$ is defined as :\\
 when $w_\phi<5/9$
 \bea\label{trwthm}
T_{re}=\left(\frac{6M^2_p(1+w_\phi)m^{end}_\phi H_{end}}{8\pi\epsilon(5-9w_\phi)} (h^r)^2\right)^{1/4}A_{re}^{\frac{-3(1+3w_\phi}{8}}\quad;\quad A_{re}=\left(\frac{8\pi(5-9w_\phi)H_{end}}{2(1+w_\phi)m^{end}_\phi}(h^r)^2\right)^{\frac{2}{3-3w_\phi}}\,,
\eea
and when $w_\phi>5/9$
\bea
T_{re}=\left(\frac{6M^2_p(1+w_\phi)m^{end}_\phi H_{end}}{8\pi\epsilon(9w_\phi-5)} (h^r)^2\right)^{1/4}A_{re}^{-1}\quad;\quad A_{re}=\left(\frac{8\pi(9w_\phi-5)H_{end}}{2(h^r)^2(1+w_\phi)m^{end}_\phi}\right)^{\frac{-1}{1-3w_\phi}}\,.
\eea
\subsection{With thermal effect}
Here we are mainly focused on the situation where $T_{rad}>m_\phi$ and the analysis for $T_{rad}<m_\phi$ will be the same as without thermal effect. In the limit, $T_{rad}> m_\phi$, the thermal effect significantly impacts the reheating dynamics due to the explicit temperature dependence on the  decay rate. The expression for decay rate in this limit can be written as (see, for instance, Eq.\ref{t1})
\begin{equation}{\label{mdr}}
   \Gamma_{\phi\rightarrow \bar ff}=\frac{(h^r)^2}{8\pi}\frac{m^2_\phi(t)}{4\,T_{rad}}\simeq \frac{(h^r)^2}{8\pi}\frac{(m^{end}_\phi)^2}{4\,T_{rad}} A^{-6w_\phi}
\end{equation}
Inserting the above decay rate in Eq.\ref{B2}, one can find the evolution of the radiation energy density for fermionic coupling as
\begin{equation}\label{radferth}
    \rho_f^r(A)=\left[\frac{\zeta(w_\phi)\,(h^r)^2\epsilon^{1/4}M^2_P}{A^5}(m^{end}_\phi)^2H_{end}\left(A^{\frac{7-15w_\phi}{2}}-1\right)\right]^{4/5}\,,
\end{equation}
$\zeta(w_\phi)=\frac{15(1+w_\phi)}{64\pi(7-15w_\phi)}$. The aforementioned equation clearly suggests that the radiation energy density behaves differently depending on the inflaton equation state, whether it is greater than or less than $7/15$. If the coupling strength $h^r>\mathscr{H}_c$, fermionic coupling always defines reheating temperature and eventually equating $\rho_\phi=\rho_f^r$, we find reheating temperature as
\bea 
T_{re}=\left(\frac{6M^2_p(1+w_\phi)m^{end}_\phi H_{end}}{8\pi\epsilon(5-9w_\phi)} (h^r)^2\right)^{1/4}A_{re}^{\frac{-3(1+3w_\phi)}{8}}\quad;\quad A_{re}=\left(\frac{8\pi(5-9w_\phi)H_{end}}{2(1+w_\phi)m^{end}_\phi}(h^r)^2\right)^{\frac{2}{3-3w_\phi}}
\eea
The above equation is valid when $w_\phi<7/15$. However, for $w_\phi>7/15$ we have 
 \begin{equation}
    T_{re}=\left[\frac{\zeta(w_\phi)\,(h^r)^2M^2_P}{\epsilon}(m^{end}_\phi)^2H_{end}A^{-5}_{re}\right]^{1/5},\,A_{re}=\left[\frac{\zeta(w_\phi)\,(h^r)^2\epsilon^{1/4}M^2_p(m^{end}_\phi)^2H_{end}}{(3M^2_pH^2_{end})^{5/4}}\right]^{\frac{5(1-3w_\phi)}{4}}.
    \end{equation}
Moreover, in the limit of $h^r>\mathscr{H}_c$, the maximum radiation energy density is also defined by the non-gravitational sector, which can be expressed as 
\begin{equation}\label{fdthermal}
\rho^{r,\,max}_{f}=\left[\zeta(w_\phi)\,(h^r)^2\epsilon^{1/4}M^2_P(m^{end}_\phi)^2H_{end}\left(A_{max}^{\frac{-3(1+5w_\phi}{2}}-A_{max}^{-5}\right)\right]^{4/5}\,,
\end{equation}
where, $A_{max}=\left(\frac{10}{3+15w_\phi}\right)^{2/{7-15w_\phi}}$. On the other hand, when $h^r<\mathscr{H}_c$ and $w_\phi>0.65$, the gravitational sector drives the reheating dynamics and the reheating temperature to be the same as Eq.\ref{tgr}.
\section{Analytical expressions of co-moving number density for FIMP which are produced from thermal bath}{\label{analytical exp DM}}
For freeze in production of DM from the thermal bath, the DM can never reach thermal equilibrium i.e. $n^r_
x\ll n^r_{x,eq}$, hence the co-moving number density $N^r_x = n^r_xA^3$ follow the following simple equation,
\begin{equation}
    \frac{N^r_x(A)}{dA}=\frac{A^2}{H}\langle\sigma v\rangle (n^{eq}_x)^2 .
\end{equation}
For this case, the DM production continue to happen untill the radiation temperature equals the DM mass $T_{rad} \simeq m_x $. Therefore, in order to solve the above equation, DM can be safely assumed to be relativistic, and in the limit $m_x<T_{rad}$, the equilibrium number density becomes, 
\begin{equation}
    n^{r}_{x,eq}=\frac{j_xT^3_{rad}}{\pi^2}
\end{equation}
The Hubble parameter can also be written as 
\begin{equation}{\label{hrh}}
    H(A)=H(A_{re})\left(\frac{A}{A_{re}}\right)^{-\frac{3}{2}(1+w_\phi)}
\end{equation}
where $H(A_{re})=\frac{\sqrt{\epsilon}T^2_{re}}{\sqrt3M_p}$ hubble parameter at the end of reheating. So, combining above three equations, one can gate
\begin{equation}{\label{cnd2}}
    \frac{N^r_x(A)}{dA}=\frac{\langle\sigma v\rangle j_x^2A^2_{re}}{\pi^4H(A_{re})} \left(\frac{A}{A_{re}}\right)^{\frac{7+3w_\phi}{2}}T^6_{rad}
\end{equation}.
\subsection{Bosonic Reheating}
\subsubsection{Without thermal effect}
From Eq. (\ref{s1}), the bath temperature can be written as
\begin{eqnarray}
T_{rad}=&&\left\{\begin{array}{ll}
     & T_{re}(A/A_{re})^{\frac{-3(1-w_\phi)}{8}}~~~\mbox{for}~~g_1^r\phi s^2 \\
     & T_{re}(A/A_{re})^{\frac{-9(1-w_\phi)}{8}}~~~\mbox{for}~~g_1^r\phi^2 s^2 
\end{array}\right.
\end{eqnarray}
Using above equation in Eq. (\ref{cnd2}), one can find the below solution of $N^r_x(A)$ at arbitrary point $A\leq A_{re}$
\begin{eqnarray}
N^r_x(A)=\frac{4\sqrt3M_p\langle\sigma v\rangle j^2_xT^4_{re}}{3\sqrt\epsilon\pi^4}A^3_{re}&&\left\{\begin{array}{ll}
     & \frac{1}{(3+5w_\phi)}\left[\left(\frac{A}{A_{re}}\right)^{\frac{3(3+5w_\phi)}{4}}-\left(\frac{A}{A_{max}}\right)^{\frac{3(3+5w_\phi)}{4}}\right]~~~\mbox{for}~~g_1^r\phi s^2 \\
     &  \frac{1}{(11w_\phi-3)}\left[\left(\frac{A}{A_{re}}\right)^{\frac{3(11w_\phi-3)}{4}}-\left(\frac{A}{A_{max}}\right)^{\frac{3(11w_\phi-3)}{4}}\right]~~~\mbox{for}~~g_2^r\phi^2 s^2 
\end{array}\right.
\end{eqnarray}
The above equation shows that most of the production of DM happens at the beginning of reheating for $\phi\phi\rightarrow ss$ reheating process for $w_\phi<3/11$.
\subsubsection{With thermal effect}
From Eq. (\ref{the1}), the bath temperature can be written as

\begin{eqnarray}
T_{rad}=&&\left\{\begin{array}{ll}
     & T_{re}(A/A_{re})^{\frac{-(1-3w_\phi)}{2}}~~~\mbox{for}~~g_1^r\phi s^2 \\
     & T_{re}(A/A_{re})^{\frac{-(3-5w_\phi)}{2}}~~~\mbox{for}~~g_1^r\phi^2 s^2 
\end{array}\right.
\end{eqnarray}
Utilising above equation in Eq.(\ref{cnd2}), we have obtained following solution of $N^r_x(A)$ at arbitrary point $A\leq A_{re}$
\begin{eqnarray}
N^r_x(A)=\frac{2\sqrt3M_p\langle\sigma v\rangle j^2_xT^4_{re}}{3\sqrt\epsilon\pi^4}A^3_{re}&&\left\{\begin{array}{ll}
     & \frac{1}{(1+7w_\phi)}\left[\left(\frac{A}{A_{re}}\right)^{\frac{3(1+7w_\phi)}{2}}-\left(\frac{A}{A_{max}}\right)^{\frac{3(1+7w_\phi)}{2}}\right]~~~\mbox{for}~~g_1^r\phi s^2 \\
     &  \frac{1}{(11w_\phi-3)}\left[\left(\frac{A}{A_{re}}\right)^{\frac{3(11w_\phi-3)}{2}}-\left(\frac{A}{A_{max}}\right)^{\frac{3(11w_\phi-3)}{2}}\right]~~~\mbox{for}~~g_2^r\phi^2 s^2 
\end{array}\right.
\end{eqnarray}
Like earlier,  most of the production of DM happens at the beginning of reheating for $\phi\phi\rightarrow ss$ reheating process for $w_\phi<3/11$
\subsection{Fermionic reheating}
\subsubsection{Without thermal effect}
\begin{itemize}
    \item \underline{For $w_\phi>5/9$:} The bath temperature can be written as 
    \begin{equation}
T_{rad}=T_{re}(A/A_{re})^{-1}
\end{equation}
Upon substituting the above equation in Eq.\ref{cnd2} along with Eq. (\ref{hrh}), the co-moving number density at any point $A$ during reheating takes the following form
\begin{equation}
    N^r_x(A)=\frac{\sqrt{3}M_p\langle\sigma v\rangle j^2_xT^4_{re}}{\sqrt{\epsilon}\pi^4}A^2_{re}\int^{A}_{A_{max}}(A/{A_{re}})^{\frac{-5+3w_\phi}{2}}dA=\frac{2\sqrt{3}M_p\langle\sigma v\rangle j^2_xT^4_{re}}{3\sqrt{\epsilon}\pi^4}A^3_{re}(A_{max}/A_{re})^{{3(1-w_\phi)}/2}
\end{equation}
The co-moving number density is constant from the beginning of reheating. 
\end{itemize}
\begin{itemize}
    \item \underline{For $w_\phi<5/9$:} The bath temperature can be written as 
      \begin{equation}
T_{rad}=T_{re}(A/A_{re})^{-\frac{3}{8}(1+3w_\phi)}
\end{equation}
Upon substituting the above equation in Eq.\ref{cnd2} along with Eq. (\ref{hrh}), the co-moving number density at any point $A$ during reheating takes the following form
\begin{equation}
\begin{aligned}
    N^r_x(A)&=\frac{\sqrt{3}M_p\langle\sigma v\rangle j^2_xT^4_{re}}{\sqrt{\epsilon}\pi^4}A^2_{re}\int^{A}_{A_{max}}(A/{A_{re}})^{\frac{5-21w_\phi}{4}}dA\\
    &=\frac{4\sqrt{3}M_p\langle\sigma v\rangle j^2_xT^4_{re}}{3\sqrt{\epsilon}(3-7w_\phi)\pi^4}A^3_{re}\left[(A/A_{re})^{\frac{{3(3-7w_\phi)}}{4}}-(A_{max}/A_{re})^{\frac{{3(3-7w_\phi)}}{4}}\right]
    \end{aligned}
\end{equation}
When $w_\phi>3/7$, the DM production happens instantaneously just at the end of inflation, as a result, the co-moving number density is constant.
\end{itemize}
\subsection{With thermal effect :} When $w_\phi>7/15$, the same situation will be happen as we discuss for without thermal effect for $w_\phi>5/9$. Now  for $w_\phi<7/15$, the bath temperature is
  \begin{equation}
T_{rad}=T_{re}(A/A_{re})^{-\frac{3}{10}(1+5w_\phi)}
\end{equation}
Upon substituting the above equation in Eq.\ref{cnd2} along with Eq. \ref{hrh}, the co-moving number density at any point $A$ during reheating takes the following form
\begin{equation}
\begin{aligned}
    N^r_x(A)&=\frac{\sqrt{3}M_p\langle\sigma v\rangle j^2_xT^4_{re}}{\sqrt{\epsilon}\pi^4}A^2_{re}\int^{A}_{A_{max}}(A/{A_{re}})^{\frac{17-75w_\phi}{4}}dA\\
    &=\frac{10\sqrt{3}M_p\langle\sigma v\rangle j^2_xT^4_{re}}{3\sqrt{\epsilon}(9-25w_\phi)\pi^4}A^3_{re}\left[(A/A_{re})^{\frac{{3(9-25w_\phi)}}{10}}-(A_{max}/A_{re})^{\frac{{3(9-25w_\phi)}}{10}}\right]
    \end{aligned}
\end{equation}
When $w_\phi>9/25$, the DM production happens instantaneously just at the end of inflation. As a result, the co-moving number density is constant.
%%%%%%%%%%%%%%%%%%%%%%%%%%%%%%%%%%%%%%%%%%
\section{Analytical expressions of minimum critical mass where both freeze-in and freeze-out mechanism coincides}{\label{critcal}}
When the DM mass $m_x$ is smaller than $T_{re}$, the DM abundance follows simple relation $\Omega_xh^2\propto m_x\langle\sigma v\rangle$ for the freeze-in scenario for a fixed $T_{re}$. Therefore, the cross-section becomes inversely proportional to the DM mass. This suggests the existence of a critical $\langle\sigma v\rangle$ and the associated mass for which the DM equilibrates with the thermal bath and freeze out happens. At that critical $\langle\sigma v\rangle_{crit}$, DM number density must equate with the equilibrium number density at the end of reheating. Corresponding to these critical $\langle\sigma v\rangle_{crit}$, there exists a minimum critical mass $m_{x,min}$ which satisfies present-day abundance. Below this mass, no mass will be available, which can give the correct abundance. Equating the solution of the DM number density $n^r_x$ with the equilibrium number density $n^r_{x,eq}$ at the end of reheating $A_{re}$, one can find following expressions of $\langle\sigma v\rangle_{crit}$ without thermal effect
\begin{eqnarray}
\langle\sigma v\rangle_{crit}=\frac{\sqrt{3\epsilon}\pi^2}{4M_pj_x}&&\left\{\begin{array}{ll}
     &\frac{3+5w_\phi}{T_{re}}~~~~~~~~~~~~~~~~~~~~~~~~~~\mbox{for}~~~g^r_1\phi s^2\\
      & \frac{11w_\phi-3}{T_{re}}~~~~~~~~~~~~~~~~~~~~~~~~~\mbox{for}~~~g^r_2\phi^2 s^2~~~\mbox{with}~~~ w_\phi>w^c_\phi\\
     &\frac{3-11w_\phi}{T_{re}}\left(\frac{T_{re}}{T_{max}}\right)^{\frac{2(3-11w_\phi)}{3(1-w_\phi)}}~~~\mbox{for}~~~g^r_2\phi^2 s^2~~~\mbox{with}~~~ w_\phi<w^c_\phi\\
    &\frac{2(1-w_\phi)}{T_{re}}\left(\frac{T_{re}}{T_{max}}\right)^{\frac{3(w_\phi-1)}{2}}~~~~\mbox{for}~~~h^r\phi\bar ff ~~~\mbox{with}~~~ w_\phi>5/9\\
     &\frac{7w_\phi-3}{T_{re}}\left(\frac{T_{re}}{T_{max}}\right)^{\frac{2(7w_\phi-3)}{1+3w_\phi}}~~~\mbox{for}~~~h^r\phi\bar ff~~~~~\mbox{with}~~~ 3/7<w_\phi<5/9\\
       &\frac{3-7w_\phi}{T_{re}}~~~~~~~~~~~~~~~~~~~~~~~~~~\mbox{for}~~~h^r\phi\bar ff ~~~\mbox{with}~~~ w_\phi<3/7
\end{array}\right.
\end{eqnarray}
and with thermal effect
\begin{eqnarray}{\label{critsigma}}
\langle\sigma v\rangle_{crit}=\frac{\sqrt{3\epsilon}\pi^2}{4M_pj_x}&&\left\{\begin{array}{ll}
     &\frac{2(1+7w_\phi)}{T_{re}}~~~~~~~~~~~~~~~~~~~~~~~~~~\mbox{for}~~~g^r_1\phi s^2\\
      & \frac{2(11w_\phi-3)}{T_{re}}~~~~~~~~~~~~~~~~~~~~~~~~~\mbox{for}~~~g^r_2\phi^2 s^2~~~\mbox{with}~~~ w_\phi>w^c_\phi\\
     &\frac{2(3-11w_\phi)}{T_{re}}\left(\frac{T_{re}}{T_{max}}\right)^{\frac{2(3-11w_\phi)}{3(1-w_\phi)}}~~~\mbox{for}~~~g^r_2\phi^2 s^2~~~\mbox{with}~~~ w_\phi<w^c_\phi\\
    &\frac{2(1-w_\phi)}{T_{re}}\left(\frac{T_{re}}{T_{max}}\right)^{\frac{3(w_\phi-1)}{2}}~~~~~~\mbox{for}~~~h^r\phi\bar ff ~~~\mbox{with}~~~ w_\phi>7/15\\
     &\frac{2(25w_\phi-9)}{T_{re}}\left(\frac{T_{re}}{T_{max}}\right)^{\frac{(25w_\phi-9)}{1+5w_\phi}}~~~\mbox{for}~~~h^r\phi\bar ff~~~~\mbox{with}~~~ 9/25<w_\phi<7/15\\
       &\frac{2(9-25w_\phi)}{5T_{re}}~~~~~~~~~~~~~~~~~~~~~~~~~~\mbox{for}~~~h^r\phi\bar ff ~~~\mbox{with}~~~ w_\phi<9/25\\
\end{array}\right.
\end{eqnarray}
Above this critical $\langle\sigma v\rangle_{crit}$, the DM can produce only from the freeze-out mechanism. Using the above critical $\langle\sigma v\rangle_{crit}$ in the DM abundance expressions, one can find the following expression of minimum critical mass
\begin{equation}{\label{critmass}}
    m_{x,min}=\frac{\Omega_xh^2}{\Omega_rh^2}\frac{\epsilon\pi^2T_{now}}{j_x}\simeq \frac{2\times10^{-7}( GeV)}{j_x}
\end{equation}
The minimum mass is independent of background reheating dynamics, i.e., reheating temperature and inflation equation of states.
\section{Modified inflaton decay width incorporating inflation oscillation}{\label{modifieddecay}}
After inflation ends, the inflaton field $\phi$ starts to oscillate around the minima of its potential with decreasing amplitude. The character of the oscillations strongly depends on the shape of the potential $V(\phi)$ in the vicinity of the
minimum. The solution of the inflation field $\phi(t)$ can be written as $\phi(t)=\phi_0(t).\mathcal{P}(t)$ \cite{Mambrini:2021zpp}, where $\mathcal{P}(t)$ is a quasi-periodic, fast-oscillating function and $\phi_0(t)$ is the envelope function which is a slowly time-varying function. After taking care of the zero mode inflaton oscillations,  the expression of the decay width for the non-gravitational couplings would be the same; the only difference is that now our mentioned coupling parameters $g^r_i$ and $h^r$ behave as an effective coupling and can be related with the actual coupling parameter $g^r_\textit{i,\,act}$ and $h^r_{act}$ as \cite{Garcia:2020wiy,Ahmed:2022tfm} (see, for instance, Eqn.\ref{t1})
\begin{eqnarray}{\label{t2}}
&&(g^r_{1})^2=(g^r_\textit{1,\,act})^2(2n+2)(2n-1)\gamma\sum^{\infty}_{k=1}k\lvert\mathcal P_k\rvert^2\nonumber\\
&&(g^r_2)^2=(g^r_{\textit{2,\,act}})^22n(2n+2)(2n-1)^2\gamma\sum^{\infty}_{k=1}k\lvert\mathcal P^2_{k}\rvert^2\\\
&&(h^r)^2=(h^r_{act})^2(2n+2)(2n-1)\gamma^3\sum^{\infty}_{k=1}k^3\lvert\mathcal P_k\rvert^2\nonumber\\
&&\gamma=\sqrt{\frac{\pi n}{2n-1}}\frac{\Gamma\left(\frac{1}{2}+\frac{1}{2n}\right)}{\Gamma\left(\frac{1}{2n}\right)}\nonumber\\
&&\omega=\gamma\times m_\phi(t)\nonumber
\end{eqnarray}
In the case of gravitational interactions, the oscillation effect slightly modifies the production rate; however, all of our predictions quantitatively remain the same. The following relation can follow the modified decay rate \cite{Clery:2021bwz,Ahmed:2022tfm}
\begin{eqnarray}
&&\Gamma_{\phi\phi\rightarrow{ss}}^{gr}=\frac{\rho_\phi(t)\omega}{8\pi(1+w_\phi)M^4_p}(1+2f_B(m_\phi/T))\sum^{\infty}_{k=1}k\lvert\mathcal P^{2n}_k \rvert^2\\ &&\Gamma_{\phi\phi\rightarrow{\bar ff}}^{gr}=\frac{\rho_\phi(t)m^2_f}{2\pi(1+w_\phi)\omega M^4_p}(1-2f_F(m_\phi/T))\sum^{\infty}_{k=1}\frac{1}{k}\lvert\mathcal P^{2n}_k \rvert^2
\end{eqnarray}
%\frac 1 8 \Gamma_{\phi\phi\rightarrow{XX}}=
 Where $w$ is the frequency of inflation oscillations and $\mathcal{P}_k$ is the Fourier coefficients in
the expansion of $\mathcal{P}(t)$. In Table-\ref{fouriersum}, we have provided the Fourier summations for different values of $n$.
\begin{table}[h!]\caption{Numerical values of the Fourier summations for different values of $n$:}\label{fouriersum}
\centering
 \begin{tabular}{||c c c c c c||} 
 \hline
 $n(w_\phi)$ & $\sum k\lvert\mathcal P_k\rvert^2$ &$\sum k\lvert\mathcal P^2_k\rvert^2$ & $\sum k^3\lvert \mathcal{P}_k\rvert^2$ & $\sum k\lvert\mathcal P^{2n}_k\rvert^2$ & $\sum\frac{1}{k}\lvert\mathcal P^{2n}_k\rvert^2$ \\ [0.5ex] 
 \hline\hline
 1 (0.0) & $\frac{1}{4}$ & $\frac{1}{8}$ & $\frac{1}{4}$ & $\frac{1}{8}$ & $\frac{1}{32}$ \\ 
 2 (1/3) & 0.229 & 0.125 & 0.241 & 0.141 & 0.030 \\
 3 (0.50)& 0.218 & 0.124 & 0.244& 0.146 & 0.024 \\
 10 (0.82) & 0.191 & 0.114 & 0.286 & 0.149 & 0.007\\
 200 (0.99) & 0.174 & 0.100 & 0.358 & 0.140 & 0.0001 \\ [1ex] 
 \hline
 \end{tabular}
\end{table}
%\begin{equation}
 %   T^{sr}_{re}=\left(\frac{3\alpha b^2}{2\epsilon}\right)^{1/4}\left[\frac{2}{\sqrt{6\alpha}}\frac{\Delta k}{k}d\delta\frac{n+1}{\lvert4-2n\rvert}\right]^{n/2}
%\end{equation}
\begin{figure}[t!] 
 	\begin{center}
 		\includegraphics[width=16.0cm,height=5.50cm]{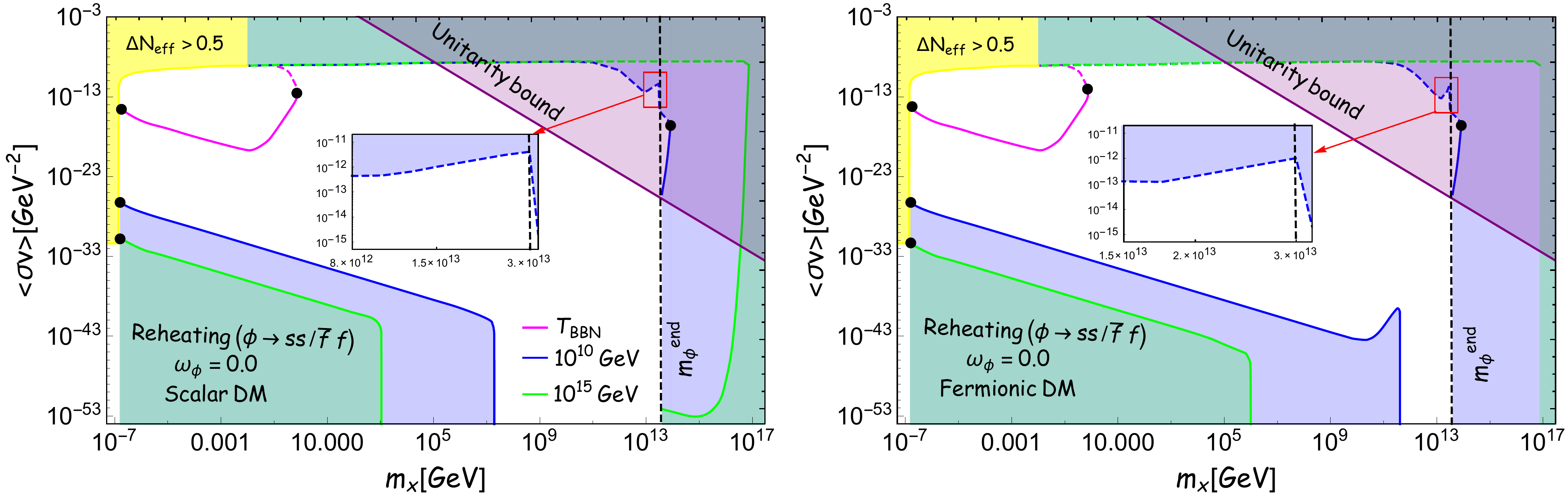}\quad
   	\includegraphics[width=16.0cm,height=5.50cm]{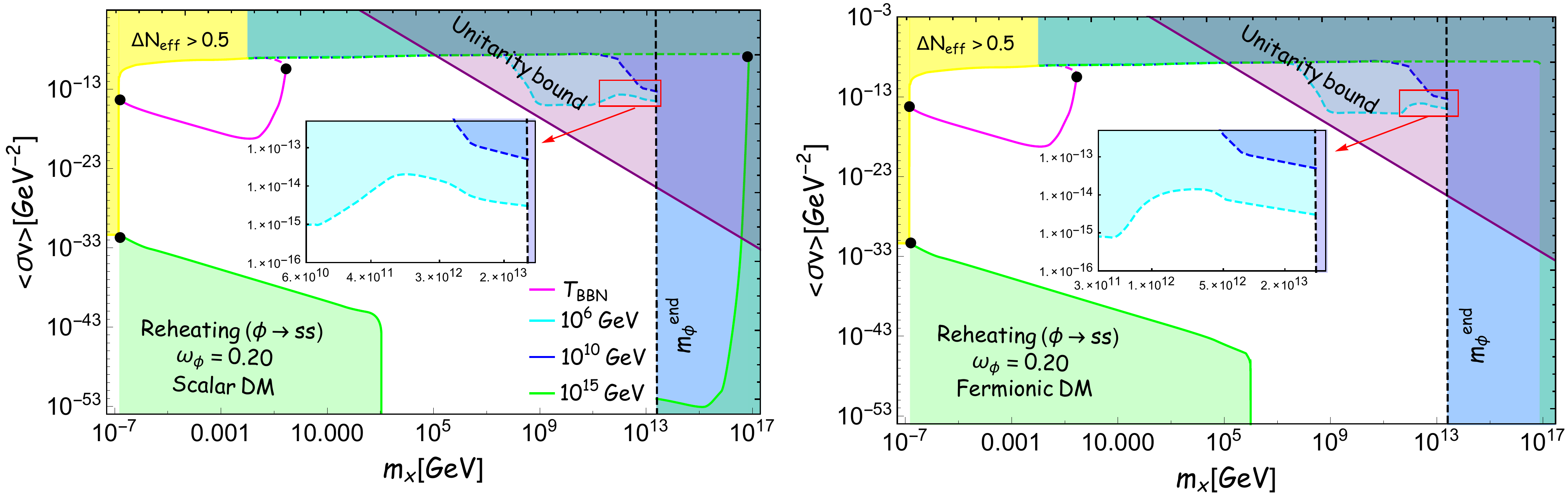}\quad
 		\caption{The description of this plot is the same as Fig.\ref{Fig.12}. Here we have shown the DM parameter space with considering the DM re-annihilation.}
   \label{reanni}
 	\end{center}
 	\end{figure}
\section{The re-annihilation of DM and its effects on DM parameter space}\label{appren}
Near the freeze-out temperature, the thermal production of DM receives Boltzmann suppression, and during this time the gravitational scattering may become the dominant channel for the DM production. For such case following conditions must satisfy
\begin{equation}
    \frac{\Gamma_{\phi\phi\rightarrow SS/FF}(t)\rho_\phi(t)}{m_\phi(t)}\geq\langle\sigma v\rangle(n^r_{x,eq}(t))^2.
\end{equation}
%The DM re-annihilation due to gravitational production of DM from inflaton is important when the gravitational production rate $\frac{\Gamma_{\phi\phi\rightarrow SS/FF}\rho_\phi}{m_\phi}$ is comparable to the DM production rate from radiation bath $\langle\sigma v\rangle(n^r_{x,eq})^2$. 
In our analysis, we ignored such an effect, and it is also important to note that such  an effect is not important when the DMs freeze out during the RD era. However, when the DM freeze-out takes place during reheating, the DM re-annihilation turns out to be important within a very narrow DM mass range ($10^{11}-m_\phi^{end}$) GeV, and its evolution till the Freeze-out point governed by,
\begin{eqnarray}
 \frac{d(a^3 n^r_x)}{dt} = -a^3\langle\sigma v\rangle (n^r_{x})^2 + \frac{a^3 \Gamma_{\phi\phi\rightarrow xx}\rho_\phi (t)}{m_\phi(t)} .
    \label{reannihilation}
\end{eqnarray}
From the equation, it is clear that DM re-annihilation to thermal bath plays a crucial role along with gravitational production. Such re-annihilation is therefore, expected to be important for higher reheating temperatures and large DM masses. The final DM abundance clearly becomes $\Omega_x h^2 \propto \Gamma_{\phi\phi\rightarrow xx}/\langle\sigma v\rangle$. Gravitational production rate of DM increases with DM mass. Therefore, with increasing DM mass, the cross-section has to increase to maintain the correct abundance, which is indeed observed in Fig.\ref{reanni} for $w_\phi =0$, for which the inflaton mass $m_\phi$ is constant. It turns out that most of the parameter regions we found are above the unitarity bound shown in Fig.\ref{reanni}.
For $w_\phi=0$, we observed the effect of re-annihilation for $T_{re}=10^{10}$ GeV. For this temperature, the corresponding DM mass range where this effect modifies the cross-section turns out to be within $(8\times10^{12}-m_\phi^{end})$ GeV for scalar DM and $(1.5\times10^{13}-m_\phi^{end})$ GeV for fermionic DM.

For the equation of state other than zero, the situation becomes different due to time-varying inflaton mass. For higher DM mass possibility may arise that its gravitational production becomes kinematically suppressed due to decreasing inflaton mass. For such cases, the re-annihilation stops, and standard DM freeze-out occurs, which we discussed in detail.
As an example for $w_\phi=0.20$, we indeed see such an effect which we have shown in Fig.\ref{reanni}. One particularly notices the curve associated with $T_{re}=10^6$ in the inset. The raising part of the curve corresponds to the re-annihilation phase for the mass range within $(6\times 10^{10}, 9\times10^{11})$ GeV for scalar DM and $(3\times 10^{11}, 4\times10^{12})$ GeV for fermionic DM. For higher mass values, kinematic suppression comes into play, and the cross-section needs to be reduced, as discussed before. For $T_{re} = 10^{10}$, we found that the re-annihilation is not prominent due to the kinetic suppression effect.

\hspace{0.5cm}
 
\end{document}